\documentclass{jfm}
\usepackage{ulem}
\usepackage{graphicx}
\usepackage[justification=justified, width=\textwidth]{caption} 
\usepackage{newtxtext}
\usepackage{newtxmath}
\usepackage{natbib}
\usepackage{hyperref}
\usepackage{bm}
\usepackage{xcolor}
\hypersetup{
    colorlinks = true,
    urlcolor   = blue,
    citecolor  = black,
}

\newcommand{\be}[1]{\begin{equation}\label{#1}}
\newcommand{\ee}{\end{equation}}
\newcommand{\ba}[1]{\begin{eqnarray}\label{#1}}
\newcommand{\ea}{\end{eqnarray}}
\newcommand{\rf}[1]{(\ref{#1})}
\newcommand{\nn}{\nonumber}

\newcommand{\RomanNumeralCaps}[1]
\linenumbers

\title{Instabilities in visco-thermodiffusive swirling flows}

\author{Oleg N. Kirillov\aff{1}
  \corresp{\email{oleg.kirillov@northumbria.ac.uk}},
  Innocent Mutabazi\aff{2}}

\affiliation{\aff{1}Northumbria University, Newcastle upon Tyne, NE1 8ST, United Kingdom
\aff{2}Laboratoire Ondes et Milieux Complexes, UMR-6294 CNRS, Universit\'e Le Havre Normandie, Normandie Universit\'e, 53 Rue de Prony, 76058 Le Havre C\'edex, France}

\begin{document}
\maketitle

\begin{abstract}

\textcolor{black}{An analytical theory is presented for linear, local, short-wavelength instabilities in swirling flows, in which axial shear, differential rotation, radial thermal stratification, viscosity, and thermal diffusivity are all taken into account. A geometrical optics approach is applied to the Navier–Stokes equations, coupled with the energy equation, leading to a set of amplitude transport equations. From these, a dispersion relation is derived, capturing two distinct types of instability: a stationary centrifugal instability and an oscillatory, visco-diffusive McIntyre instability. Instability regions corresponding to different axial or azimuthal wavenumbers are found to possess envelopes in the plane of physical parameters, which are explicitly determined using the discriminants of polynomials. As these envelopes are shown to bound the union of instability regions associated with particular wavenumbers, it is concluded that the envelopes correspond to curves of critical values of physical parameters, thereby providing compact, closed-form criteria for the onset of instability. The derived analytical criteria are validated for swirling flows modelled by a cylindrical, differentially rotating annulus with axial flow induced by either a sliding inner cylinder, an axial pressure gradient, or a radial temperature gradient combined with vertical gravity. These criteria unify and extend, to viscous and thermodiffusive differentially heated swirling flows, the Rayleigh criterion for centrifugally driven instabilities, the Ludwieg–Eckhoff–Leibovich–Stewartson criterion for isothermal swirling flows, and the Goldreich–Schubert–Fricke criterion for non-isothermal azimuthal flows. Additionally, they predict oscillatory modes in swirling, differentially heated, visco-diffusive flows, thereby generalising the McIntyre instability criterion to these systems.}

\end{abstract}



\section{Introduction}
\label{sec:Introduction}

The stability of swirling flows and their transition to turbulence represent a significant scientific challenge due to their occurrence in diverse natural and industrial settings.

These flows appear in natural phenomena such as tropical cyclones and tornadoes \citep{EM1984,E2018}, \textcolor{black}{rotating convection \citep{L2012,ODD2017,ODD2018,HA2018,CNH2021,SOD2022}}, swirling vortex rings \citep{LH1993,HH2010}, meandering rivers \citep{PA2019}, and astrophysical flows, including magnetic tornadoes in the solar atmosphere \citep{KS1982,BP2006,W2012,LMA2013}.

In engineering applications, swirling flows develop behind aircraft wingtips as trailing vortices \citep{ES1978,LS1983,EC1984,LNOD2001,DPA2010,BG2013,FM2016} and play a crucial role in industrial processes such as combustion \citep{C2014}, isotope separation via centrifugation \citep{L2017}, cooling and lubrication of rotating machinery \citep{K1969,LM1989,F2011,SLW2022}, wastewater purification \citep{OPS1991}, everyday piping systems and physiological flows  \citep{A2016,BA2020}, oil drilling operations \citep{MA1993}, metal solidification \citep{V1988}, and crystal growth \citep{DBPD2010}.

Hydrodynamic modeling describes swirling flows \textcolor{black}{as a result of }
the combined effects of rotation and shear in two orthogonal directions \textcolor{black}{(e.g. azimuthal and axial in cylindrical geometry)}.
\textcolor{black}{Swirling flows can occur either in confined configurations common in engineering applications (e.g. heat transfer systems, motor turbines, etc.) or in open geometries typical of natural phenomena such as atmospheric flows.}

The \textit{circular Couette flow} (CCF), in which two coaxial cylinders can rotate independently, is the simplest system for experimental and theoretical studies of swirling flows \citep{LHS2023, HLS2023}. Axial flow in this system can be induced by various mechanisms, including the sliding motion of the inner cylinder, as in sliding Couette flow \citep{DN2011}; an external pressure gradient, as in annular Poiseuille flow \citep{MFN2008, H2008}; or a radial temperature gradient combined with axial gravity, as in baroclinic convection \citep{BMA2000, LP2007, WC2022}. These mechanisms yield different variants of CCF with axial flow \textcolor{black}{that replicate swirling flows}:

\begin{itemize}
    \item [(SCF)] \textit{Spiral Couette Flow} – where axial flow is driven by inner-cylinder sliding \citep{L1960, L1964, NT1982, AW1993, MM2000}.
    \item [(SPF)] \textit{Spiral Poiseuille Flow} – where axial flow is induced by an external pressure gradient \citep{TJ1981, CP2004, MM2002, MM2005, CM2005, BEAL2023}.
    \item [(BCF)] \textit{Baroclinic Couette Flow} – where a radial temperature gradient and axial gravity cause baroclinic convection \citep{SK1964, AW1990, LG2008, YNM2013, G2015, Kang2015, Kang2023}.
\end{itemize}

\textcolor{black}{Note that in short Couette–Taylor cells and rotating cylindrical tanks, Ekman pumping induced by stationary or rotating endwalls can generate a secondary axial flow in the sidewall Stewartson boundary layer, which in turn gives rise to a spiralling motion within this region \citep{HK1996,LM2010}; see also \cite{GBS1996} and \cite{VS2020}. In the present study, however, we focus on Couette–Taylor cells with an infinite aspect ratio.}

The stability of CCF has been extensively studied experimentally, theoretically, and numerically \citep{LHS2023, HLS2023}, as it provides a well-controlled system for investigating the onset of turbulence. Rayleigh’s inviscid stability criterion states that a flow with curved streamlines becomes unstable if angular momentum decreases outward. However, this criterion must be supplemented by linear stability analysis to account for viscosity and determine the instability threshold as a function of the radii ratio and the relative rotation velocities of the cylinders \citep{C1961}. Additionally, axial flows in a stationary cylindrical annulus are susceptible to wall-driven shear instabilities, with thresholds that depend on the radii ratio \citep{BMA2000, LP2007, MFN2008, H2008, DN2011,  WC2022}.

The presence of a radial temperature gradient and thermal diffusion introduces new destabilization mechanisms in differentially rotating azimuthal flows, such as CCF. These include the visco-thermodiffusive extension of Rayleigh's centrifugal instability, known as the Goldreich-Schubert-Fricke (GSF) instability \citep{AG1978, M2013, KM2017, JT2023}, and the visco-thermodiffusive oscillatory McIntyre instability \citep{M1970, LK2021}. Both instabilities can destabilize Rayleigh-stable flows, including quasi-Keplerian ones, depending on the Prandtl number and the direction of heating \citep{KM2017, MMY2021}.

A radial temperature gradient naturally arises in many applications of swirling flows. For example, recent simulations \citep{ODD2017} of rotating convection between two differentially heated horizontal plates in a shallow cylindrical domain revealed the formation of a vertical swirling base flow with a naturally developed hot core and an outward radial temperature gradient, consistent with observations of tropical cyclones \citep{E2018}. In such non-isothermal visco-thermodiffusive swirling flows, all the previously mentioned linear instabilities may be present, along with \textcolor{black}{algebraically growing} and spatio-temporal instabilities \citep{ES1978,GS2003,GS2003jfm,HP2006,H2008a,MSL2009} and nonlinear effects \citep{SH1988}, potentially leading to complex new interactions.

For isothermal, inviscid, incompressible swirling flows, the following three general analytical criteria for stationary centrifugal instability have been developed in the literature:
\begin{itemize}
\item[-] The Ludwieg criterion for SCF in the narrow-gap limit \citep{L1960,L1964};
\item[-] The Leibovich-Stewartson criterion for swirling jets \citep{LS1983};
\item[-] The Eckhoff criterion, derived using a local geometrical optics approach \citep{EC1984}.
\end{itemize}

\textcolor{black}{The equivalence between the local and global approaches leading to these instability conditions—commonly referred to as the Ludwieg–Eckhoff–Leibovich–Stewartson (LELS) criterion—was formally established by \cite{LL2005}. \cite{EM1984} rederived the LELS criterion by identifying a plane that locally contains all the velocity shear relative to a rotating coordinate system and performing computations in the resulting helical coordinate system. In a later experimental study, \cite{HK1996} used physical arguments to justify a result similar to that of \cite{EM1984}, in the context of centrifugal instability of the helical secondary flow within the Stewartson sidewall boundary layer in a rotating cylinder with a more rapidly corotating upper lid.}

However, the LELS criterion applies only to isothermal, non-stratified, and inviscid swirling flows. In particular, by this reason it does not capture, e.g., the folded neutral stability surface observed in SCF and SPF through numerical linear stability analysis by \cite{MM2000,MM2002,MM2005}, nor does it account for non-isothermal \citep{CM2005,YNM2013} or stratification \citep{DPA2010,YMM2001} effects. \textcolor{black}{Thus, similar to the case of rotating plane shear flows \citep{OK2024}, there is a strong need for a unified theory of instabilities in swirling flows that incorporates axial shear, differential rotation, thermal stratification, viscosity, and thermal diffusion and aligns with numerical and experimental results}.

The present work is a comprehensive theoretical investigation of both isothermal and non-isothermal viscous and thermodiffusive swirling flows; \textcolor{black}{ it presents general criteria of instability in analytical form. 
It provides the complete development of our recent brief communication (\cite{KM2024})}, 
which focused solely on \textcolor{black}{centrifugal instability. More importantly, in the present work} we extend this analytical approach to identify the conditions for oscillatory visco-thermodiffusive McIntyre instability in non-isothermal swirling flows.

The methodology of this study is based on the local geometrical optics approach, initially introduced in hydrodynamics for inviscid flows \citep{ES1978,LH1991, LH1993, FV1991, MODC2014, IK2017} and later extended to visco-diffusive flows \citep{KS2013, KSF2014, KM2017,KM2024,LK2021,K2017,K2021,K2025} and \textcolor{black}{to multiple-diffusive flows \citep{SM2019, SM2021,V2019}.} We apply this approach to helical stationary solutions of the Navier-Stokes equations coupled with the energy equation in the Boussinesq-Oberbeck approximation, which—depending on boundary conditions—represent SCF, SPF, or BCF.

We establish that the neutral stability curves in the plane \textcolor{black}{spanned by the azimuthal and axial Reynolds numbers}—corresponding to both stationary (for isothermal and non-isothermal flows) and oscillatory (for non-isothermal flows only) instabilities—form families that possess envelopes, regardless of whether they are parameterized by the azimuthal or axial wavenumber. Since the axial wavenumber can take any real value, these envelopes define the boundaries of the unions of individual stationary and oscillatory instability domains for specific \textcolor{black}{values of} wavenumbers.

We find that the envelopes, and thus the instability domains, undergo splitting during the transition from Rayleigh-unstable to Rayleigh-stable flows. Notably, this splitting occurs at different Rossby number values for stationary and oscillatory instabilities of non-isothermal flows. This discrepancy offers a predictive tool for determining whether stationary or oscillatory instability will dominate in a visco-thermodiffusive swirling flow.

For Rayleigh-stable flows, as the azimuthal Reynolds number approaches infinity, an asymptotic line to the envelope of stationary instability domains corresponds to the inviscid LELS criterion for isothermal flows and provides an analytical expression for the unified LELS-GSF criterion in non-isothermal flows. In the isothermal case, we derive a compact closed-form expression for the envelope, extending the inviscid LELS criterion to viscous swirling flows and broadening its applicability across a wide range of azimuthal Reynolds numbers, from infinity to small but finite values. For non-isothermal flows, we establish explicit equations that define the asymptotic lines of the instability domains, offering a complete characterization of their boundaries.

The paper is organized as follows. In Section~\ref{sec:Mathematical_Setting}, we present the dimensionless nonlinear equations governing swirling flows in the Boussinesq-Oberbeck approximation, identify three main helical base states (detailed in Appendix~\ref{appA}), and derive the linearized equations about these base states. Section~\ref{Sec3} develops the geometrical optics \textcolor{black}{asymptotic} solution to the linearized equations, leading to the amplitude transport equations for the localized wavepacket evolving along the helical streamlines and the eikonal equation for its wavevector. In Section~\ref{Sec4}, we derive the dispersion relation and demonstrate its connection to previous works in Appendix~\ref{AppB}. Section~\ref{Sec5} introduces the new instability criteria, applying them to viscous isothermal swirling flows in Section~\ref{Sec5_1} and to visco-thermodiffusive non-isothermal swirling flows in Section~\ref{Sec5_2}. \textcolor{black}{Appendix~\ref{AppBa} establishes the connection between our results and the classical inviscid LELS criterion. Appendix~\ref{AppCC} relates our results to the criterion of \cite{DPA2010} for centrifugal instability in an inviscid swirling flow with radial density stratification but without mass diffusivity.} Finally, Section~\ref{Sec6} presents the conclusions.

\section{Mathematical setting}\label{sec:Mathematical_Setting}

\begin{figure}
  \centerline{\includegraphics[width=.5\textwidth]{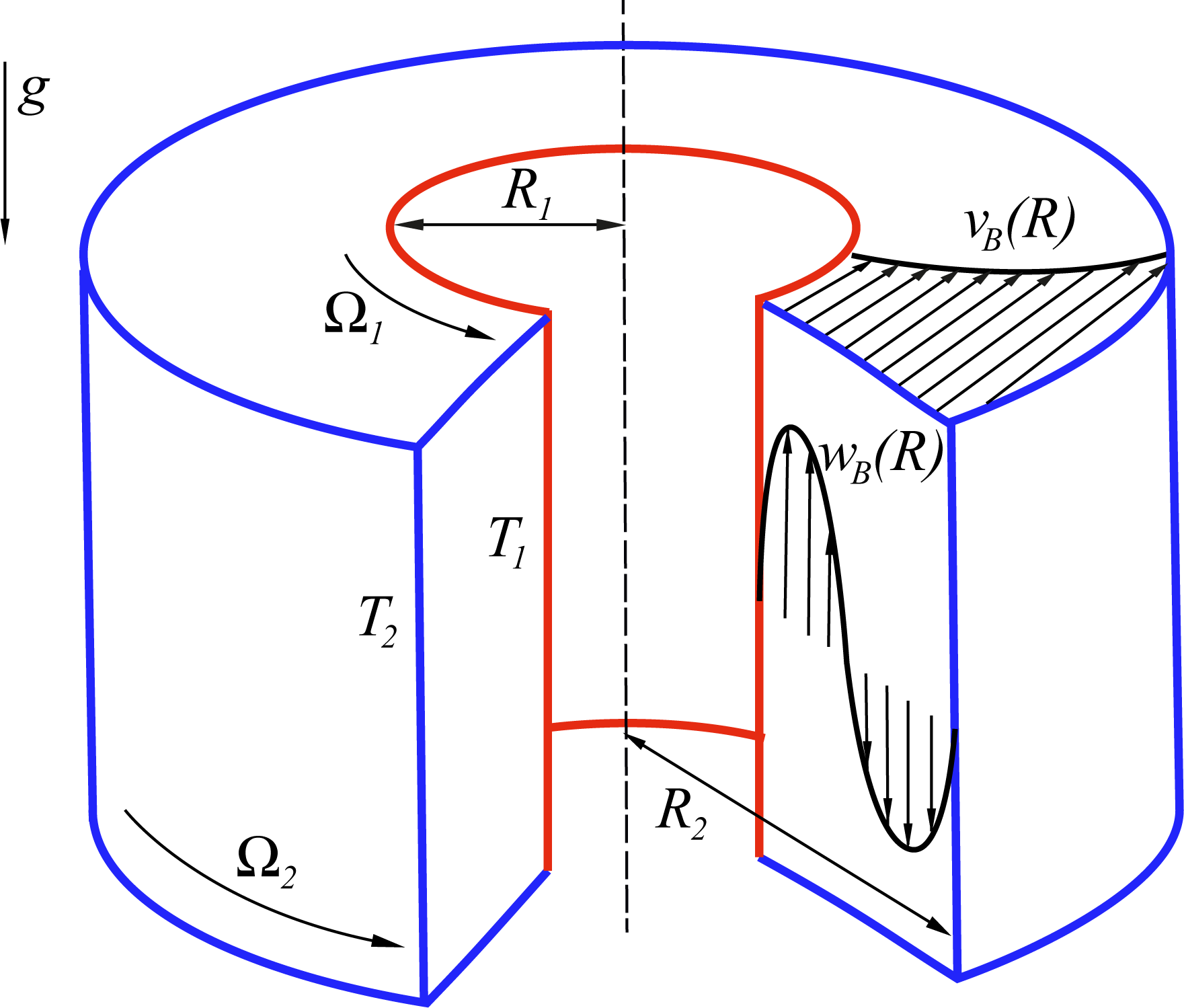}}
\caption{The helical base state as a superposition of the circular Couette flow $v_B(R)$ and an axial annular flow $w_B(R)$ in a differentially heated \textcolor{black}{infinitely long} cylindrical annulus.}
\label{fig:1}
\end{figure}

\subsection{Nonlinear governing equations}

We consider an incompressible Newtonian fluid with constant reference density $\rho$ as well as constant thermal expansion coefficient $\alpha$, kinematic viscosity $\nu$, and thermal diffusivity $\kappa$. This fluid is confined within an infinitely long cylindrical annulus with a gap width $d = R_2 - R_1$, where $R_1$ is the radius of the inner cylinder at temperature $T_1$, rotating with angular velocity $\Omega_1$, and $R_2$ is the radius of the outer cylinder at temperature $T_2 = T_1 - \Delta T$, rotating with angular velocity $\Omega_2$, see figure~\ref{fig:1}. We denote the ratios of radii and angular velocities as follows
\be{etamu}
\eta=\frac{R_1}{R_2}, \quad \mu=\frac{\Omega_2}{\Omega_1}.
\ee
The system is subjected to a uniform gravity field with acceleration $g$ along the $Z$-axis of the cylindrical coordinates $(R, \varphi, Z)$, which aligns with the common rotation axis of the cylinders, figure~\ref{fig:1}.

\textcolor{black}{We choose $V_0 = \Omega_1 R_1$ as the velocity scale, $d$ as the length scale, $\frac{d}{V_0}$ as the time scale, and $\rho V_0^2$ as the pressure scale. Under the Boussinesq–Oberbeck approximation, in which fluid properties are assumed constant except for density, which varies linearly with temperature in the driving forces, the dimensionless governing equations can be written as follows:}

\begin{subequations}\label{eq1:nlge}
\ba{}
&\boldsymbol{\nabla} \bcdot \boldsymbol{u}=0,&\label{eq1:nlgea}\\
&\frac{d \boldsymbol{u}}{d t}+\boldsymbol{\nabla} p-\frac{1}{Re}\bnabla^2\boldsymbol{u}+
\left(\gamma \frac{v^2}{r}\boldsymbol{e}_r-Ri\boldsymbol{e}_z\right)\theta=0,&\label{eq1:nlgeb}\\
&\frac{d \theta}{d t}-\frac{1}{RePr}\bnabla^2\theta=0,&\label{eq1:nlgec}
\ea
\end{subequations}
where \( \frac{d}{dt} = \frac{\partial}{\partial t} + \boldsymbol{u} \bcdot \boldsymbol{\nabla} \), \( r = \frac{R}{d} \), and \( z = \frac{Z}{d} \). Here, \( p \) is the pressure, \( \boldsymbol{u} = (u, v, w) \) is the velocity field, and \( \theta = \frac{T - T_2}{\Delta T} \) represents the temperature deviation from the reference temperature \( T_2 \). The parameter \( \gamma = \alpha \Delta T > 0 \) corresponds to outward heating (\( T_1 > T_2 \)), while \( \gamma < 0 \) corresponds to inward heating (\( T_1 < T_2 \)). \textcolor{black}{The velocity field satisfies the no-slip boundary conditions at the surfaces of the cylinders.}

The last two terms in Eq.~\eqref{eq1:nlgeb} represent the centrifugal buoyancy (\( \gamma \frac{v^2}{r} \boldsymbol{e}_r \)) and Archimedean buoyancy (\( -Ri \boldsymbol{e}_z \)), respectively.

The dimensionless control parameters in Eq.~\eqref{eq1:nlge} are defined as:
\ba{eq2:dlp}
\quad Re=\frac{V_0 d}{\nu},
\quad Pr=\frac{\nu}{\kappa},
\quad S=\frac{V_0}{W_0},
\quad Ri=\frac{W_T}{W_0}\frac{1}{S Re},
\ea
where  \( Re \) is the Reynolds number associated with the rotation of the inner cylinder, \( Pr \) is the Prandtl number, \textcolor{black}{$W_0$ is the characteristic axial velocity of the flow}, \( S \) is the swirl parameter \citep{AW1990,AW1993}, and \( Ri \) is the Richardson number. The Richardson number is defined using the characteristic thermal velocity \citep{CK1980}:
\be{eq4:wtga}
W_T=\frac{\gamma g d^2}{\nu}.
\ee

Using \( W_T \), we can also introduce the Grashof number (\( Gr \)), which characterizes the strength of the baroclinic flow:
\be{gn}
Gr=\frac{W_T d}{\nu}.
\ee
It is worth noting that if \( W_0 = W_T \), then \( S = \frac{Re}{Gr} \), \textcolor{black}{and consequently the Richardson and Grashof numbers are related as follows:}
\be{rigrre2}
Ri = \frac{1}{SRe} = \frac{Gr}{Re^2}.
\ee

\subsection{Helical base state}

\textcolor{black}{In an infinitely long cylindrical annulus, the base flow is a steady helical flow, invariant in both the azimuthal and axial directions}.
The temperature and velocity field depend only on the radial coordinate \(r\), while the pressure varies with both \(r\) and \(z\) \citep{KM2024}:
\refstepcounter{equation}
$$
\boldsymbol{u}_B(r)=(0,V(r),S^{-1}W(r)), \quad \theta_B(r)=\Theta(r), \quad p_B(r,z)=p_1(r)+z p_2.
\eqno{(\theequation{\mathit{a},\mathit{b},\mathit{c}})}\label{eq3:bse}
$$

Explicit expressions for the radial distributions of temperature  $\Theta(r)$, azimuthal velocity $V(r)$, axial velocity $W(r)$, and the characteristic axial velocity $W_0$ for various particular helical base flows—including isothermal and non-isothermal cases, such as spiral Couette flow (SCF), spiral Poiseuille flow (SPF), and baroclinic Couette flow (BCF)—are derived in Appendix~\ref{appA}. \textcolor{black}{Table \ref{tab:1} presents the base state profiles and their values at the dimensionless geometric mean radius \(r_g\) \rf{mgr}.}

\begin{table}
  \begin{center}
\def~{\hphantom{0}}
  \begin{tabular}{lccc}
  \vspace{10pt}
      Flow  & Base state profile   &   Values at $ r_{g}=\frac{\sqrt{\eta}}{1-\eta} $\\
  \vspace{10pt}
       CCF   & $\scriptstyle  V(r)=\frac{\eta}{1+\eta}\left( \frac{1-\mu}{(1-\eta)^2}\frac{1}{r}-\frac{\eta^2 - \mu}{\eta^2}r\right)$ & $\scriptstyle \Omega_{g}=\left(\frac{V}{r}\right)_{\scriptstyle r_{g}}=\frac{1-\eta}{\eta} \frac{\eta+\mu}{1+\eta}$, $\scriptstyle Ro_{g}= -\frac{\eta}{\eta + \mu}\frac{1 - \mu}{1 - \eta}$  \\
    \vspace{10pt}
       BCF   & \(
\begin{array}{rcl}
\vspace{2pt}
&\scriptstyle
V(r) \scriptstyle = \scriptstyle \frac{\eta}{1+\eta}\left( \frac{1-\mu}{(1-\eta)^2}\frac{1}{r}-\frac{\eta^2 - \mu}{\eta^2}r\right)& \\
\vspace{2pt}
&\scriptstyle
W(r) \scriptstyle = \scriptstyle C_1\left[(r_2^2-r_1^2)\frac{\ln(r/r_2)}{\ln \eta} + r^2 - r_2^2\right]-(r^2 - r_1^2)\frac{\ln(r/r_2)}{4\ln\eta}& \\
\vspace{2pt}
&\scriptstyle
C_{1} \scriptstyle = \scriptstyle \frac{ (1-3\eta^2)(1-\eta^2)-4\eta^4\ln \eta }{16( (1-\eta^2)^2+(1-\eta^4)\ln \eta)}&\\
\vspace{2pt}
&\scriptstyle
\Theta(r) \scriptstyle = \scriptstyle \frac{\ln[r(1-\eta)]}{\ln \eta}&
\end{array}
\) & \(
\begin{array}{rcl}
\vspace{2pt}
&\scriptstyle
W_g  \scriptstyle = \scriptstyle  \frac{4\eta(\eta^2 + \eta + 1)\ln \eta - (\eta^2 + 4\eta + 1)(\eta^2 - 1)}{32(\eta^2 - 1)(\eta^2\ln\eta - \eta^2 + \ln\eta + 1)}& \\
\vspace{2pt}
&\scriptstyle
DW_g \scriptstyle = \scriptstyle  \frac{ -4\eta(\eta^4 + 1)(\ln \eta)^2 + 2\eta(\eta^2 - 1)(3\eta^2 - 2\eta + 3)\ln \eta}{16\sqrt{\eta}(1 - \eta)^2(1 + \eta)((\eta^2 + 1)\ln \eta - \eta^2 + 1)\ln \eta}&\\
\vspace{2pt}
&\scriptstyle
+
\frac{  (\eta^2 - 4\eta + 1)(1-\eta^2)^2}{16\sqrt{\eta}(1 - \eta)^2(1 + \eta)((\eta^2 + 1)\ln \eta - \eta^2 + 1)\ln \eta}&\\
\vspace{2pt}
&\scriptstyle
\Theta_{g} \scriptstyle = \scriptstyle \frac{1}{2}, \quad D\Theta_{g}=\frac{1 - \eta}{\sqrt{\eta}\ln \eta}& \\
\end{array}
\) \\
  \vspace{10pt}
       SCF   & \(
\begin{array}{rcl}
\vspace{2pt}
&\scriptstyle
V(r) \scriptstyle = \scriptstyle \frac{\eta}{1+\eta}\left( \frac{1-\mu}{(1-\eta)^2}\frac{1}{r}-\frac{\eta^2 - \mu}{\eta^2}r\right)& \\
\vspace{2pt}
&\scriptstyle
W(r) \scriptstyle = \scriptstyle \frac{1+C_2(1+\eta)}{\ln \eta}\ln\left(\frac{r}{r_2}\right)+C_2(1-\eta)(r^2 - r_2^2)& \\
\vspace{2pt}
&\scriptstyle
C_{2} \scriptstyle = \scriptstyle -\frac{2\eta^2\ln\eta + 1 - \eta^2}{(1+\eta)[(1+\eta^2) \ln \eta  + 1 - \eta^2]}&
\end{array}
\) & \(
\begin{array}{rcl}
\vspace{2pt}
&\scriptstyle
C_2=0:~~\qquad W_g \scriptstyle = \scriptstyle \frac{1}{2}, \quad  DW_g=\frac{1 - \eta}{\sqrt{\eta}\ln\eta} & \\
\vspace{2pt}
&\scriptstyle
C_2\ne 0:~~  W_g \scriptstyle = \scriptstyle \frac{(1-\eta^3 + 3\eta^2 + \eta)\ln\eta + 2(1-\eta^2)}{2[(1+\eta^2)\ln\eta+1 - \eta^2](1+\eta)}& \\
\vspace{2pt}
&\scriptstyle
 \qquad~ DW_g \scriptstyle = \scriptstyle \frac{ (\eta^2 - 1)(\eta^2 + 2\eta - 1)-4\eta^3 \ln\eta}{[(1+\eta^2)\ln\eta +1- \eta^2](1+\eta)\sqrt{\eta}}&
\end{array}
\) \\
  \vspace{10pt}
       SPF  & \(
\begin{array}{rcl}
\vspace{2pt}
&\scriptstyle
V(r) \scriptstyle = \scriptstyle \frac{\eta}{1+\eta}\left( \frac{1-\mu}{(1-\eta)^2}\frac{1}{r}-\frac{\eta^2 - \mu}{\eta^2}r\right)& \\
\vspace{2pt}
&\scriptstyle
W(r) \scriptstyle = \scriptstyle \frac{2(1-\eta)^2\ln\eta }{\eta^2-(\eta^2+1)\ln\eta- 1 }\left[r^2 - r_2^2 + \frac{1}{\ln\eta}\frac{1+\eta}{1-\eta}\ln\left(\frac{r}{r_2}\right)\right]&
\end{array}
\) & \(
\begin{array}{rcl}
\vspace{2pt}
&\scriptstyle
W_g \scriptstyle = \scriptstyle \frac{(1-\eta)^2\ln\eta}{(1+\eta^2)\ln\eta +1 - \eta^2}& \\
\vspace{2pt}
&\scriptstyle
DW_g \scriptstyle = \scriptstyle \frac{2(\eta^2-2\eta\ln\eta - 1)(1-\eta)}{[(1+\eta^2)\ln\eta +1- \eta^2 ]\sqrt{\eta}}&
\end{array}
\) \\
  \end{tabular}
  \caption{\textcolor{black}{Expressions of base flow in cylindrical annulus} and their values evaluated at the geometric mean radius
$r_{g} = \sqrt{r_{1} r_{2}}$, where $r_1=\frac{\eta}{1-\eta}$ and $r_2=\frac{1}{1-\eta}$, for circular Couette flow (CCF), baroclinic Couette flow (BCF),
spiral Couette flow (SCF), and spiral Poiseuille flow (SPF).
}
  \label{tab:1}
  \end{center}
\end{table}

\subsection{Linearization}

To test the stability of the base state \rf{eq3:bse}, we introduce small three-dimensional perturbations $\left(\boldsymbol{u}', p',\theta'\right)$ and linearize the nonlinear equations \rf{eq1:nlge} about the base state and write the linearized equations in the matrix form \citep{K2017,K2021}:
\be{eq1m}
\left(
  \begin{array}{cc}
    \partial_t+ \mathcal{U}+\frac{2V}{r}\gamma\Theta \boldsymbol{ e}_r\boldsymbol{ e}_{\varphi}^T+\boldsymbol{ u}_B\bcdot \bnabla -\frac{1}{Re}\bnabla^2 & \gamma\frac{V^2}{r}\boldsymbol{ e}_r-Ri\boldsymbol{ e}_z \\
    (\bnabla \Theta)^T  & \partial_t+\boldsymbol{ u}_B\bcdot \bnabla -\frac{1}{RePr}\bnabla^2\\
  \end{array}
\right)\left(
         \begin{array}{c}
           \boldsymbol{ u}' \\
           \theta' \\
         \end{array}
       \right)=-\left(
                 \begin{array}{c}
                   \bnabla p' \\
                   0 \\
                 \end{array}
               \right),
\ee
where $\boldsymbol{u}'$ is subject to the incompressibility constraint
\be{eq1i}
\bnabla \bcdot  \boldsymbol{u}'=0.
\ee

The gradients of the base state are given by \citep{KM2024}
\be{Um}
\mathcal{U}=\bnabla \boldsymbol{ u}_B=\left(
  \begin{array}{ccc}
    0 & -\Omega & 0 \\
     (1+2Ro)\Omega & 0 & 0 \\
    \frac{1}{S} DW & 0 & 0 \\
  \end{array}
\right), \quad \bnabla \Theta= \left(\begin{array}{c}
D\Theta\\
0\\
0\\
\end{array}\right)
\ee
with $D=\frac{d}{dr}$, $\Omega=\frac{V}{r}$, and the Rossby number defined as \citep{KS2013}
\be{ro}
Ro=\frac{rD\Omega}{2\Omega}.
\ee

\textcolor{black}{The perturbations must vanish at the boundaries of the cylindrical annulus; that is, $\boldsymbol{u}' = 0$ and $\theta' = 0$ at both $r = r_1$ and $r = r_2$. However, these boundary conditions are not applied in the analysis that follows, as we restrict our attention to local instabilities.}

\section{Geometrical optics equations}\label{Sec3}

Using a small parameter $0<\epsilon \ll 1$, we
represent the perturbations as asymptotic expansions \citep{ES1978,FV1991,LH1991}:
\ba{pert}
&\boldsymbol{u}^{\prime} =\left(\boldsymbol{u}^{(0)}(\boldsymbol{x}, t){+}\epsilon \boldsymbol{u}^{(1)}(\boldsymbol{x}, t)\right){e}^{\frac{{\rm i} \Phi(x, t)}{ \epsilon}}+\epsilon \boldsymbol{u}^{(r)}(\boldsymbol{x}, t,\epsilon)\textcolor{black}{~ +~ c.c.},&\nn\\
&\theta^{\prime} =\left(\theta^{(0)}(\boldsymbol{x}, t){+}\epsilon \theta^{(1)}(\boldsymbol{x}, t)\right){e}^{\frac{{\rm i} \Phi(x, t)}{ \epsilon}}+\epsilon \theta^{(r)}(\boldsymbol{x}, t,\epsilon)\textcolor{black}{~ +~c.c.},&\nn\\
&p^{\prime} =\left(p^{(0)}(\boldsymbol{x}, t){+}\epsilon p^{(1)}(\boldsymbol{x}, t)\right){e}^{\frac{{\rm i} \Phi(x, t)}{ \epsilon}}+\epsilon p^{(r)}(\boldsymbol{x}, t,\epsilon)\textcolor{black}{~ +~ c.c.}&
\ea
\textcolor{black}{Here}, ${\rm i} = \sqrt{-1}$, and $\Phi$ is generally a complex-valued scalar function representing the phase of the wave, or the eikonal. \textcolor{black}{The quantities $\boldsymbol{u}^{(i)}$, $\theta^{(i)}$, and $p^{(i)}$, for $i = 1, 2, \ldots$, are complex amplitudes, and the complex} remainder terms $\boldsymbol{u}^{(r)}$, $\theta^{(r)}$, and $p^{(r)}$ are assumed to be uniformly bounded in $\epsilon$ over any fixed time interval. \textcolor{black}{The notation c.c. denotes complex conjugate terms, included to ensure that the expansions are real-valued, consistent with the real-valued nature of the primed quantities on the left-hand side.}

\citet{M1986} observed that high-frequency oscillations $\exp\left({\rm i}\epsilon^{-1}\Phi(\boldsymbol{ x}, t)\right)$ quickly die out because of viscosity unless one assumes a quadratic dependency of viscosity and diffusivity on the small parameter $\epsilon$: \textcolor{black}{$\nu=\epsilon^2 \widehat{\nu}$ and $\kappa=\epsilon^2\widehat{\kappa}$}. Hence, following \cite{M1986}, \textcolor{black}{we have} $Re = \epsilon^{-2}\widehat{Re}$,
see also \cite{LPL1984,CC1986,SLB2003}.

\textcolor{black}{With the assumptions made, substitution of} the asymptotic series into the incompressibility condition \rf{eq1i} and  \textcolor{black}{collection of terms of the order $\epsilon^{-1}$ and $\epsilon^{0}$ yield}
\be{eq6}
\epsilon^{-1}:\quad \boldsymbol{ u}^{(0)} \bcdot \bnabla \Phi = 0,
\ee
\be{eq7}
\epsilon^{0}:\quad \bnabla \bcdot  \boldsymbol{ u}^{(0)} + {\rm i} \boldsymbol{ u}^{(1)} \bcdot \bnabla \Phi = 0.
\ee

A similar procedure applied to the linearised Navier-Stokes and 
energy equations \rf{eq1m} yields the two systems of equations

\begin{subequations}
\ba{eq9f}
\epsilon^{-1}:\quad & &\left(\begin{array}{cc}
\frac{\partial \Phi}{\partial t}+\boldsymbol{ u}_B\bcdot \bnabla \Phi & 0\\
0 & \frac{\partial \Phi}{\partial t}+\boldsymbol{ u}_B\bcdot \bnabla \Phi\\
\end{array}\right)\left(\begin{array}{c}
\boldsymbol{ u}^{(0)}\\
\theta^{(0)}
\end{array}\right)=- p^{(0)}\left(\begin{array}{c}
 \bnabla \Phi \\
0\\
\end{array} \right),\label{eq8}\\
\epsilon^{0}:\quad &{\rm i}& \left(\begin{array}{cc}
\frac{\partial \Phi}{\partial t}+\boldsymbol{ u}_B\bcdot \bnabla \Phi & 0\\
0 & \frac{\partial \Phi}{\partial t}+\boldsymbol{ u}_B\bcdot \bnabla \Phi\\
\end{array} \right)\left(\begin{array}{c}
\boldsymbol{ u}^{(1)}\\
\theta^{(1)}
\end{array}\right)=-{\rm i} p^{(1)} \left(\begin{array}{c}
\bnabla \Phi\\
0\\
\end{array} \right)\nn\\
&-&\left(\begin{array}{cc}
\frac{d}{d t}+\mathcal{U} +\frac{1}{\widehat{Re}}\left(\bnabla \Phi\right)^2 & 0\\
0 & \frac{d}{d t} +\frac{1}{\widehat{Re}Pr}\left(\bnabla \Phi\right)^2\\
\end{array}\right)\left(\begin{array}{c}
\boldsymbol{ u}^{(0)}\\
\theta^{(0)}
\end{array}\right)\nn\\
&-&\left( \begin{array}{cc}
\frac{2V}{r}\gamma\Theta\boldsymbol{ e}_r{\bf e}_{\varphi}^T &
\gamma  \frac{V^2}{r}\boldsymbol{ e}_r-{Ri}\boldsymbol{ e}_z\\
\left(\bnabla\Theta\right)^T & 0\\
\end{array}\right)\left(\begin{array}{c}
\boldsymbol{ u}^{(0)}\\
\theta^{(0)}
\end{array}\right)- \left(\begin{array}{c}
\bnabla p^{(0)}\\
0\\
\end{array} \right),\label{eq9}
\ea
\end{subequations}
\textcolor{black}{where \( \frac{d}{dt} = \frac{\partial}{\partial t} + \boldsymbol{u}_B \bcdot \boldsymbol{\nabla} \).}

Taking the dot product of the first of the equations in the system \rf{eq8} with $\bnabla \Phi$ under the constraint \rf{eq6} we find that for $\bnabla \Phi \ne 0$
\be{eq10}
p^{(0)}=0.
\ee
Under the condition \rf{eq10} the system \rf{eq8} has a non-trivial solution if the determinant of the $4\times 4$ matrix in its left-hand side is vanishing. This gives us a 4-fold characteristic root corresponding to the Hamilton-Jacobi equation
\be{eq11}
\frac{\partial \Phi}{\partial t}+\boldsymbol{ u}_B\bcdot \bnabla \Phi=0,
\ee
with the initial data: $\Phi(\boldsymbol{ x},0)=\Phi_0(\boldsymbol{ x})$. Taking the gradient of \rf{eq11} yields the eikonal equation
\be{eq12}
\frac{d}{dt}\bnabla \Phi=-\bnabla \boldsymbol{ u}_B \bcdot \bnabla \Phi=-\mathcal{U}^T \bnabla \Phi
\ee
with the initial condition $\bnabla\Phi(\boldsymbol{ x},0)=\bnabla\Phi_0(\boldsymbol{ x})$, where $\mathcal{U}^T$ denotes the transposed $3 \times 3$ matrix $\mathcal{U}$ defined by equation \rf{Um}.

Relations \rf{eq11} and \rf{eq12} allow us to reduce the system \rf{eq9} to
\begin{subequations}
\ba{eq14}
&\left(\frac{d}{dt}+\mathcal{U}+\frac{1}{\widehat{Re}}\left(\bnabla \Phi \right)^2 \textcolor{black}{\mathcal{I}}+\frac{2V}{r}\gamma \Theta\boldsymbol{ e}_r\boldsymbol{ e}_{\varphi}^T\right)\boldsymbol{ u}^{(0)}+\left(\gamma \frac{V^2}{r}\boldsymbol{ e}_r-Ri\boldsymbol{ e}_z\right)\theta^{(0)}=-{\rm i}\bnabla \Phi p^{(1)},& \label{eq14a}\\
&\left(\bnabla\Theta\right)^T\boldsymbol{ u}^{(0)}+\left(\frac{d}{dt} +\frac{1}{\widehat{Re}Pr}\left(\bnabla \Phi\right)^2\right)\theta^{(0)}=0,& \label{eq14b}
\ea
\end{subequations}
\textcolor{black}{where $\mathcal{I}$ is the $3\times 3$ identity matrix, $\boldsymbol{ e}_r\boldsymbol{ e}_{\varphi}^T$ is a $3\times 3$ matrix, and $\left(\bnabla\Theta\right)^T\boldsymbol{ u}^{(0)}=\bnabla\Theta \bcdot \boldsymbol{ u}^{(0)} $.}

Multiplying equation \rf{eq14a} with the vector $\bnabla \Phi$ from the left and taking into account the relation \rf{eq6}, we isolate the pressure term
\be{eq15}
p^{(1)}={\rm i}\frac{\bnabla \Phi}{(\bnabla \Phi)^2}\bcdot\left[ \left(\frac{d}{dt}+\mathcal{U}+\frac{2V}{r}\gamma \Theta\boldsymbol{ e}_r\boldsymbol{ e}_{\varphi}^T\right)\boldsymbol{ u}^{(0)}+\left(\gamma \frac{V^2}{r}\boldsymbol{ e}_r-Ri\boldsymbol{ e}_z\right)\theta^{(0)} \right].
\ee

Taking into account the identity
\be{eq16}
\frac{d}{dt}(\bnabla \Phi \bcdot \boldsymbol{ u}^{(0)})=\frac{d \bnabla \Phi}{dt}\bcdot \boldsymbol{ u}^{(0)}+\bnabla \Phi \bcdot \frac{d \boldsymbol{ u}^{(0)}}{dt}=0
\ee
we modify \rf{eq15} in the following way
\ba{eq17}
p^{(1)}&=&{\rm i}\frac{\bnabla \Phi}{(\bnabla \Phi)^2}\bcdot\left[ \left(\mathcal{U}+\frac{2V}{r} \gamma \Theta\boldsymbol{ e}_r\boldsymbol{ e}_{\varphi}^T\right)\boldsymbol{ u}^{(0)}+\left(\gamma \frac{V^2}{r}\boldsymbol{ e}_r-Ri\boldsymbol{ e}_z\right)\theta^{(0)} \right]\nn\\
&-&{\rm i} \frac{1}{(\bnabla \Phi)^2}\frac{d \bnabla \Phi}{d t}\bcdot \boldsymbol{ u}^{(0)}.
\ea
Now using the eikonal equation \rf{eq12} we transform the last term in equation \rf{eq17} to obtain
\ba{eq18}
p^{(1)}&=&{\rm i}\frac{2V}{r}\gamma \Theta\frac{(\bnabla \Phi)^T\boldsymbol{ e}_r{\bf e}_{\varphi}^T}{(\bnabla \Phi)^2}\boldsymbol{ u}^{(0)}+{\rm i}\left(\gamma \frac{V^2}{r}\frac{(\bnabla \Phi)^T\boldsymbol{ e}_r}{(\bnabla \Phi)^2}-Ri\frac{(\bnabla \Phi)^T\boldsymbol{ e}_z}{(\bnabla \Phi)^2}\right)\theta^{(0)}\nn\\
&+&2{\rm i} \frac{(\bnabla \Phi)^T\mathcal{U}}{(\bnabla \Phi)^2} \boldsymbol{ u}^{(0)}.
\ea

Substituting \rf{eq18} into \rf{eq14a} we finally arrive at the transport equations for the leading-order amplitudes $\boldsymbol{ u}^{(0)}$ and $\theta^{(0)}$ of the localized wave packet moving along the streamlines of the base flow written in the stationary frame \citep{KM2024}:
\ba{eq19}
\frac{d {\bm u}^{(0)}}{dt}+\frac{|{\bm k}|^2}{\textcolor{black}{\widehat{Re}}}{\bm u}^{(0)}
&=&-\left(\mathcal{I}-\frac{{\bm k}{\bm k}^T}{|{\bm k}|^2}\right)\left(\gamma\frac{V^2}{r}{\bm e}_r-Ri{\bm e}_z\right)\theta^{(0)}\nn\\
&&-\left(\mathcal{I}-2 \frac{{\bm k}{\bm k}^T}{|{\bm k}|^2}\right) \mathcal{U}{\bm u}^{(0)}-2\gamma \Theta\Omega \left(\mathcal{I}-\frac{{\bm k}{\bm k}^T}{|{\bm k}|^2}\right){\bm e}_r{\bm e}_{\varphi}^T{\bm u}^{(0)},\nn\\
\frac{d \theta^{(0)}}{dt}+\frac{|{\bm k}|^2}{\textcolor{black}{\widehat{Re}}Pr}\theta^{(0)}&=&-\left(\bnabla\Theta\right)^T{\bm u}^{(0)},
\ea
where \textcolor{black}{${\bm k}{\bm k}^T$ is a $3\times 3$ matrix} and we denoted \textcolor{black}{the wavenumber of the perturbations as} $\boldsymbol{ k}=\bnabla \Phi$. In this notation the eikonal equation \rf{eq11} determines evolution of the wavevector in the stationary frame:
\be{eq13}
\frac{d\boldsymbol{ k}}{d t}=-\mathcal{U}^T\boldsymbol{ k}
\ee
under the constraint (following from \rf{eq6})
\be{const}
\boldsymbol{ k}\bcdot \boldsymbol{ u}^{(0)}=0.
\ee

Notice that the equations \rf{eq19} differ from the amplitude equations derived earlier in \citep{KM2017} by the Archimedean buoyancy term $-Ri\boldsymbol{e}_z$ and the term $S^{-1}DW$ in the matrix $\mathcal{U}$.

\section{Dispersion relation} \label{Sec4}

The derivative $\frac{\partial {\bm k}}{\partial t}$ of  the wavevector ${\bm k}=(k_r,k_{\varphi},k_z)$  in the frame of the wave packet, rotating about the vertical axis with the angular velocity \(\Omega\) is related to the derivative of this vector in the stationary frame as \citep{EY1995}:
\be{veck}
\frac{d{\bm k}}{dt}=\frac{\partial {\bm k}}{\partial t}+\mathcal{J}{\bm k}, \quad
\mathcal{J}=\left(
  \begin{array}{rrr}
    0 & -\Omega & 0 \\
    \Omega & 0 & 0 \\
    0 & 0 & 0 \\
  \end{array}
\right).
\ee
Taking into account \rf{veck} in \rf{eq13}, we get
\be{eike}
\frac{\partial {\bm k}}{\partial t}=\left(
              \begin{array}{ccc}
                0 & -2\Omega Ro & -S^{-1}DW \\
                0  & 0 & 0 \\
                0 & 0 & 0 \\
              \end{array}
            \right)\left(
                     \begin{array}{c}
                       k_r \\
                       k_{\varphi} \\
                       k_z \\
                     \end{array}
                   \right).
\ee
Under the condition \citep{ES1978,LS1983,EM1984}
\refstepcounter{equation}
$$
k_\varphi=-\overline{DW} k_z, \quad {\rm with} \quad \overline{DW}=\frac{DW}{2\Omega RoS},
\eqno{(\theequation{\mathit{a},\mathit{b}})}\label{ec}
$$
the components of the wave vector $\boldsymbol{k}$ are time-independent in the rotating frame: $k_r=const.$, $k_{\varphi}=const.$, $k_z=const.$
\textcolor{black}{Note that, since the vorticity of the helical base flow (\ref{eq3:bse}\textcolor{red}{a}) has two components, $\omega_B = (0, -S^{-1}DW, 2\Omega(1 + Ro))$, we can conclude from \rf{ec} that the azimuthal wavenumber of the perturbations is linked to their axial wavenumber through the azimuthal vorticity component ($-S^{-1}DW$) of the base flow. These perturbations exhibit helical symmetry \citep{EM1984}, remaining invariant along circular helices with a pitch of $2\pi r \overline{DW}$, and correspond to the most unstable and exponentially growing three-dimensional perturbations, as demonstrated in the works of \cite{ES1978,EC1984,LS1983,BG2013}.}

\textcolor{black}{Note that perturbations violating the constraint \rf{ec} were shown already by \cite{ES1978} to grow algebraically and to be weaker than those satisfying the constraint. For this reason, the former have been a less popular subject of investigation in the context of swirling flows; however, see the works of \cite{HP2006,DPA2010}. These weaker instabilities also lie outside the scope of the present paper.}

\textcolor{black}{Under the condition \rf{ec}} the amplitude equations \rf{eq19} are autonomous in the rotating frame. \textcolor{black}{Following the standard procedure, see e.g. \cite{EY1995} and \cite{FV1995}, we write explicitly the material derivative \textcolor{black}{$\frac{d}{dt} = \frac{\partial}{\partial t} + \boldsymbol{u}_B \bcdot \boldsymbol{\nabla}$} in the left side of equations \rf{eq19} and  explicitly compute their right side, exploiting the relation
\be{normk}
|\bm{k}|^2=k_r^2+k_z^2\left[1+\overline{DW}^2\right].
\ee
Since $\boldsymbol{u}_B(r)=(0,V(r),S^{-1}W(r))$ the left hand side of the resulting linear equations with constant coefficients does not contain the derivative with respect to $r$. This allows us to seek for their solution in the modal form: $\bm{u}^{(0)}, \theta^{(0)}\sim e^{s t+{\rm i} m\varphi+{\rm i}k_z z}$,} where $s=\sigma + {\rm i} \omega$, $\sigma,\omega \in \mathbb{R}$, is the complex growth rate, and $m=k_\varphi r$ and $k_z$ are the integer azimuthal and real axial wavenumbers. This yields the system of linear algebraic equations:
\begin{subequations}\label{aespft}
\ba{}
\lambda u_r^{(0)} &=&  -  \frac{DW}{S Ro } \frac{k_r k_z}{|\boldsymbol{ k}|^2} u_r^{(0)} -\frac{|\boldsymbol{ k}|^2}{\widehat{Re}} u_r^{(0)} + 2\Omega (1{-}\gamma\Theta)\left(1 - \frac{k_r^2}{|\boldsymbol{ k}|^2}\right)u_{\varphi}^{(0)}\nn\\
& &- \left( \gamma r  \Omega^2\left(1 - \frac{k_r^2}{|\boldsymbol{ k}|^2}\right)+Ri \frac{k_rk_z}{|\boldsymbol{ k}|^2}\right)\theta^{(0)},\label{aespft1a}\\
\lambda u_{\varphi}^{(0)} &=& - 2 \Omega\left(Ro + \frac{k_r^2 + k_z^2}{|\boldsymbol{ k}|^2}\right)u_r^{(0)} -\frac{|\boldsymbol{ k}|^2}{\widehat{Re}} u_{\varphi}^{(0)} +  (1-\gamma \Theta)\frac{DW}{S Ro }\frac{k_rk_z}{|\boldsymbol{ k}|^2}u_{\varphi}^{(0)}\nn\\
& &- \left( \gamma r\Omega^2-Ri  \frac{k_z}{k_r}\right)\frac{1}{2\Omega}\frac{DW}{S Ro }\frac{k_r k_z}{|\boldsymbol{ k}|^2}\theta^{(0)}, \label{aespft2}\\
\lambda u_z^{(0)}  &=& -\frac{|\boldsymbol{ k}|^2}{\widehat{Re}} u_z^{(0)} - \frac{DW}{S Ro} \left(Ro + \frac{k_z^2}{|\boldsymbol{ k}|^2}\right) u_r^{(0)} - 2 \Omega (1-\gamma\Theta)\frac{k_rk_z}{|\boldsymbol{ k}|^2}u_{\varphi}^{(0)} \nn\\
& &+ \gamma r\Omega^2\frac{k_r k_z}{|\boldsymbol{ k}|^2}\theta^{(0)}+ Ri\left(1 - \frac{k_z^2}{|\boldsymbol{ k}|^2}\right)\theta^{(0)},\label{aespft3}\\
\lambda \theta^{(0)} &=&-D\Theta u_r^{(0)}
-\frac{|\boldsymbol{ k}|^2}{Pr \widehat{Re}}\theta^{(0)},\label{aespft4}
\ea
\end{subequations}
where
\be{eigp}
\lambda=\sigma+{\rm i}\left(\omega+m\Omega+\frac{k_zW}{S} \right).
\ee

\textcolor{black}{The generalized eigenvalue $\lambda$ includes the Doppler frequency $\omega_D=m\Omega+\frac{k_z W}{S}$ of the fluid particle moving along the helix around the cylindrical annulus at the radius $r\in[r_1,r_2]$}.
One can see that multiplying the equation \rf{aespft1a} with $\frac{k_r}{k_z}$,
subtracting the result from the equation \rf{aespft2} multiplied with $\overline{DW}$,
and taking into account \rf{const} yields the equation \rf{aespft3},
which is therefore decoupled from the first two. The remaining equations \rf{aespft1a}, \rf{aespft2}, and \rf{aespft4}
result in the matrix eigenvalue problem for the $3 \times 3$ matrix $\mathcal{H}$ with the eigenvalue parameter $\lambda$ given by \rf{eigp}, where \citep{KM2024}
\be{mahc}
\mathcal{ H}=\left(
          \begin{array}{ccc}
            -\frac{ DW}{ SRo}\frac{k_r k_z}{|\boldsymbol{ k}|^2}-\frac{|\boldsymbol{ k}|^2}{\widehat{Re}} & 2\Omega(1-\gamma \Theta)\left(1-\frac{k_r^2}{|\boldsymbol{ k}|^2}\right) & -\left(r \gamma \Omega^2
            \left(1-\frac{k_r^2}{|\boldsymbol{ k}|^2} \right) + Ri\frac{k_r k_z}{|\boldsymbol{ k}|^2}\right)  \\
            -2\Omega\left(Ro+\frac{k_r^2+k_z^2}{|\boldsymbol{ k}|^2} \right) & (1-\gamma\Theta)\frac{ DW}{ S Ro}\frac{k_r k_z}{|\boldsymbol{ k}|^2} -\frac{|\boldsymbol{ k}|^2}{\widehat{Re}}& -\left(r  \gamma \Omega^2   - Ri\frac{k_z}{k_r}\right)\frac{1 }{2\Omega} \frac{ DW}{S Ro}\frac{k_r k_z}{|\boldsymbol{ k}|^2}  \\
            -D\Theta  & 0 & \frac{|\boldsymbol{k}|^2}{\widehat{Re}} \frac{Pr-1}{Pr}-\frac{|\boldsymbol{k}|^2}{\widehat{Re}} \\
          \end{array}
        \right).
\ee

We find the dispersion relation by computing the characteristic polynomial of $\mathcal{H}$:
\be{poly}
p(\lambda)=-\det( \mathcal{ H}-\lambda \mathcal{ I})=a_3\lambda^3+a_2\lambda^2+a_1\lambda+a_0,
\ee
with the real coefficients
\begin{subequations}\label{polpr}
\ba{}
a_3&=&1,\label{polpra3}\\
a_2&=&\frac{k_r k_z}{|\boldsymbol{ k}|^2} \frac{DW}{SRo}\gamma \Theta+\frac{2Pr+1}{Pr}\frac{|\boldsymbol{ k}|^2}{\widehat{Re}},\label{polpra2}\\
a_1&=&  \Omega^2\left(1 - \frac{k_r^2}{|\boldsymbol{ k}|^2}\right)\left\{4\left[Ro + \frac{k_r^2 + k_z^2}{|\boldsymbol{ k}|^2}\right](1-\gamma \Theta ) - \gamma r D\Theta \right\}+\frac{Pr+2}{Pr}\frac{|\boldsymbol{ k}|^4}{\widehat{Re}^2} \nn\\
&-& \frac{(DW)^2}{S^2Ro^2}\frac{k_r^2k_z^2}{|\boldsymbol{ k}|^4}(1-\gamma \Theta) - Ri \frac{k_r k_z}{|\boldsymbol{ k}|^2}D\Theta +\frac{Pr+1}{Pr}\frac{DW}{SRo}\frac{k_r k_z}{|\boldsymbol{ k}|^2}\frac{|\boldsymbol{ k}|^2}{ \widehat{Re}}\gamma \Theta ,\label{polpra1}\\
a_0&=& \frac{D\Theta DW}{SRo} \frac{k_z^2}{|\boldsymbol{ k}|^2}  Ri  (1-\gamma \Theta )-\frac{k_r k_z}{|\boldsymbol{ k}|^2}\frac{|\boldsymbol{ k}|^2}{\widehat{Re}}RiD\Theta\nn\\
&+&\frac{|\boldsymbol{ k}|^2\Omega^2}{\widehat{Re} Pr}\left(1 - \frac{k_r^2}{|\boldsymbol{ k}|^2}\right)\left\{4\left[Ro + \frac{k_r^2 + k_z^2}{|\boldsymbol{ k}|^2}\right](1-\gamma \Theta ) -  \gamma  r D\Theta Pr \right\}\nn\\
&+&\frac{|\boldsymbol{ k}|^2}{ \widehat{Re} Pr}\left( \frac{|\boldsymbol{ k}|^2}{ \widehat{Re}}-\frac{k_rk_z}{|\boldsymbol{ k}|^2}\frac{DW}{SRo}(1-\gamma\Theta) \right)\left(\frac{|\boldsymbol{ k}|^2}{ \widehat{Re}} + \frac{k_rk_z}{|\boldsymbol{ k}|^2}\frac{DW}{SRo}\right).\label{polpra0}
\ea
\end{subequations}
 In view of $Ri=\frac{W_T}{W_0}\frac{1}{S\widehat{Re}}$ the coefficients \rf{polpr} and the matrix \rf{mahc} reduce to those derived by \citet{KM2017} in the limit $S \rightarrow \infty$, as is shown in Appendix~\ref{AppB}.

\section{\textcolor{black}{Local linear} stability analysis of swirling flows}\label{Sec5}

In the following, we derive general stability conditions of the base flow by applying the Li\'enard-Chipart stability criterion \citep{KM2017,K2021} to \textcolor{black}{the characteristic polynomial} \rf{poly} with the coefficients \rf{polpr}. \textcolor{black}{This criterion 
guarantees} $\lambda$ \rf{eigp} to have only negative real part
\refstepcounter{equation}
$$
a_0 >0, \quad a_2 >0,\quad a_1a_2-a_0>0.
\eqno{(\theequation{\,\mathit{a},\mathit{b},\mathit{c}})}\label{lsh}
$$

We first examine isothermal flows \textcolor{black}{($\gamma=0$, $Ri=0$)}, such as Spiral Couette and Spiral Poiseuille flows, where the coefficients \rf{polpr} simplify enough to allow for a fully analytical treatment. Next, we explore the most general scenario, involving both radial heating and Archimedean buoyancy, which includes Baroclinic Couette flow. \textcolor{black}{For the sake of simplicity of notation in Section~\ref{Sec5} we omit the hat over the Reynolds numbers.}

\subsection{Viscous isothermal swirling flows}\label{Sec5_1}

\subsubsection{\textcolor{black}{Viscous extension of the LELS centrifugal instability criterion}}

Assuming $\gamma=0$ and $Ri=0$ in equation \rf{polpr} automatically satisfies the condition $a_2>0$. This leaves only two inequalities, (\textcolor{red}{\ref{lsh}a}) and (\textcolor{red}{\ref{lsh}c}), to determine stability. The condition $a_0=0$ following from the inequality (\textcolor{red}{\ref{lsh}a}) defines the neutral stability curve as follows:
\ba{ita0}
4\Omega^2\left(1 - \frac{k_r^2}{|\boldsymbol{ k}|^2}\right)\left[Ro + \frac{k_r^2 + k_z^2}{|\boldsymbol{ k}|^2}\right]-\frac{k_r^2k_z^2}{|\boldsymbol{ k}|^4}\frac{DW^2}{S^2Ro^2} +\frac{|\boldsymbol{ k}|^4}{ Re^2}=0,
\ea
where $|\boldsymbol{ k}|$ is given  by equation \rf{normk}. The swirl parameter
\be{spscf}
S=\frac{Re}{Re_z}
\ee
is now defined via the axial Reynolds number \(Re_z = \frac{W_0 d}{\nu}\), based on the \textcolor{black}{characteristic axial flow velocity $W_0$ for isothermal flows. The details of its definition for SCF and SPF base flows are given in Appendix~\ref{appA}}.

Taking into account relations \rf{ec} and \rf{normk}, we can rewrite equation \rf{ita0} as
\be{ita01}
\frac{4\Omega^2k_z^2(\overline{DW}^2Ro + Ro + 1)Re^2 + (\overline{DW}^2k_z^2 + k_r^2 + k_z^2)^3}{4\Omega^2Re^2(\overline{DW}^2k_z^2 + k_r^2 + k_z^2)}=0,
\ee
which, given the positivity of its denominator, leads to an equation
\be{ita02}
q(k_z) = 0,
\ee
where
\ba{p0lykz}
q(k_z) &=& (\overline{DW}^2 + 1)^3k_z^6 + 3k_r^2(\overline{DW}^2 + 1)^2k_z^4 \nn\\
&+& \left[3(\overline{DW}^2 + 1)k_r^4 + 4\Omega^2 Re^2(\overline{DW}^2Ro + Ro + 1)\right]k_z^2 + k_r^6
\ea
is a real polynomial of degree six in \( k_z \). Using the explicit expression (\textcolor{red}{\ref{ec}b}) for \( \overline{DW} \) and \rf{spscf} for the swirl parameter \( S \), we can interpret the equation \rf{ita02} as defining a family of neutral stability curves in the \( (Re_z, Re) \)-plane.

\textcolor{black}{\cite{MFN2008} observed, using global numerical linear stability analysis, that the neutral stability curves of three-dimensional perturbations in plane Poiseuille flow with streamwise system rotation share a common envelope in the plane defined by the axial Reynolds number and the rotation number. This envelope exhibits both vertical and horizontal asymptotes at large values of these parameters. They further hypothesised that a similar envelope is likely to exist for spiral Poiseuille flow (SPF) between concentric cylinders with $\eta=0.5$ \citep{MM2002}. Since equation \rf{ita02} describes the family of neutral stability curves corresponding to the onset of centrifugal instability in isothermal swirling flows—including both SCF and SPF—we aim to test this hypothesis and search for an envelope of the family defined by \rf{ita02}.}

To determine the envelope, we can compute the discriminant of the polynomial \rf{p0lykz}, using a standard tool in any modern computer algebra system:
\ba{dispolykz}
{\rm Disc}_{k_z}(q) &=& \left[27(\overline{DW}^2 + 1)k_r^4 + 16\Omega^2 Re^2(\overline{DW}^2 Ro + Ro + 1)\right]^2 \nn\\
&\times& (\overline{DW}^2Ro + Ro + 1)^4(\overline{DW}^2 + 1)^9(-16384)Re^8\Omega^8k_r^6.
\ea

Since the envelope corresponds to a portion of the discriminant set of the polynomial, given by the equation \( {\rm Disc}_{k_z}(q) = 0 \) \citep{HW1953,BG1992}, we focus only on the first factor in \rf{dispolykz}. Introducing the Rayleigh discriminant
\be{rd}
N_{\Omega}^2 = 4(1 + Ro)\Omega^2,
\ee
we express \( Ro \) in terms of \( N_{\Omega}^2 \) and set the first factor in \rf{dispolykz} to zero, obtaining a compact analytical expression that defines the envelope of the neutral stability curves \rf{ita0} for isothermal viscous swirling flows \citep{KM2024}:
\be{masterisc}
E(Re_z, Re) = \frac{N_{\Omega}^2}{\Omega^2} - \frac{4\overline{DW}^2}{1 + \overline{DW}^2} + \frac{27}{4\Omega^2} \frac{k_r^4}{Re^2} = 0.
\ee



Alternatively, the envelope \rf{masterisc} can be derived using the classical approach \citep{HW1953,BG1992} by first differentiating \rf{ita02} with respect to \( k_z \), resulting in a biquadratic equation in $k_z$. Solving this equation explicitly and substituting the solution back into \rf{ita02} leads directly to the envelope equation \rf{masterisc}.

\textcolor{black}{According to \cite{H2020} (see also \cite{BS2000}, \cite{MS2002}), the determination of an envelope can be formulated as a parametric optimization problem. To illustrate this third approach, we present an example of such a problem that leads to the envelope \rf{masterisc}.}

\textcolor{black}{First, we notice that equation \rf{ita01} can be rewritten as
\be{ita02a}
Re = \frac{\left(k_r^2 +\left(1+\overline{DW}^2\right)k_z^2\right)^{3/2}}{2\Omega k_z\left(-1-Ro\left(1+\overline{DW}^2\right)\right)^{1/2}}.
\ee
}

\textcolor{black}{Since the right-hand side of \rf{ita02a} contains \( Re \) through the swirl parameter \rf{spscf}, which itself is embedded in \( \overline{DW} \) (\textcolor{red}{\ref{ec}b}), this equation implicitly defines a neutral stability curve in the \( (Re_z, Re) \)-plane. Alternatively, for a fixed \( S \), we can treat \rf{ita02a} as defining \( Re \) as a function of \( k_z \) and seek to minimize it with respect to the axial wavenumber.}

\textcolor{black}{At the minimizer, given by}
\be{minzer}
\textcolor{black}{k_{z,c} = k_r \left[2 (1+ \overline{DW}^2)\right]^{-1/2},
}
\ee
\textcolor{black}{the minimal Reynolds number is}
\be{ita03}
\textcolor{black}{Re_c = \frac{3\sqrt{3}k_r^2}{2\Omega} \left( \frac{4\overline{DW}^2}{1+\overline{DW}^2} - \frac{N_{\Omega}^2}{\Omega^2} \right)^{-1/2}=\\
\frac{6\sqrt{3}k_r^2}{\Omega} \left(\frac{\overline{DW}^2}{1+\overline{DW}^2} - (1+Ro) \right)^{-1/2}.
}
\ee

\textcolor{black}{Through straightforward algebraic manipulations, the condition \textcolor{black}{\( Re = Re_c \)} leads directly to the envelope equation \rf{masterisc}, in which we, again, should take into account \rf{spscf} and (\textcolor{red}{\ref{ec}b}).}

\textcolor{black}{The azimuthal component, $-S^{-1}DW$, of vorticity $\omega_B$ due to the axial flow destabilizes the azimuthal flow in the annulus: it decreases the threshold of the Rayleigh-unstable flows ($1+Ro<0$) whereas the  Rayleigh-stable flows are destabilized by the axial flow, if
\be{lels0}
0< 1+Ro <\frac{\overline{DW}^2}{1+\overline{DW}^2}
\ee
or
\be{lels}
\frac{N_\Omega^2}{\Omega^2}-\frac{4{\overline{DW}}^2}{1+{\overline{DW}}^2} < 0.
\ee}

\textcolor{black}{In \rf{lels} we recover the inviscid LELS instability criterion of isothermal swirling flow at $Re\rightarrow \infty$, see Appendix~\ref{AppBa}.}

\textcolor{black}{In the absence of the axial flow ($\overline{DW}=0$)}, the inequality $E(Re_z, Re)<0$ exactly reproduces the result of
\cite{EY1995} for the centrifugal instability of the viscous Couette-Taylor flow:
\be{ey95}
\frac{N_{\Omega}^2}{\Omega^2} + \frac{27}{4\Omega^2} \frac{k_r^4}{Re^2} < 0.
\ee

\subsubsection{Growth rate of the \textcolor{black}{centrifugal LELS instability of isothermal viscous swirling flows}}
The dispersion relation \rf{poly} factorizes for isothermal \textcolor{black}{viscous} swirling flows as follows:
\be{pdrf}
\left(\lambda^2 + 2\frac{|\bm{k}|^2}{Re}\lambda+ \frac{|\bm{k}|^4}{Re^2} + 4\Omega^2\frac{k_z^2}{|\bm{k}|^2}\left(1+Ro\left(1+\overline{DW}^2\right) \right) \right)\left(\lambda+\frac{|\bm{k}|^2}{RePr}\right),
\ee
where $|\bm{k}|^2$ is defined in equation \rf{normk}. The simple root associated with the second factor is always negative and corresponds to damped modes of the perturbation. The other two roots:
\be{gri}
\lambda=-\frac{|\bm{k}|^2}{Re}\pm 2\Omega \frac{k_z}{|\bm{k}|}\sqrt{-1-Ro\left(1+\overline{DW}^2\right)},
\ee
determined by the quadratic polynomial in the first factor, can exhibit positive real parts if and only if $Re$ exceeds the marginal value specified on the right-hand side of Eq.~\rf{ita02a}, which was derived from the condition $a_0 = 0$. In this case, one mode is damped, and the other is amplified with a frequency  $m\Omega+\frac{k_zW}{S}$, according to \rf{eigp}.

This analysis confirms that $a_0 < 0$ is the sole instability condition following from the three inequalities of the criterion \rf{lsh} for viscous isothermal swirling flows. The envelope \rf{masterisc} of the individual instability domains specified by the inequality $a_0 < 0$ yields viscous extension of the inviscid LELS instability criterion \rf{lels} for isothermal swirling flows.

\textcolor{black}{The growth rate of the centrifugal instability can be estimated analytically at any position within the annulus using equation \rf{gri}, particularly at the geometric mean radius, for which the expressions for $DW_g$ are provided in Table 1.}

\textcolor{black}{Specifically, for CCF ($W = 0$), we obtain
\be{}
\lambda = -\frac{|\bm{k}|^2}{Re} \pm 2\Omega \frac{k_z}{|\bm{k}|}\sqrt{-1 - Ro}.
\ee
Hence, the marginal stability condition for Rayleigh-unstable CCF $(1 + Ro < 0)$ is given by
\be{}
Re_m^{CCF} = \pm \frac{|\bm{k}|^3}{k_z}\frac{1}{2\Omega \sqrt{-1 - Ro}}.
\ee
For Rayleigh-stable CCF $(1 + Ro > 0)$, the epicyclic frequency is given by $N_{\Omega} = 2\Omega \sqrt{1 + Ro}$.}

\subsubsection{Spiral Couette Flow (SCF)}

To illustrate the inviscid LELS criterion \rf{lels} and its viscous extension, $E(Re_z, Re)<0$, with $E$ given by Eq.~\rf{masterisc},
we consider the \textit{enclosed Spiral Couette Flow} as a base flow. Then, $W_0=W_1$, i.e. the sliding speed of the inner cylinder, the dimensionless axial velocity $W(r)$ is given by equations \rf{wscf} and \rf{C2e}, while $V(r)$ is defined by \rf{ssv}, \textcolor{black}{see Table~\ref{tab:1}.}

\begin{figure}
\centering
\includegraphics[width=.8\textwidth]{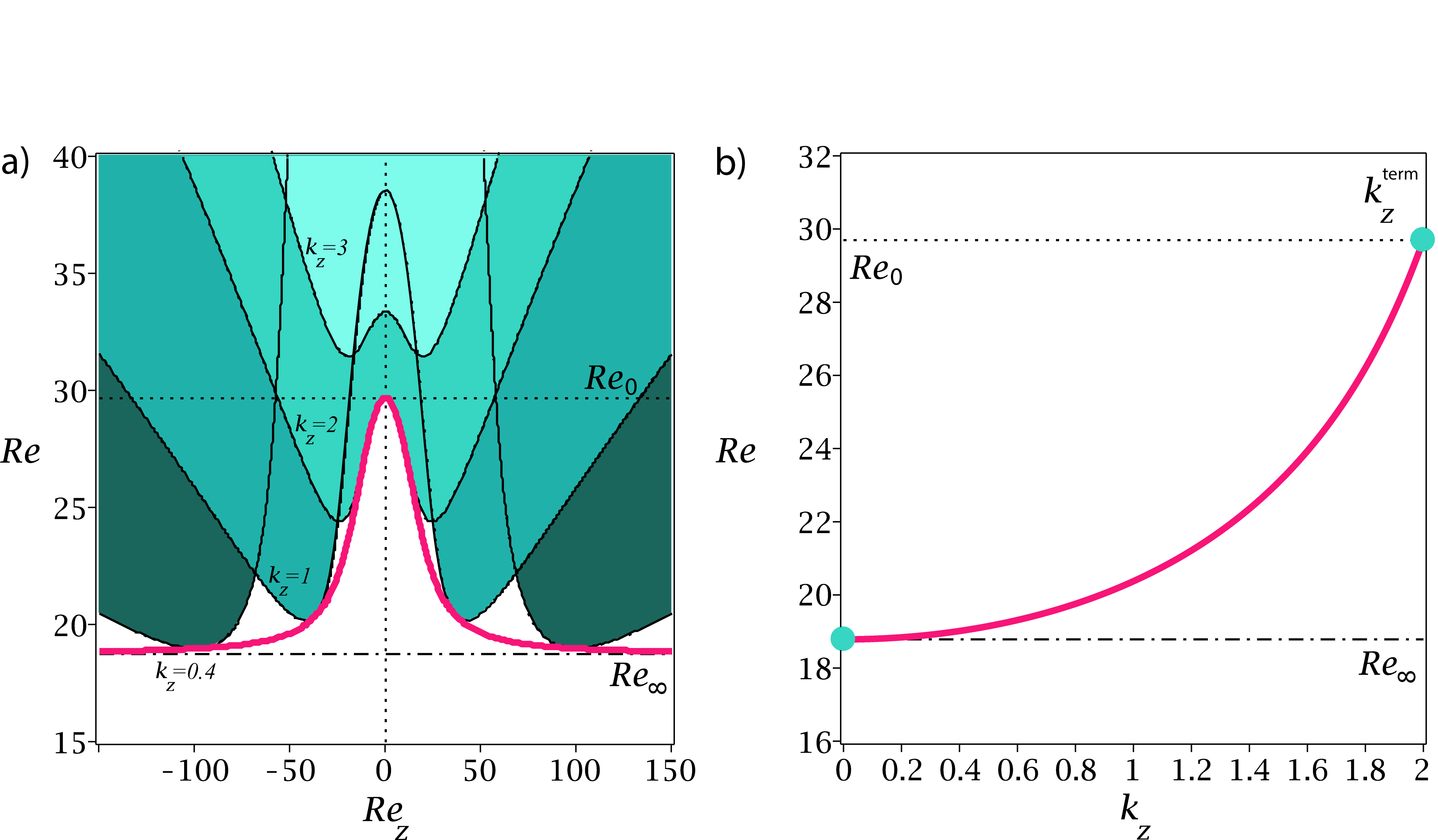}
\includegraphics[width=.8\textwidth]{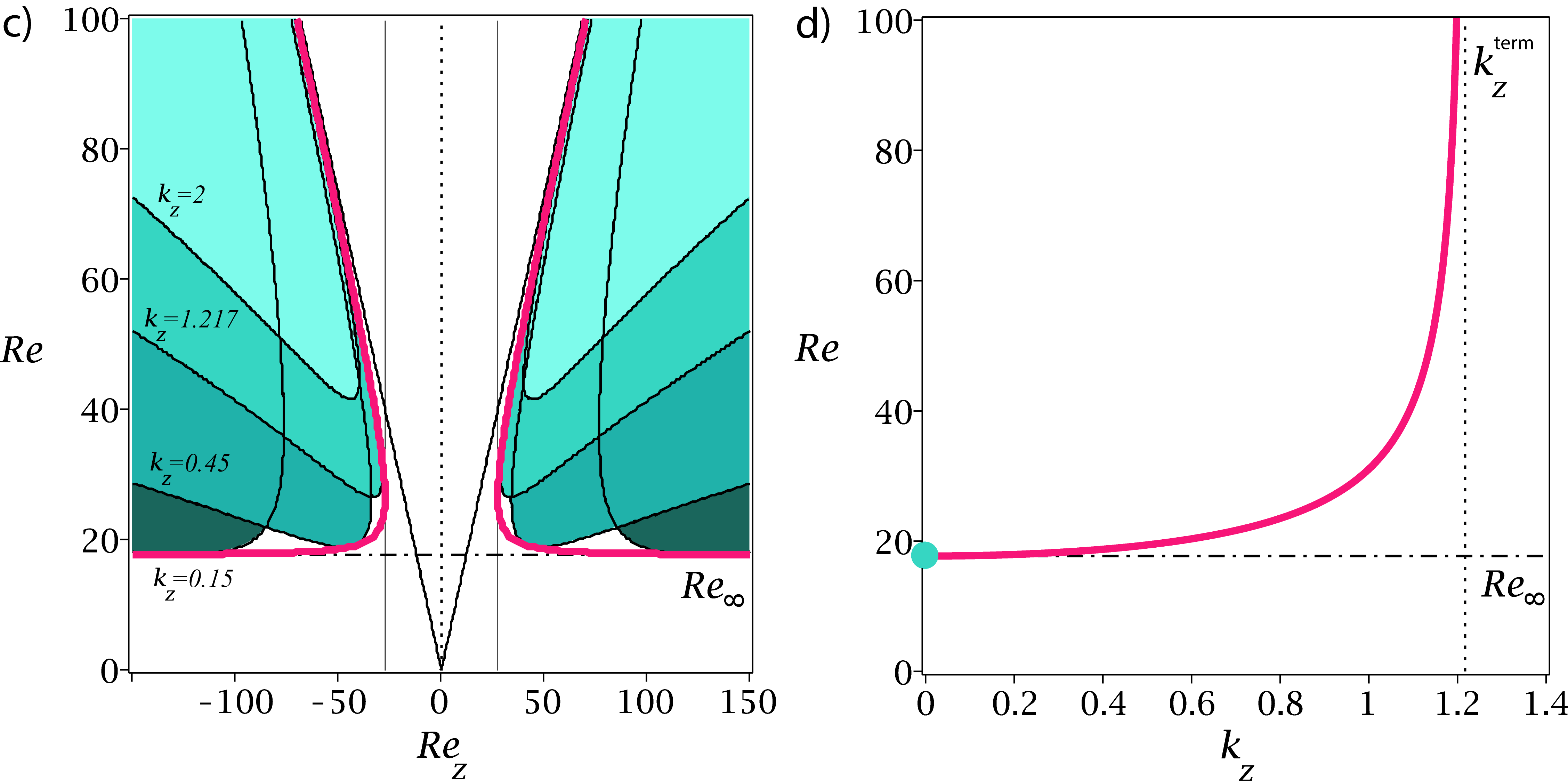}
\caption{ \textcolor{black}{Isothermal enclosed SCF with \(\eta = 0.4\), and \(k_r = 2\sqrt{2}\). (a, c) Greenish centrifugal LELS instability domains in the \((Re_z, Re)\)-plane, parameterized by \(k_z\), and (thick red curves) their envelope \(\rf{masterisc}\) for (a) \(\mu = 0\) and $Re_{\infty}\approx 18.78$ and (c) \(\mu = 0.5\) and $Re_{\infty}\approx 17.71$. (b) Variation of \(k_z\) from $0$ to the terminal value \(k_z^{\rm term} = \frac{\sqrt{2}}{2} k_r = 2\) according to \(\rf{kze}\) as \(Re\) increases from \(Re_{\infty}\approx 18.78\) (dot-dashed line) to \(Re_0\approx 29.70\) (dotted line) for $\mu=0$. (d) Variation of \(k_z\) from $0$ to the terminal value \(k_z^{\rm term} = \frac{\sqrt{-2Ro}}{2} k_r \approx 1.217\) according to \(\rf{kze}\) as \(Re\) increases from \(Re_{\infty}\approx 17.71\) (dot-dashed line) to infinity for $\mu=0.5$. }}
\label{fig:2abcd}
\end{figure}

\begin{itemize}
\item \textcolor{black}{\uline{Parameterisation by the axial wavenumber $k_z$}}. \textcolor{black}{The neutral stability curves \rf{ita02} for the enclosed SCF with $W(r)$ given by \rf{wscf} and \rf{C2e}, \(\eta = 0.4\), and \(k_r = 2\sqrt{2}\) are shown in the $(Re_z,Re)$-plane in figure~\ref{fig:2abcd} for different values of $k_z$ and $\mu$ as black solid curves bounding the greenish centrifugal LELS instability domains. All computations for the SCF are performed at the geometric mean radius \rf{mgr}.}
\end{itemize}

    For Rayleigh-unstable flows ($N_{\Omega}^2<0$ or $Ro < -1$), their envelope \rf{masterisc} is a single curve, \textcolor{black}{shown as a red thick curve in figure~\textcolor{red}{\ref{fig:2abcd}a} for $\mu=0$.} The envelope has a maximum $Re=Re_0$ when $Re_z=0$ and a horizontal asymptote $Re=Re_{\infty}$ as $|Re_z|\rightarrow \infty$, where:
\refstepcounter{equation}
$$
     Re_0=\frac{3\sqrt{3}k_r^2}{4\Omega\sqrt{-Ro-1}}, \quad
     Re_\infty=\frac{3\sqrt 3 k_r^2}{4\Omega\sqrt{-Ro}}.
\eqno{(\theequation{\mathit{a},\mathit{b}})}\label{re0i}
$$
\textcolor{black}{The asymptotes $Re=Re_{\infty}$ are shown in figure~\ref{fig:2abcd} by the black dot-dashed lines.}

As $Ro \rightarrow -1$,  $Re_0 \rightarrow \infty$, and for Rayleigh-stable flows $(-1 < Ro < 0)$, the envelope splits into two curves, each with vertical tangents at $Re_z = \pm Re_z^{\min}$, where
  \be{rezm}
  Re_z^{\min}= \frac{3}{2}\frac{\sqrt{3}k_r^2}{DW}\left(1 + \sqrt{Ro+1}\right)
  \ee
is the minimal critical axial Reynolds number destabilizing Rayleigh-stable isothermal azimuthal flows. The vertical solid lines in figure~\textcolor{red}{\ref{fig:2abcd}c} show $\pm Re_z^{\min}\approx \pm27.43$.

Additionally, in figure~\textcolor{red}{\ref{fig:2abcd}c}, the classical inviscid \textcolor{black}{LELS} criterion \rf{lels} is represented by the oblique black solid lines touching the upper parts of the envelope \rf{masterisc} as $Re \rightarrow \infty$. While \rf{lels} does not apply to Rayleigh-unstable flows, its viscous extension, \textcolor{blue}
{$E(Re_z, Re)<0$}, defines the stability boundary in this case, as given by the envelope \rf{masterisc}, see figure~\textcolor{red}{\ref{fig:2abcd}a}.

\textcolor{black}{We can interpret these results as indicating that the axial flow induced by the sliding inner cylinder introduces a destabilising mechanism into the Couette-Taylor flow. The value $Re_{\infty}$ corresponds to the minimum rotation rate below which the sliding of the inner cylinder is no longer able to destabilise the SCF.}

Expressing $\overline{DW}$ from the envelope equation \rf{masterisc}, substituting the result into \rf{minzer}, and subsequently utilizing the explicit expressions provided in \rf{rd} and \rf{re0i} for isothermal swirling flows, \textcolor{black}{we derive the following analytical expression, which describes the dependence of the critical axial wavenumber $k_z$ of the perturbations on the critical Reynolds number $Re$ along the envelope} \textcolor{black}{\rf{masterisc}:
 \be{kze}
     |k_z|(Re)=k_r\frac{\sqrt{2}}{2}\sqrt{\frac{1-\left(\frac{Re_{\infty}}{Re}\right)^2}{1-\left(\frac{Re_{\infty}}{Re_0}\right)^2}}.
 \ee}
\textcolor{black}{In particular, it implies}  $|k_z|\rightarrow 0$  as $Re \rightarrow Re_{\infty}$ and reaches its terminal value at $Re = Re_0$,
\be{kzt0}
k_z^{\rm term}=\frac{\sqrt{2}}{2} k_r,
\ee
for Rayleigh-unstable flows, figure~\textcolor{red}{\ref{fig:2abcd}b}, and
\be{kzti}
k_z^{\rm term}=\frac{\sqrt{-2Ro}}{2} k_r,
\ee
as $Re \rightarrow \infty$ for Rayleigh-stable flows, figure~\textcolor{red}{\ref{fig:2abcd}d}, agreeing with numerical studies by \cite{L1964}, \cite{NT1982}, \cite{AW1993}, and \cite{MM2000}.

Additionally, figure~\textcolor{red}{\ref{fig:2abcd}b,d} shows that the spectrum of axial wavenumbers for perturbations differs between Rayleigh-unstable and Rayleigh-stable flows. In Rayleigh-unstable flows \textcolor{black}{(figure~\textcolor{red}{\ref{fig:2abcd}b})}, the destabilizing centrifugal force broadens the wavenumber range, whereas in Rayleigh-stable flows \textcolor{black}{(figure~\textcolor{red}{\ref{fig:2abcd}d})}, it is limited to lower \( k_z \) values. Long axial wavelength perturbations are favored as \( Re \) approaches \( Re_{\infty} \) (and $Re_z \rightarrow \infty$).

\begin{figure}
\centering
\includegraphics[width=.8\textwidth]{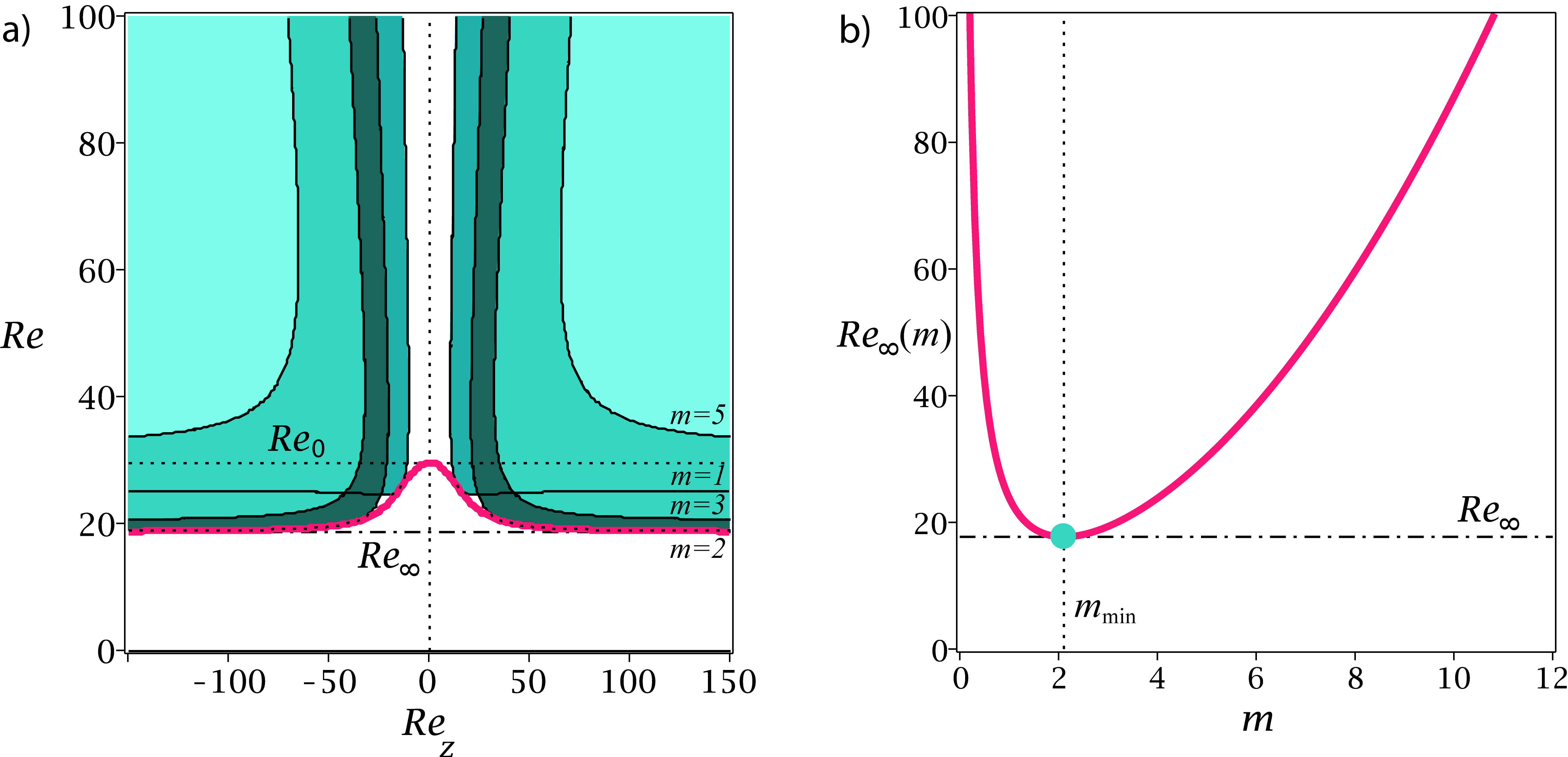}
\includegraphics[width=.8\textwidth]{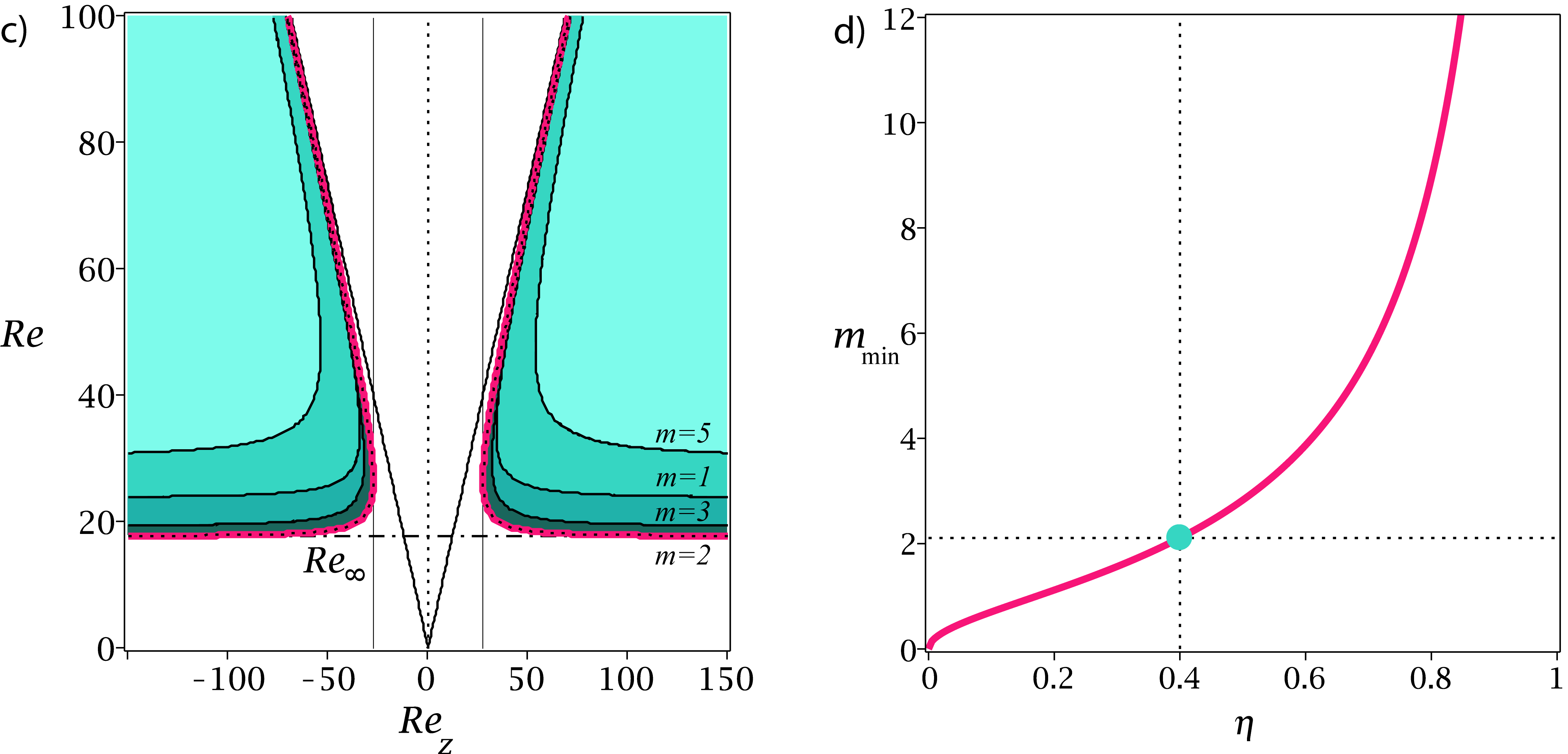}
\caption{\textcolor{black}{Isothermal enclosed SCF with $\eta = 0.4$ and $k_r = 2\sqrt{2}$. (a, c) Greenish centrifugal LELS instability domains in the $(Re_z, Re)$-plane, parameterised by $m$, and (thick red curves) their envelope $\rf{masterisc}$ for (a) $\mu = 0$, with $Re_{\infty} \approx 18.78$, and (c) $\mu = 0.5$, with $Re_{\infty} \approx 17.71$. (b) The asymptotic value $Re_\infty(m)$ $\rf{reterm}$ for individual neutral stability curves when $\mu = 0.5$, showing (green circle) a minimum equal to $Re_{\infty} \approx 17.71$, given by (\textcolor{red}{\ref{re0i}b}), at $m = m_{\min} \approx 2.108$, as given by $\rf{maw}$ or $\rf{mawe}$. (d) Variation of $m_{\min}$ with $\eta$ for $\mu = 0.5$, according to $\rf{mawe}$, where the green circle indicates the minimal value $m \approx 2.108$ at $\eta = 0.4$.}}
\label{fig:3abcd}
\end{figure}

From Eq. (\ref{masterisc}), we determine that at the points of the envelope,
\begin{equation} \label{rgi}
    \textcolor{black}{
    Re = Re_{\infty}\left(1 + \frac{27k_r^4}{8DW^2Re_z^2}+ \ldots\right)
    }
\end{equation}
as \( Re_z \to \infty \), where \( Re_{\infty} \) is defined by Eq. (\textcolor{red}{\ref{re0i}b}).

Substituting Eq. (\ref{rgi}) into Eq. (\ref{kze}), we find the relation between axial and radial wavenumbers on the stability boundary,
\begin{equation} \label{kzgr}
    k_z = \frac{3}{4} \frac{\sqrt{-6Ro}}{DW} \frac{k_r^3}{Re_z} + O(Re_z^{-3}),
\end{equation}
as \( Re_z \to \infty \), which confirms the numerical result from \cite{AW1993} that \( k_z \sim \frac{1}{Re_z} \) in this limit; \textcolor{black}{cf. figure 11 of their work.} Unlike \cite{AW1993}, however, we have explicitly derived the coefficient of \( Re_z^{-1} \), showing its dependence on the flow properties.
Indeed, substituting \( Ro \) evaluated at the mean geometric radius for \( \mu = 0 \) from Eq.~\rf{omro} to Eq. (\ref{kzgr}), we obtain
\begin{equation} \label{kzgreta}
    k_z = \frac{3\sqrt{6}}{4DW\sqrt{1 - \eta}} \frac{k_r^3}{Re_z} + O(Re_z^{-3}).
\end{equation}
\textcolor{black}{Thus, we confirm that the coefficient of $Re_z^{-1}$ increases as $\eta \to 1$, in agreement with the numerical findings of \cite{AW1993}; cf. equation 9 and table II of their work.}

\begin{itemize}
\item \textcolor{black}{\uline {Parameterisation by azimuthal wavenumber $m$}}. Utilizing the relationship \rf{ec} between the azimuthal and axial wavenumbers, with $m = r k_{\varphi}$, we can interpret Eq. \rf{ita02} as defining a family of curves parameterized by the azimuthal wavenumber $m \in \mathbb{R}$. Although in this parametrization the individual neutral stability curves \rf{ita02} in the $(Re_z, Re)$-plane look differently (cf.  figures~\ref{fig:2abcd} and \ref{fig:3abcd}), they have the same envelope  \rf{masterisc}, as verified either through direct calculation of the envelope or by minimizing the critical Reynolds number in Eq. \rf{ita02a} with respect to $m$. Specifically, considering the relation \rf{spscf}, we find that the neutral stability curves \rf{ita02}, parameterized by $m$, exhibit horizontal asymptotes in the $(Re_z,Re)$-plane as $Re_z \to \infty$:
\be{reterm}
Re_{\infty}(m) = \frac{(k_r^2 r^2 + m^2)^{3/2}}{2|m|\Omega r^2 \sqrt{-Ro}}.
\ee
\end{itemize}

\textcolor{black}{In figure~\textcolor{red}{\ref{fig:3abcd}a,c} the greenish domains of centrifugal instability bounded by the solid black neutral stability curves \rf{ita02}, parameterized by \(m\), are shown in the \((Re_z, Re)\)-plane for the isothermal enclosed SCF with $W(r)$ given by \rf{wscf} and \rf{C2e}, \(\eta = 0.4\), and \(k_r = 2\sqrt{2}\) for various values of $m$ and $\mu$. The neutral stability curves are flattening as $Re_z \to \infty$, thus confirming the asymptotes \rf{reterm}. The oblique black solid lines in figure~\textcolor{red}{\ref{fig:3abcd}c} indicate the inviscid LELS criterion \rf{lels}. Vertical solid lines in figure~\textcolor{red}{\ref{fig:3abcd}c} show $\pm Re_z^{\min}\approx \pm27.43$ given by \(\rf{rezm}\). All computations for the SCF are performed at the mean geometric radius \rf{mgr}.}

By differentiating Eq. \rf{reterm} with respect to $m$, we determine the minimizer of $Re_{\infty}(m)$:
\be{maw}
m_{\rm min} = \pm\frac{\sqrt{2}}{2} rk_r,
\ee
which thus defines the terminal azimuthal wavenumber as the floor of $m_{\rm min}$: $m_{\rm term}=\pm\left\lfloor |m_{\rm min}|\right\rfloor \in \mathbb{Z}$, previously known only through numerical computations \textcolor{black}{\citep{AW1993} (cf. figure 5 and table II of their work) and \cite{MM2000} (cf. figure 2 of their work).} Substituting \rf{maw} into Eq. \rf{reterm} yields the minimal value of $Re_{\infty}(m)$, exactly matching the limiting value $Re_{\infty}$ in Eq.~(\textcolor{red}{\ref{re0i}b}), see figure~\textcolor{red}{\ref{fig:3abcd}a,b}.

At the mean geometric radius \rf{mgr}, Eq. \rf{maw} takes the form
\be{mawe}
m_{\min} = \pm\frac{k_r}{2}\frac{\sqrt{2\eta}}{1 - \eta},
\ee
allowing to find the terminal azimuthal wavenumber $m_{\rm term}=\pm\left\lfloor |m_{\rm min}|\right\rfloor$ at different ratios of the cylinder radii, $\eta$, see figure~\textcolor{red}{\ref{fig:3abcd}d}. \textcolor{black}{Comparing figure~\textcolor{red}{\ref{fig:3abcd}d} with figure 13 from \cite{AW1993}, one can see full qualitative and very good quantitative agreement between our analytical formula \rf{mawe} and their earlier numerical results.}

\textcolor{black}{For example, using the parameters from figure~\ref{fig:3abcd}, namely $\eta = 0.4$ and $k_r = 2\sqrt{2}$, in equation~\rf{mawe}, we find that $m_{\min} \approx 2.108$. This yields a terminal axial wavenumber of $m_{\rm term} = 2$. In figure~\ref{fig:3abcd}, the neutral stability curve for azimuthal wavenumber $m = 2$ is nearly indistinguishable from the envelope at large $Re_z$, unlike the curves for $m = 1, 3,$ and $5$. This is because the terminal Reynolds number $Re_{\infty}(2) \approx 17.74$, as given by equation~\rf{reterm}, is very close to the horizontal asymptote of the envelope: $Re_{\infty} \approx 17.71$ for the Rayleigh-stable case with $\mu = 0.5$ (figure~\textcolor{red}{\ref{fig:3abcd}c}), and $Re_{\infty} \approx 18.78$ for the Rayleigh-unstable case with $\mu = 0$ (figure~\textcolor{red}{\ref{fig:3abcd}a}), as given by (\textcolor{red}{\ref{re0i}b}). The horizontal asymptotes $Re=Re_{\infty}$ are shown in figure~\textcolor{red}{\ref{fig:3abcd}a,c} by black dot-dashed lines.}

\begin{itemize}
\item \textcolor{black}{\uline {Explicit expression for neutral stability surface in the $(Re_z, Re_2, Re)$-space}}.
The envelope equation \rf{masterisc} is universal for all isothermal swirling flows. For example, for flows between differentially rotating cylinders, the following relationship between the ratio of the angular velocities of the cylinders and the ratio of their radii, defined in \rf{etamu}, holds:
\refstepcounter{equation}
$$
\mu = \eta \frac{Re_2}{Re}, \quad Re = \frac{R_1 \Omega_1 d}{\nu}, \quad Re_2 = \frac{R_2 \Omega_2 d}{\nu},
\eqno{(\theequation{\mathit{a},\mathit{b},\mathit{c}})}\label{murere2}
$$
where $Re$ and $Re_2$ are the Reynolds numbers of the inner and outer cylinders, respectively \citep{MM2000, MM2002}.
This reformulates the azimuthal velocity \rf{ssv} and the expressions \rf{omro} evaluated at the mean geometric radius \rf{mgr}, which incorporate
the parameter $\mu$, in terms of $Re_2$. As a result, Eq.~\rf{masterisc} yields \textcolor{black}{the explicit analytical expression for} the
neutral stability surface within the $(Re_z, Re_2, Re)$-space:
\be{oscfrr2}
\left[\frac{(1 + \eta)DW_gRe_z}{2(Re-\eta Re_2)}\right]^2 +\frac{27k_r^4(1 + \eta)^2 - 16(1 - \eta)(Re + Re_2)(\eta Re - Re_2)}
{27k_r^4(1 + \eta)^2 - 16(1 - \eta)(Re + Re_2)(Re-\eta Re_2)}=0.
\ee
\end{itemize}
\textcolor{black}{To the best of our knowledge, such a closed-form equation has not previously been reported in the literature on isothermal swirling flows, where numerical results are prevailing.}

\begin{figure}
\centering
\includegraphics[width=.44\textwidth]{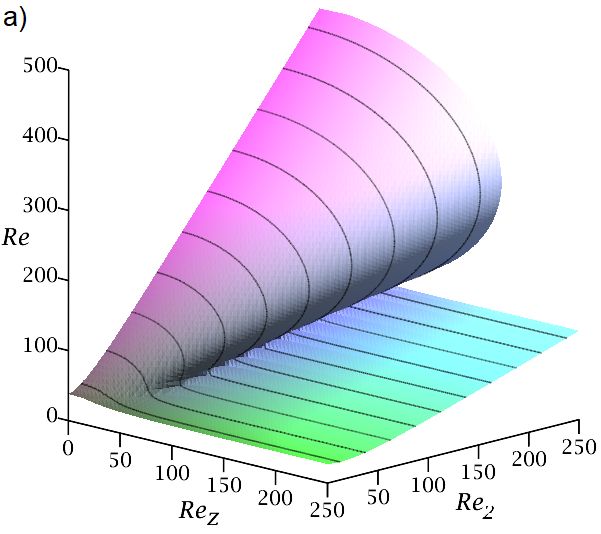}
\includegraphics[width=.39\textwidth]{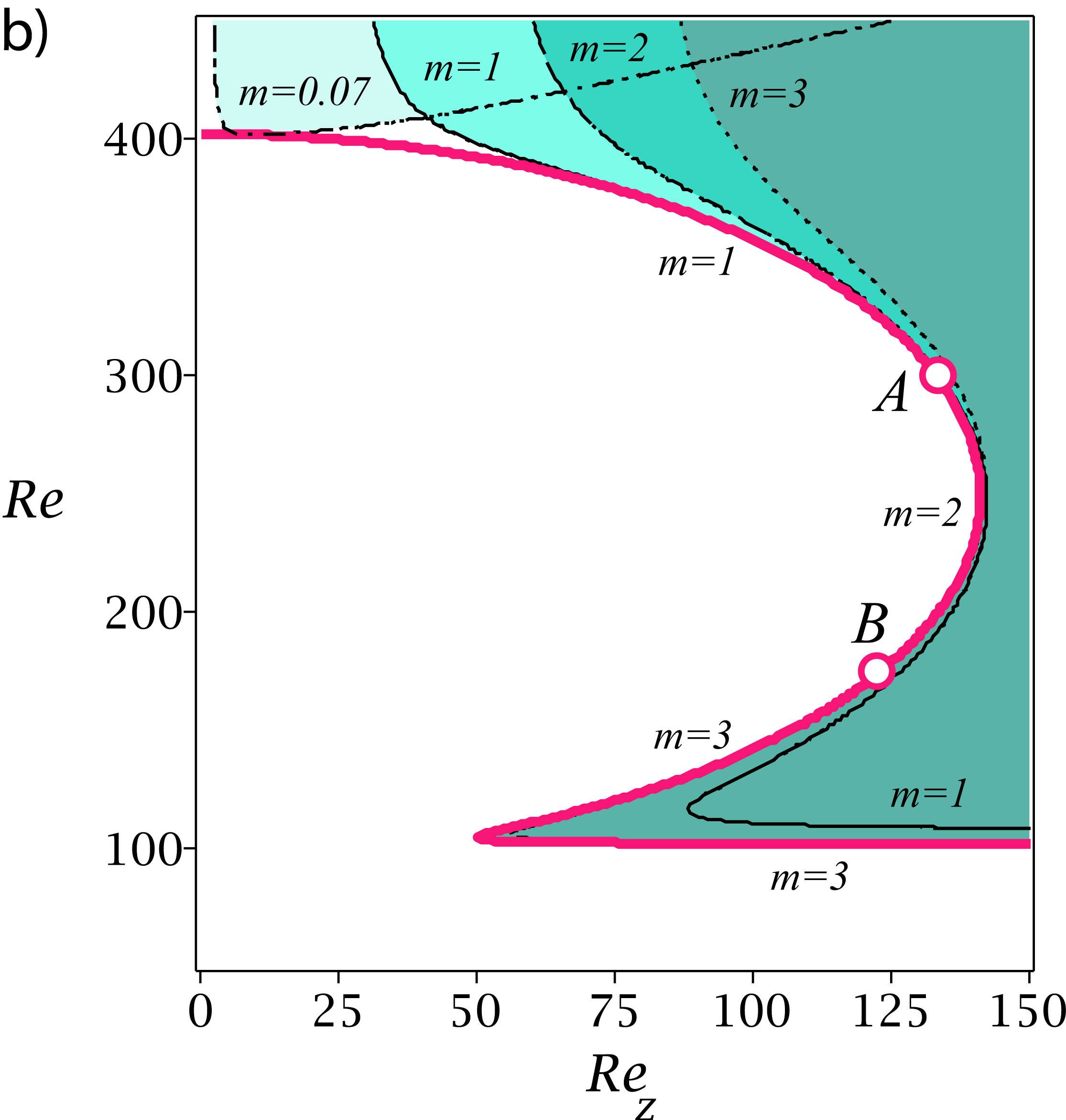}
\includegraphics[width=.84\textwidth]{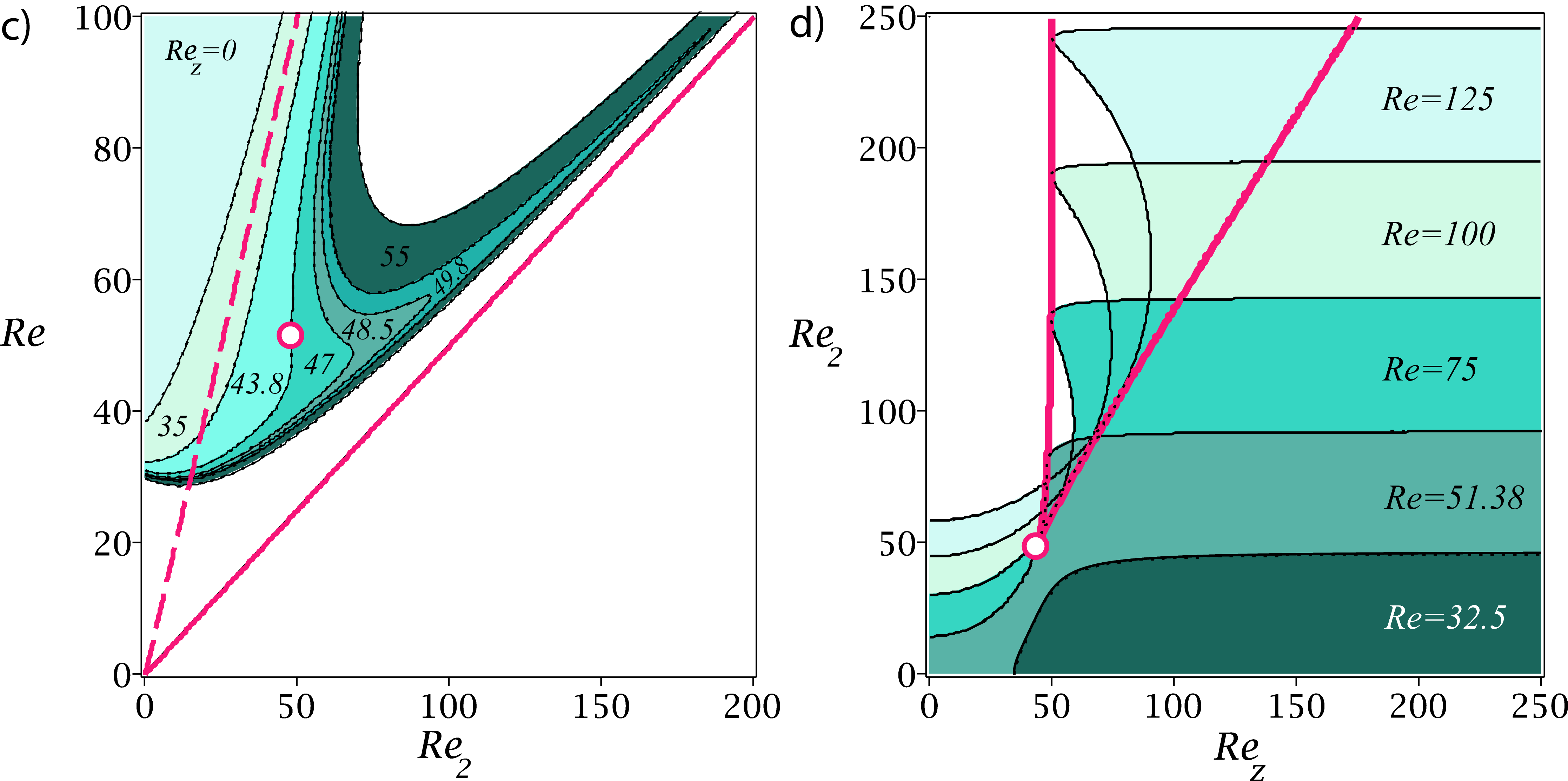}
\caption{\textcolor{black}{For the open SCF with \( DW_g \) given by Eq.~\rf{wgdwgscfo}, with \( k_r = \pi \) and \( \eta = 0.5 \):
(a) The folded surface of the envelope \rf{oscfrr2} in \((Re_z, Re_2, Re)\)-space.
(b) For \( Re_2 = 200 \), the thick solid red line shows the envelope \rf{oscfrr2} in the \((Re_z, Re)\)-plane. Shaded areas indicate individual instability domains defined by Eqs.~\rf{ita02} and \rf{ec}, corresponding to \( m = 0.07, 1, 2, 3 \).
(c) Cross-sections of the instability domains (shaded regions) defined by the envelope surface \rf{oscfrr2} in the \((Re_2, Re)\)-plane, covering a range of \( Re_z \) values from 0 to 55.
(d) Shaded cross-sections of the surface \rf{oscfrr2} indicating instability domains at various \( Re \) values in the \((Re_z, Re_2)\)-plane, which for $Re>51.28$ touch the thick red curve \rf{cusp} that features a cusp point at \((43.79, 48.21)\) shown by the open circle. \label{Figure3}}}
\end{figure}

Note that the ``individuality'' of an isothermal swirling flow enters the equation \rf{oscfrr2} only by means of the radial derivative, $DW_g$, of its
axial velocity evaluated at the mean geometric radius \rf{mgr}.
\textcolor{black}{The expressions for $DW_g$ derived in Appendix A are presented in Table~\ref{tab:1} for the spiral Poiseuille flow \rf{dwgspf}, as well as for both open \rf{wgdwgscfo} and enclosed \rf{dwgcfe} spiral Couette flows.}

In the case of the open Spiral Couette Flow the expression for $DW_g$ given
by Eq.~\rf{wgdwgscfo} is especially simple: \textcolor{black}{$DW_g=\frac{1 - \eta}{\sqrt{\eta}\ln\eta}$}. The surface \rf{oscfrr2} and its cross-sections are shown in figure~\ref{Figure3} for the open SCF with \(k_r = \pi\) and \(\eta = 0.5\) to
facilitate comparison with the numerical results of \cite{MM2000}.  All computations for the open SCF are performed at the mean geometric radius \rf{mgr}.

At $Re_z = 0$, the surface yields the neutral stability curve of the Couette-Taylor flow obtained by \textcolor{black}{the same} local stability analysis by \cite{EY1995}:
\be{eyrr2}
(Re + Re_2)(\eta Re - Re_2)=\frac{27}{16}\frac{(1 + \eta)^2}{1-\eta}k_r^4,
\ee
which yields at large $Re_2$:
\be{eyrr3}
Re=\frac{1}{\eta}Re_2 + \frac{27}{16}\frac{1+\eta}{1-\eta}\frac{k_r^4}{Re_2} + O(Re_2^{-3})
\ee
meaning that \rf{eyrr2} has the inviscid Rayleigh line
\be{rl}
Re=\eta^{-1}Re_2
\ee
as its asymptote \textcolor{black}{shown by the thick dashed straight line} in figure~\textcolor{red}{\ref{Figure3}c}.

In another limit, $Re_z \rightarrow \infty$, the surface \rf{oscfrr2} produces different neutral stability curve
\be{eyrr4}
(Re + Re_2)(Re-\eta Re_2)=\frac{27}{16}\frac{(1 + \eta)^2}{1-\eta}k_r^4,
\ee
which yields at large $Re_2$:
\be{eyrr5}
Re=\eta Re_2 + \frac{27}{16}\frac{1 + \eta}{1-\eta }\frac{k_r^4}{Re_2} + O(Re_2^{-3})
\ee
providing the line of solid body rotation
\be{sbr}
Re=\eta Re_2
\ee
as the asymptote to the curve \rf{eyrr4} \textcolor{black}{shown by the thick solid straight line} in figure~\textcolor{red}{\ref{Figure3}c}.

\textcolor{black}{Cross-sections of the instability domains (shaded regions), defined by the envelope surface \rf{oscfrr2} in the $(Re_2, Re)$-plane and shown in Figure~\textcolor{red}{\ref{Figure3}c}, span a range of $Re_z$ values from 0 to 55. They demonstrate that the axial velocity component destabilises the Rayleigh-stable base flow and elucidate, in detail, the transition of the asymptotes to the stability boundaries from the Rayleigh line to the solid-body rotation line, as illustrated in Figure~\textcolor{red}{\ref{Figure3}c}.}

\textcolor{black}{A similar phenomenon occurs in the standard magnetorotational instability of magnetohydrodynamics, known as the Velikhov–Chandrasekhar paradox \citep{BH1991}. In that case, the stability boundary of magnetised Couette–Taylor flow drops below the Rayleigh line and approaches the solid-body rotation line when an axial magnetic field is applied, provided that the magnetic Prandtl number (the ratio of kinematic viscosity to magnetic diffusivity) differs from unity \citep{WB2002, KPS2011, KS2011}.}
\begin{itemize}
\item \textcolor{black}{\uline {Folds, pleat, and cusp of the envelope surface for the open SCF}}.
\textcolor{black}{The transition between the Rayleigh line and the solid-body rotation line arises from the two folds of the surface \rf{oscfrr2}, visible in Figure~\textcolor{red}{\ref{Figure3}a}, which converge at the pleat located at the point $(43.79, 48.21, 51.38)$ in the $(Re_z, Re_2, Re)$-space. These folds are projected onto a thick red curve in the $(Re_z, Re_2)$-plane, which exhibits a cusp at $(43.79, 48.21)$, highlighted in Figure~\textcolor{red}{\ref{Figure3}d} by the open circle. The open circle at $(48.21, 51.38)$ in Figure~\textcolor{red}{\ref{Figure3}c} marks the projection of the pleat onto the $(Re_2, Re)$-plane. The neutral stability surface with a pleat for the open SCF was first computed numerically by \cite{MM2000}; our model now provides an explicit equation \rf{oscfrr2}, allowing for the analytical computation of this surface’s projections, pleat, and cusp points.}
\end{itemize}

Indeed, differentiating \rf{oscfrr2} with respect to \( Re \) and setting the result to zero yields the projection of the surface folds onto the \((Re_2, Re)\)-plane:
\ba{foldc}
&  729(1 + \eta)^4k_r^8 + 432(1 - \eta)(1 + \eta)^2(Re + Re_2)(Re_2\eta^2 - 2Re\eta + Re_2)k_r^4& \nn\\
&- 128(1 - \eta)^2(Re + Re_2)^2(Re-Re_2\eta)(Re_2\eta^2 - 2Re\eta + Re_2)=0.&
\ea

Interpreting \rf{oscfrr2} as a family of curves in the \((Re_2, Re_z)\)-plane parameterized by \( Re \), we compute its discriminant as a polynomial in \( Re \) to obtain the folds and pleat projection onto the \((Re_2, Re_z)\)-plane for the open SCF flow \rf{wgdwgscfo}:
\ba{cusp}
& 9\left(k_r^4 + \frac{2Re_2^2}{27}\right)Re_z^8
+ 126(\ln2)^2Re_2^2\left(k_r^4 + \frac{5Re_2^2}{63}\right)Re_z^6&\nn\\
&+2(\ln2)^4(27k_r^4-4Re_2^2)\left(243k_r^8 - Re_2^2(27k_r^4 + 4Re_2^2)\right)Re_z^4&\\
&- 118098(\ln2)^6Re_2^2\left(k_r^4 + \frac{4Re_2^2}{27}\right)^2\left(k_r^4 + \frac{4Re_2^2}{243}\right)Re_z^2+ 4782969(\ln2)^8\left(k_r^4 + \frac{4Re_2^2}{27}\right)^4k_r^4= 0.&\nn
\ea

By differentiating \rf{foldc} with respect to \( Re \), solving the resulting cubic equation, and substituting the roots back into \rf{foldc}, we find \( Re_2 \) at the cusp (and pleat) point for \( \eta = 0.5 \):
\be{redc}
Re_2=\frac{3\sqrt{(48\sqrt{3} + 207)^{2/3} - (48\sqrt{3} + 207)^{1/3} + 33}}{2(48\sqrt{3} + 207)^{1/6}}k_r^2\approx 4.88k_r^2,
\ee
which, for \( k_r = \pi \), gives \( Re_2 \approx 48.21 \). Then, from \rf{cusp}, \( Re_z \approx 43.79 \), and from \rf{foldc}, \( Re \approx 51.38 \), locating the cusp and pleat points for the open SCF with \( k_r = \pi \) and \( \eta = 0.5 \) as shown in figure~\textcolor{red}{\ref{Figure3}}.

\textcolor{black}{In addition, figure~\textcolor{red}{\ref{Figure3}b} shows, in the $(Re_z, Re)$-plane, the greenish individual instability domains defined by Eqs.~~\rf{ita02} and \rf{ec} for azimuthal wavenumbers $m = 0.07, 1, 2, 3$. The boundaries of these domains are tangent to the envelope \rf{oscfrr2} shown as a thick red line, but intersect in such a way that point $A$ at $(133.67, 299.53)$ lies at the intersection of the neutral stability curves for $m = 1$ and $m = 2$, while point $B$ at $(122.59, 174.45)$ lies at the intersection of those for $m = 2$ and $m = 3$. This implies that the $m = 1$ mode is critical before point $A$; $m = 2$ is critical between points $A$ and $B$; and $m = 3$ becomes critical after point $B$, as illustrated in Figure~\textcolor{red}{\ref{Figure3}b}. The azimuthal mode $m = 3$ is terminal, as indicated by Eq.~~\rf{mawe} and consistent with \citep{MM2000}. The instability domain corresponding to $m = 0.07$ is also shown; as $m \rightarrow 0$, this domain reduces to the semi-infinite interval $[Re_0, \infty)$, where $Re_0$ is given by Eq.~(\textcolor{red}{\ref{re0i}a}).}


\begin{figure}
\centering
\includegraphics[width=0.9\textwidth]{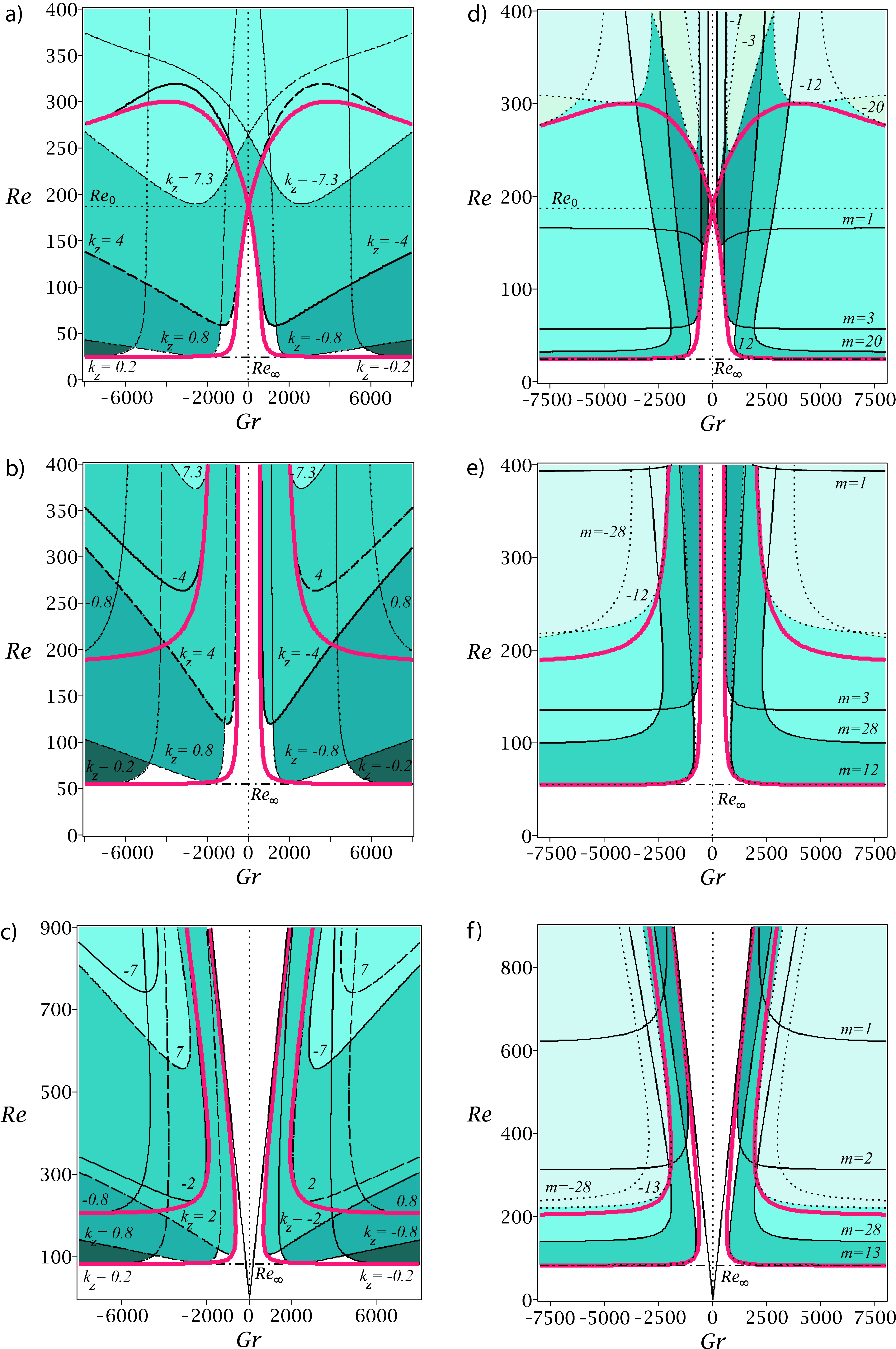}
\caption{For BCF with $W(r)$ given by \rf{WBCF} and \rf{C1}, and $\eta=0.8$, $Pr=5.5$, $\gamma=0.0004$, and \(k_r = 4\sqrt{2}\) the neutral stability curves $a_0=0$ with $a_0$ given by Eq.~\rf{polpra0} in the \((Gr, Re)\)-plane, parameterized (a-c) by \(k_z\) or (d-f) by $m$ for the three different $\mu$: (a), (d) Rayleigh unstable flow, \(\mu = 0\); (b), (e) modified Rayleigh line, \(\mu = \mu_R \approx 0.63935\) (from \(\rf{murl}\)); and (c), (f) Rayleigh stable flow, \(\mu = 0.8\). The thick red curves show their envelope that has a horizontal asymptote $Re=Re_{\infty}$ for $m>0$ with (a), (d) $Re_{\infty}\approx24.66$, (b), (e) $Re_{\infty}\approx55.13$, and (c), (f) $Re_{\infty}\approx 83.99$, cf. figure~\ref{Figure8}. In (a), (d) the two envelope branches intersect at $Re=Re_0$, where $Re_0\approx187.14$ is given by Eq.~\rf{re0new}. The oblique black solid lines in (c), (f) indicate the new unified visco-thermodiffusive extension \rf{elsscf1} of the LELS criterion. All computations for the BCF are performed at the mean geometric radius \rf{mgr}. \label{Figure6}}
\end{figure}

\subsection{Visco-diffusive non-isothermal swirling flows - baroclinic Couette flow (BCF)}\label{Sec5_2}
\subsubsection{Stationary \textcolor{black}{centrifugal} instability as a combination of GSF and LELS instabilities}

In general, equation $a_0=0$ with $a_0$ given by Eq.~\rf{polpra0}, defines a family of marginal stability curves for the Baroclinic Couette Flow (BCF) in the \((Gr, Re)\)-plane, parameterized by the axial wavenumber \( k_z \) (or equivalently \( m \) via \(\rf{ec}\)), as shown in figure~\ref{Figure6}.

\begin{itemize}
\item \textcolor{black}{\uline {Envelope of the family of centrifugal instability domains}}.
The individual marginal stability curves of the BCF have an envelope determined by computing the discriminant of the expression \rf{polpra0}, treated as a polynomial in \(k_z\) or \(m\) \citep{BG1992,HW1953,H2020}.
\end{itemize}

\begin{itemize}
\item \textcolor{black}{\uline {Modified Rayleigh line}}. The envelope, independent of the chosen parametrization, separates the domain of unstable modes from the stability zone. It consists of two distinct symmetric curves that can intersect at $Gr = 0$ and $Re=Re_0$, as shown in figure~\textcolor{red}{\ref{Figure6}a}, where
\be{re0new}
Re_0 = \frac{3\sqrt{3}k_r^2}{2\Omega}\frac{1}{\sqrt{4(Ro_R-Ro)(1-\gamma\Theta)}}
\ee
and  $Ro_R$ is the modified Rayleigh line for non-isothermal flows:
\be{mur}
 Ro_R = -1+\frac{rD\Theta\gamma Pr }{4\left(1{-}\gamma \Theta\right)}.
\ee
\end{itemize}

In the isothermal case $(\Theta \equiv 0)$, $Ro_R=-1$, and the expression \rf{re0new} reduces to (\textcolor{red}{\ref{re0i}a}).  Moreover, rewriting Eq.~\rf{re0new} in terms of the Taylor number $Ta = \frac{2Re \Omega}{3\sqrt{3}k_r^2}$ exactly reproduces the result obtained in \citep{KM2017} for circular Couette flow with a radial temperature gradient.

At the geometric mean radius \rf{mgr} the modified Rayleigh line \rf{mur} takes a convenient form for Couette-Taylor applications:
\be{murl}
\mu_R=\eta^2\frac{2(\gamma - 2)\ln\eta + (1-\frac{1}{\eta})\gamma Pr}{2(\gamma - 2)\ln\eta + (1 - \eta)\gamma Pr}.
\ee
This relation shows that due to the temperature gradient, $\mu_R$ can substantially deviate from the isothermal Rayleigh line $\mu={\eta}^2$.

\begin{itemize}
\item \textcolor{black}{\uline {Rayleigh-unstable flow and intersection of envelope branches at $Gr=0$}}.
In the Rayleigh-unstable regime ($Ro < Ro_R$ or $\mu<\mu_R$), the envelope in the $(Gr,Re)$-plane can be approximated by its tangent lines computed at the intersection point $(Gr,Re)=(0,Re_0)$:
\ba{spli}
\frac{Re}{Re_0} &=& 1 \pm \sqrt{2} \frac{2D\Theta Pr Ro - 3\gamma \Theta DW   k_r^2}{27 Ro k_r^4} Gr,
\ea
where $Re_0$ is given by \rf{re0new}.
This result demonstrates that the temperature gradient plays a pivotal role in the existence of two distinct branches of the envelope for non-isothermal flows. These branches merge into a single curve in the case of isothermal Rayleigh-unstable flows, as established in the previous Section, see figures~\ref{fig:2abcd} and \ref{fig:3abcd}.
\end{itemize}

In the absence of vertical gravity, as in the case of spiral Poiseuille flow with a radial temperature gradient (SPFRT) considered by \cite{CM2005}, the term \(2D\Theta Pr Ro\) is absent in \rf{spli}, and the splitting is primarily governed by the parameter \(\gamma\), which is often very small in practical \textcolor{black}{situations} \citep{K2025}. As shown in \citep{K2025}, when \(|\gamma| \ll Pr\), the envelope splitting becomes negligible, allowing for a zeroth-order approximation in \(\gamma\), \textcolor{black}{that yields a smooth curve closely matching} the exact envelope.

The condition \( a_0 < 0 \) defines the instability regions for each \( k_z \) (or \( m \)). The union of these regions for \( k_r = 4\sqrt{2} \) and \( k_z = \pm 0.2 \), \( \pm 0.8 \), \( \pm 4 \), and \( \pm 7.3 \) is represented as the green shaded area in figures~\textcolor{red}{\ref{Figure6}a} and \textcolor{red}{\ref{Figure6}b}. The boundaries of the individual instability domains, corresponding to the terminal values of the axial wavenumber, touch the envelope at \( Gr = 0 \) and \( Re = Re_0 \). These terminal values are \( k_z^{\rm term} = 4 \) (dashed) and \( -k_z^{\rm term}=-4 \) (solid), as determined by the equation \rf{kzt0}, which follows directly from Eq.~\rf{polpra0} evaluated at \( Gr = 0 \) and \( Re = Re_0 \), where $Re_0$ is defined by Eq.~\rf{re0new}.

For  $|k_z|>k_z^{\rm term}$ the instability regions lie entirely within the green shaded area, as shown by dashed lines for \( k_z = 7.3 \), and by solid lines for \( k_z = -7.3 \) in figure~\textcolor{red}{\ref{Figure6}a}. In figure~\textcolor{red}{\ref{Figure6}c}, the dashed and solid curves mark the boundaries of the individual instability domains (which union is shown as a shaded green area) for \( k_z = 1\) and \( 3.5\), and \( k_z = -1\) and \( -3.5\), respectively.

While intersecting for $\mu<\mu_R$, the envelope branches have vertical asymptotes at the modified Rayleigh line \rf{murl}, delimiting the zone where no instability modes can be obtained, figures~\textcolor{red}{\ref{Figure6}b,e}.

\begin{itemize}
\item \textcolor{black}{\uline {Rayleigh-stable flow and unified LELS-GSF centrifugal instability criterion}}.
In the Rayleigh-stable regime ($Ro > Ro_R$ or $\mu>\mu_R$), the asymptotes to the envelope are inclined, placing the instability domains within the half-planes \citep{KM2024}
\be{elsscf1}
\frac{N_{\Omega}^2}{\Omega^2}(1-\gamma\Theta)+Pr\frac{N^2}{\Omega^2} <
\frac{\left( \overline{DW}(2-\gamma\Theta)-\frac{Pr  D\Theta}{2k_r^2\Omega  S}\right)^2}{1 + \overline{DW}^2},
\ee
defined by the asymptotes to the external branch, shown by oblique solid lines in figure~\textcolor{red}{\ref{Figure6}c,f}:
\be{masterels2}
Re  =
\pm\frac{DW}{2\Omega Ro} \sqrt{\frac{\Omega^2\left(2-\gamma\Theta-\frac{Pr  D\Theta Ro}{k_r^2DW}\right)^2-N_{\Omega}^2(1-\gamma\Theta )-PrN^2}
{N_{\Omega}^2(1-\gamma\Theta )+PrN^2}}Gr.
\ee
Here $N_{\Omega}^2$ is the Rayleigh discriminant \rf{rd}, $S=\frac{Re}{Gr}$, and
\be{bv}
N^2= -\gamma \Omega^2 r D\Theta
\ee
is the square of the centrifugal Brunt-V\"{a}is\"{a}l\"{a} frequency \citep{KM2017}.
\end{itemize}

The inequality \rf{elsscf1} yields the inviscid LELS criterion \rf{lels}
in the isothermal case \textcolor{black}{($\gamma=0$, $D\Theta=0$)} \citep{LS1983}, and the GSF criterion
\be{gsf}
N_{\Omega}^2(1-\gamma\Theta)+Pr N^2 < 0
\ee
in the limit of azimuthal flow $S\rightarrow \infty$ \citep{KM2017,M2013,JT2023}.
Thus, \rf{elsscf1} presents a novel unified \textcolor{black}{LELS-GSF} centrifugal instability criterion for viscous and thermodiffusive swirling flows, incorporating the effects of Prandtl number ($Pr$) and temperature variation ($D\Theta$).

\textcolor{black}{Note that a similar criterion for an inviscid isothermal incompressible swirling flow with radially varying density follows from the result by \cite{ES1978} for compressible fluids, as shown by \cite{DPA2010}, see Appendix~\ref{AppCC} for detail.}

\begin{figure}
\centering
\includegraphics[width=\textwidth]{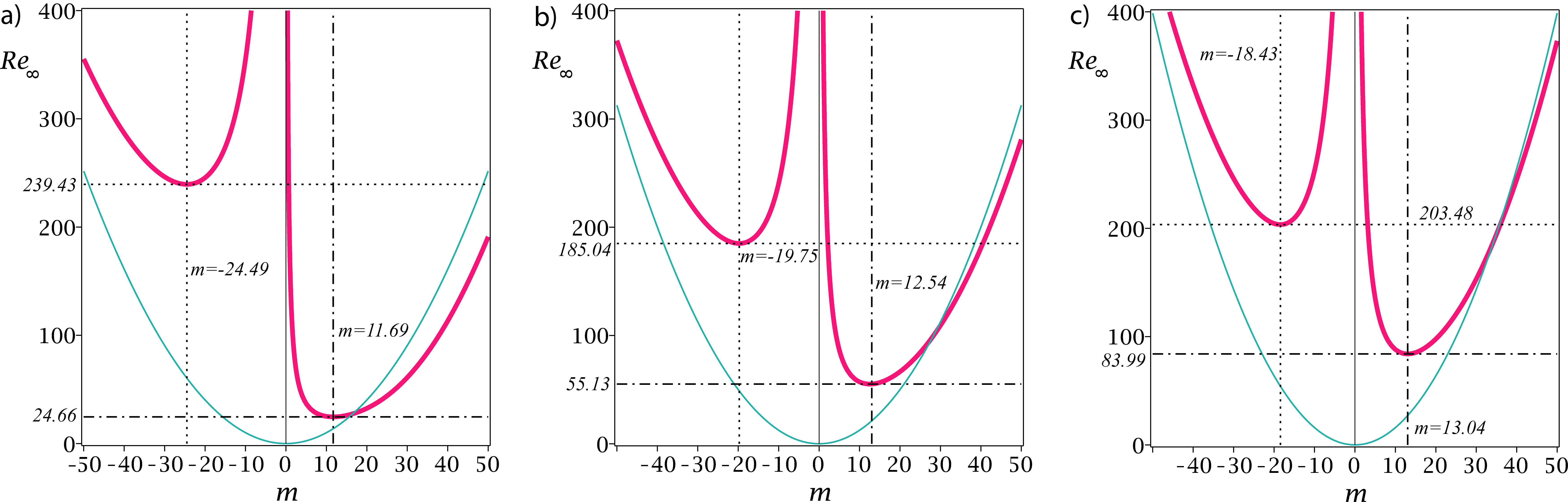}
\caption{For BCF with \( W(r) \) defined by \rf{WBCF} and \rf{C1}, and parameters \(\eta = 0.8\), \(Pr = 5.5\), \(\gamma = 0.0004\), and \(k_r = 4\sqrt{2}\), (red, thick) the asymptotic values of \(Re\) as \(|Gr| \to \infty\) are shown for the neutral stability curves parameterized by \(m\), based on Eq.~\rf{qerei}, for the following cases: (a) Rayleigh-unstable flow, \(\mu = 0\); (b) modified Rayleigh line, \(\mu = \mu_R \approx 0.63935\) (from \rf{murl}); and (c) Rayleigh-stable flow, \(\mu = 0.8\). The minimum of \(Re_{\infty}(m)\) for \(m < 0\) corresponds to the horizontal asymptote of the upper branch of the envelope, while the minimum for \(m > 0\) corresponds to the horizontal asymptote of the lower branch of the envelope, as illustrated in figure~\ref{Figure6}. The thin green parabola represents the approximation from \rf{lasf}.\label{Figure8a}}
\end{figure}

\textcolor{black}{For completeness, we also present the equation for an asymptote to the internal branch of the envelope in the Rayleigh-stable case:}
\be{inabcf}
\textcolor{black}{\frac{N_{\Omega}^2}{\Omega^2}(1-\gamma\Theta) +  Pr \frac{N^2}{\Omega^2} = 4(1-\gamma\Theta)\left(\overline{DW} - \frac{Pr D\Theta}{2k_r^2\Omega S}\right)\frac{\overline{DW}}{1+\overline{DW}^2}.
}
\ee

The  envelope in figure~\ref{Figure6} closely matches the critical states curve from both numerical linear stability analysis and experiments \citep{LG2008,YNM2013,G2015,Kang2015,Kang2023}, \textcolor{black}{e.g. cf our figure~\textcolor{red}{\ref{Figure6}a,d} and figure 1a in \citep{YNM2013}.} The envelope, unlike individual neutral stability curves, has a horizontal asymptote at \(Re = Re_{\infty}\) as $|Gr| \rightarrow \infty$. This explains the seemingly smooth and nearly \(Gr\)-independent stability boundary observed in \citep{LG2008,YNM2013,G2015,Kang2015,Kang2023} for $\mu=0$, although this boundary is actually piecewise smooth, with each neutral stability curve touching the common envelope, which flattens at large $|Gr|$ where shear instability dominates. Despite the Rayleigh-Fj\"ortoft shear instability mechanism due to an inflection point in the axial velocity profile \citep{D2002} leading to axisymmetric perturbations \citep{BMA2000,LP2007}, rotation ensures that the critical modes of the BCF are three-dimensional \textcolor{black}{with $k_z\ne 0$ and $m\ne 0$} \citep{D2005}.

\begin{itemize}
\item \textcolor{black}{\uline {Selection of unstable modes through envelope branches}}.
Figure~\ref{Figure6}, in conjunction with relation \rf{ec}, demonstrates how the envelope selects the critical modes. The lower left branch of the envelope corresponds to the left spiral modes (\(k_z > 0\), \(m > 0\)), while the lower right branch corresponds to the right spiral modes (\(k_z < 0\), \(m > 0\)). This is consistent with numerical findings by \cite{AW1990}. The modes with $m<0$ touch only the upper parts of the envelope and define the instability domains lying entirely within the green shaded area in figure~\ref{Figure6}, corresponding to the union of the instability regions with different values of $m$.
\end{itemize}

\begin{itemize}
\item \textcolor{black}{\uline {Parameterisation by $m$ and determination of horizontal asymptotes to envelope branches}}.
When parameterized by $m$, the neutral stability curves $a_0=0$ with $a_0$ given by Eq.~\rf{polpra0}, corresponding to different values of $m=k_{\varphi} r$ have horizontal (not depending on $Gr$) asymptotes in the $(Gr,Re)$-plane, see figure~\ref{Figure6}. The exact asymptotic values of the Reynolds number $Re_{\infty}(m)$ for a particular $m$ are determined by the quadratic equation, following from $a_0=0$ in the limit $|Gr| \rightarrow \infty$:
\ba{qerei}
q(m,Re_{\infty})&:=&4\Omega^2m^2r^4(1-\gamma\Theta)\left[(k_r^2r^2 + m^2)(1+ Ro_R-Ro)DW + r^2 Ro D\Theta Pr \right]Re_{\infty}^2 \nn\\
&-& 2\Omega m k_r r^3 (k_r^2 r^2 + m^2)\left[r^2 Ro D\Theta Pr  - \gamma \Theta DW (k_r^2r^2 + m^2) \right]Re_{\infty} \nn\\
&-& DW(k_r^2r^2 + m^2)^4=0.
\ea
In the isothermal case $(\gamma=0, D\Theta=0)$, $Ro_R=-1$, and \rf{qerei} reduces to the equation
\be{gerei1}
-DW(k_r^2r^2 + m^2)\left[4Ro \Omega^2 m^2r^4  Re_{\infty}^2 + (k_r^2r^2 + m^2)^3\right]=0,
\ee
which yields expression \rf{reterm} determining $Re_{\infty}(m)$ for the isothermal flows.
\end{itemize}

As \(m \to \infty\), the positive root of \(\rf{qerei}\) asymptotically behaves as:
\be{lasf}
Re_{\infty}(m)  \sim \frac{ m^2}{2\Omega r^2\sqrt{(1 + Ro_R-Ro)(1-\gamma\Theta)}}.
\ee
In contrast, as \(m \to 0\), it follows that \(Re_{\infty}(m) \sim m^{-1}\). This behavior indicates the existence of a minimum value of \(Re_{\infty}\) at some finite value of \(m\).

Solving simultaneously the equation $q(m,Re_{\infty})=0$ given by \rf{qerei} and the equation $\partial_m q(m,Re_{\infty})=0$, we can find the pairs of $m_{\min}$ and $Re_{\infty}(m_{\min})$ determining the minimizer and the minimal value of $Re_{\infty}$, which is the horizontal asymptote to the envelope in the non-isothermal case. For instance, choosing the parameters of figure~\ref{Figure6} corresponding to the BCF with $\eta=0.8$, $Pr=5.5$, $\gamma=0.0004$, and $k_r=4\sqrt{2}$, evaluating \rf{qerei} at the mean geometric radius \rf{mgr}, and applying the procedure just described, we find the pairs shown in figure~\ref{Figure8a}.

\subsubsection{Visco-thermodiffusive oscillatory McIntyre instability}

\begin{itemize}
\item \textcolor{black}{\uline {The critical Reynolds number and the modified Rayleigh line for purely azimuthal flow}}. In the absence of axial flow ($DW = 0$), we consider the equation \( a_1a_2 = a_0 \), derived from the stability condition (\ref{lsh}\textcolor{red}{c}), as a polynomial in \( k_z \). Computing its discriminant provides the envelope of the neutral stability curves for oscillatory instability in the \( (Pr, Re) \)-plane, expressed as \( Re = Re_0' \), where
\be{soi}
\textcolor{black}{
Re_0' =\frac{3\sqrt{3}k_r^2}{2\Omega} \frac{1+Pr }{ Pr} \frac{1}{\sqrt{4(Ro_R'-Ro)(1-\gamma\Theta)}}
}
\ee
and
\be{roro}
%
\textcolor{black}{
Ro_R'=-1 + \frac{1+Pr}{8Pr} \frac{\,r\,\gamma\, D\Theta}{ 1-\gamma\Theta}
}
\ee
defines the modified Rayleigh line. Notably, \( Ro_R' = Ro_R \) and \( Re_0' = 2Re_0 \) at \( Pr = 1 \).
\end{itemize}

Rewriting Eq.~\rf{soi} in terms of the Taylor number $ Ta = \frac{2Re \Omega}{3\sqrt{3}k_r^2} $ exactly reproduces the result of \cite{KM2017}, obtained for circular Couette flow with a radial temperature gradient.

At the geometric mean radius \rf{mgr}, the expression \rf{roro} takes a form well-suited for Couette-Taylor applications:
\be{murp}
\mu_R' = \eta^2\frac{4Pr(\gamma - 2)\ln \eta + (1-\frac{1}{\eta})\gamma(Pr + 1)}{4Pr (\gamma - 2)\ln \eta + (1 - \eta)\gamma(Pr + 1)}.
\ee
Similar to \rf{murl}, the presence of a temperature gradient can cause \( \mu_R' \) to deviate significantly from the isothermal Rayleigh line \( \mu={\eta}^2 \).

\begin{figure}
\centering
\includegraphics[width=0.95\textwidth]{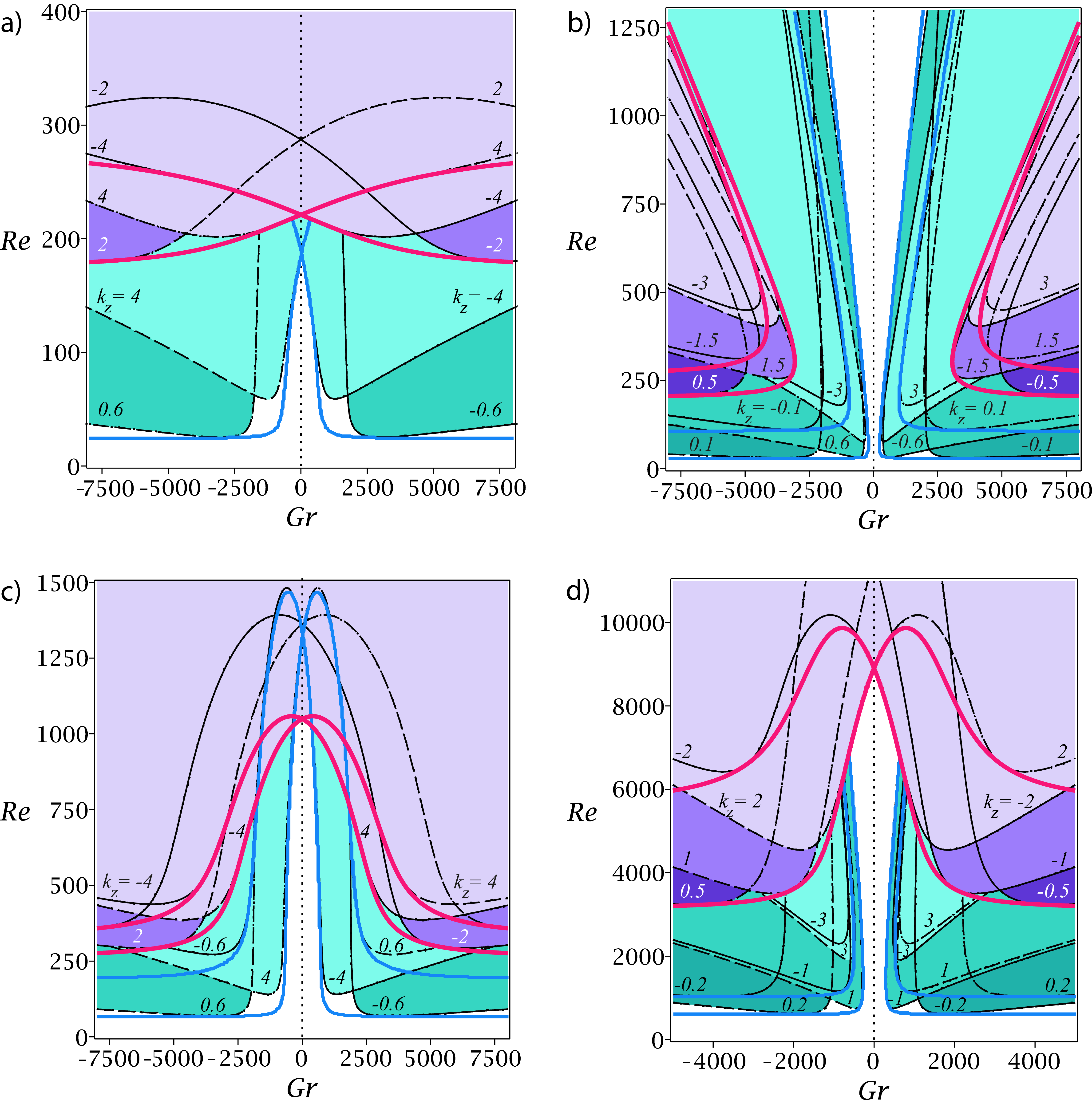}
\caption{ Green-shaded regions represent stationary (LELS-GSF) instability domains touching the thick blue envelope, while purple-shaded regions above them correspond to oscillatory (McIntyre) instability domains touching the thick red envelope. Dashed black lines indicate neutral stability curves for $k_z > 0$, and solid black lines indicate neutral stability curves for $k_z < 0$.
(a) Rayleigh-unstable BCF ($\eta = 0.8$, $\mu = 0$) with radial wavenumber $k_r = 4\sqrt{2}$, Prandtl number $Pr = 5.5$, outward heating ($\gamma = 0.0004$), and $k_z = \pm 0.6, \pm 2, \pm 4$;
(b) Rayleigh-stable BCF ($\eta = 0.8$, $\mu = 0.8$) with $k_r = 3\sqrt{2}$, $Pr = 5.5$, $\gamma = 0.01$, and $k_z = \pm 0.1, \pm 0.5, \pm 0.6, \pm 1.5, \pm 3$;
(c) Rayleigh-unstable BCF ($\eta = 0.8$, $\mu = 0.62$) with $k_r = 4\sqrt{2}$, $Pr = 4$, $\gamma = 0.01$, and $k_z = \pm 0.6, \pm 2, \pm 4$;
(d) Rayleigh-stable BCF with a quasi-Keplerian azimuthal velocity profile ($\eta = 0.99$, $\mu = \eta^{3/2}$), $k_r = 2\sqrt{2}$, $Pr = 0.5$, inward heating ($\gamma = -0.01$), and $k_z = \pm 0.2, \pm 0.5, \pm 1, \pm 2, \pm 3$.
\label{Figure8}}
\end{figure}

\begin{itemize}
\item \textcolor{black}{\uline {The envelope of oscillatory instability domains for visco-thermodiffusive swirling flows}}. For $ DW \neq 0 $, we plot the neutral stability curves $ a_1a_2 - a_0 = 0 $, which bound the oscillatory instability domains in the $ (Gr, Re) $-plane for different fixed values of $ k_z $. These we show alongside the neutral stability curves $ a_0 = 0 $, which bound the stationary instability domains in the same plane.
\end{itemize}

Figure~\ref{Figure8} presents the results of this computation for a range of Rayleigh-unstable and Rayleigh-stable baroclinic Couette flows (BCF). Both the stationary and oscillatory instability domains form families parameterized by $ k_z $, with their envelopes depicted as thick red curves (oscillatory instability) and thick blue curves (stationary instability). Each envelope consists of two branches,
which may intersect at $ Gr = 0 $ and $ Re = Re_0 $ for the stationary instability, and at $ Gr = 0 $ and $ Re = Re_0' $ for the oscillatory instability.
Here, $ Re_0 $ is given by \rf{re0new}, and $ Re_0' $ is defined by \rf{soi}. Recall that \rf{re0new} indicates that the intersection point
$ (Gr = 0, Re = Re_0) $ exists if $ Ro < Ro_R $ (or $ \mu < \mu_R $), where $ Ro_R $ (or $ \mu_R $) is determined from \rf{mur} (or \rf{murl}).

\textcolor{black}{Similarly, the intersection point \( (Gr = 0, Re = Re_0') \) of the branches forming the envelope of the oscillatory instability domains exists, as follows from \rf{soi}, if \( Ro < Ro_R' \) (or \( \mu < \mu_R' \)), where \( Ro_R' \) (or \( \mu_R' \)) is determined from \rf{roro} (or \rf{murp}). At this intersection point the linear approximation to the branches is given by the formula:
\be{splitf}
\frac{Re}{Re_0'} = 1 \pm \frac{\sqrt{2}}{243} \frac{\gamma^2 \left( Pr - 1 \right) Re_0'^2 \Omega^2 r\Theta D\Theta DW    + 9k_r^2 \left( 1+Pr  \right) \left( RoD\Theta  - 3 \gamma\Theta  k_r^2 DW \right)}{ Ro k_r^{6} \left(1+ Pr\right)^2 } PrGr.
\ee
Note that \rf{splitf} presents an analytical expression for the onset of a visco-diffusive McIntyre oscillatory instability in a visco-thermodiffusive swirling flow with a radial temperature gradient and natural gravity, valid for an arbitrary Prandtl number, which, to the best of our knowledge, has not previously been reported in the literature}.

By comparing the values of $ Re_0 $ and $ Re_0' $, we can predict which type of instability will define the critical $ Re $ near $ Gr = 0 $. For the Rayleigh-unstable flow with an outward heating ($\gamma>0$) in figure~\textcolor{red}{\ref{Figure8}a}, we find $ Re_0 \approx 187.14 $ and $ Re_0' \approx 221.10 $, indicating that the flow is more likely to experience stationary instability. This is confirmed by the domain of oscillatory instability, shown in purple in figure~\textcolor{red}{\ref{Figure8}a}, being contained entirely within the green-shaded domain of stationary instability. This result agrees with the numerical and experimental results by \cite{YNM2013} and \cite{G2015}, respectively.

In contrast, the Rayleigh-unstable BCF with an outward heating in figure~\textcolor{red}{\ref{Figure8}c} exhibits $ Re_0' \approx 1049.11 < Re_0 \approx 1326.27 $, indicating that oscillatory instability dominates over stationary instability within a finite range of $ Gr $ near $ Gr = 0 $. Notably, the red envelope of the oscillatory instability domains intersects with the blue envelope of the stationary instability domains, forming two codimension-2 points for
$Gr < 0$ and $Gr > 0$, as shown in figure~\textcolor{red}{\ref{Figure8}c}. This suggests that sufficiently large temperature gradient may favour the stationary modes.

For the Rayleigh-stable BCF with an outward heating in figure~\textcolor{red}{\ref{Figure8}b}, the stationary and oscillatory instability domains are distinctly separated. Specifically, for the parameters used in figure~\textcolor{red}{\ref{Figure8}b}, $ \mu = 0.8 > \mu_R' \approx 0.6381 $ ($ \mu = 0.8 > \mu_R \approx 0.6224 $), indicating that $ Ro > Ro_R' $ in \rf{soi} and $ Ro > Ro_R $ in \rf{re0new}. Consequently, the self-intersection points for both envelopes do not exist at $ Gr = 0 $. The domain of oscillatory instability lies entirely within the domain of stationary instability, with the latter being dominant for the BCF shown in figure~\textcolor{red}{\ref{Figure8}b}.

\begin{figure}
\centering
\includegraphics[width=0.95\textwidth]{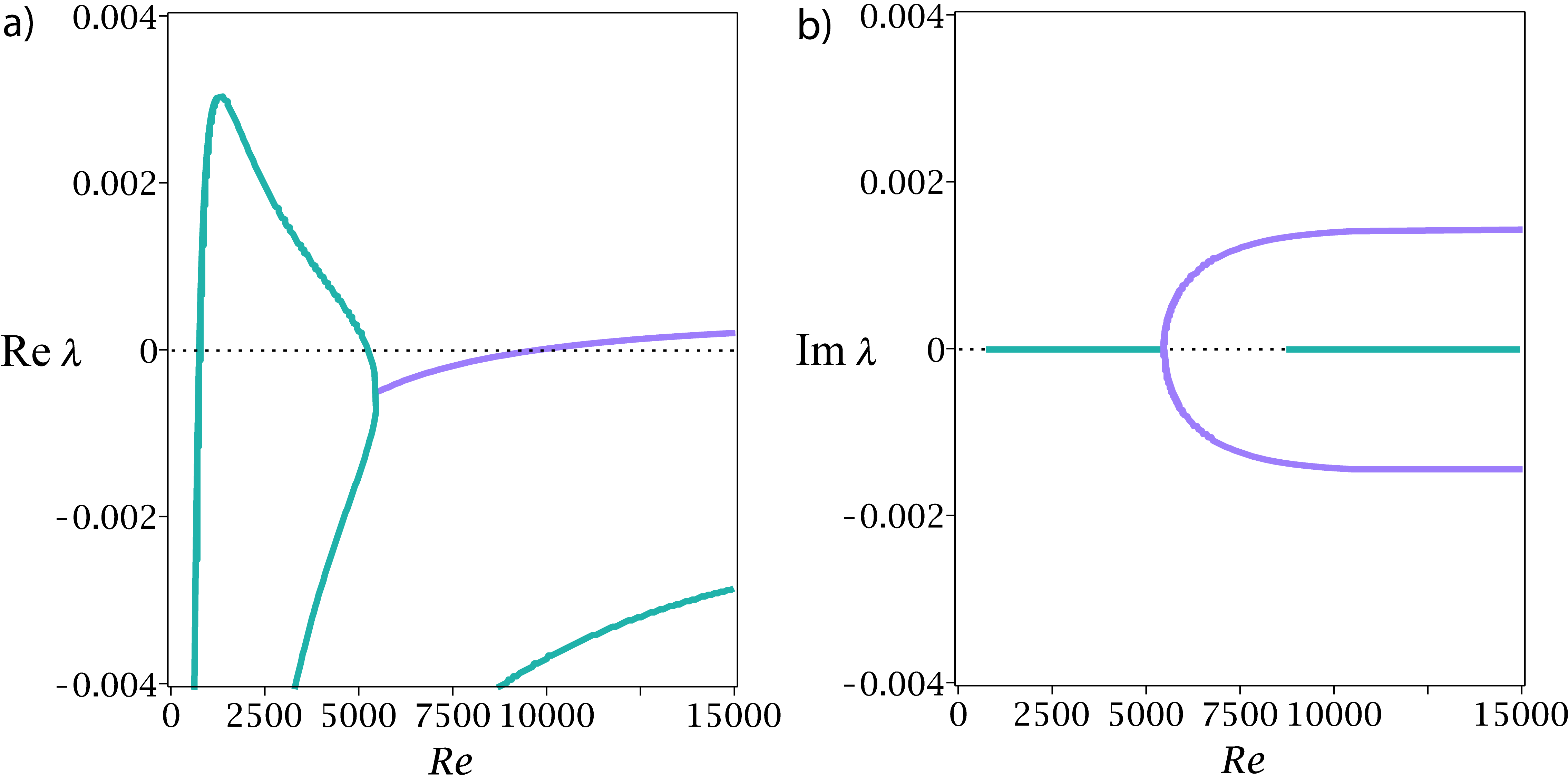}
\caption{ (a) Growth rates and (b) frequencies of (green) stationary and (purple) oscillatory McIntyre instabilities for the Rayleigh-stable BCF with a quasi-Keplerian azimuthal velocity profile ($\eta = 0.99$, $\mu = \eta^{3/2}$), $k_r = 2\sqrt{2}$, $Pr = 0.5$, inward heating ($\gamma = -0.01$), $Gr=500$, and $k_z =-1$.
\label{Figure9}}
\end{figure}

Finally, the Rayleigh-stable quasi-Keplerian BCF with inward heating ($ \gamma < 0 $) shown in figure~\textcolor{red}{\ref{Figure8}d} is characterized by the stationary instability domain being split into two parts, as $ \mu \approx 0.9850 > \mu_R \approx 0.9825 $. In contrast, the envelope of the oscillatory instability domains exhibits a self-intersection point at $ Gr = 0 $, as $ \mu \approx 0.9850 < \mu_R' \approx 0.9874 $. As a result, oscillatory instability dominates over stationary instability within a finite range near $ Gr = 0 $, and the envelopes of the oscillatory and stationary instability domains intersect, forming two codimension-2 points, as shown in figure~\textcolor{red}{\ref{Figure8}d}. This behavior qualitatively agrees with findings by \cite{CM2005} for spiral Poiseuille flow with a radial temperature gradient; see also \citep{K2025}.

\begin{figure}
\centering
\includegraphics[width=0.95\textwidth]{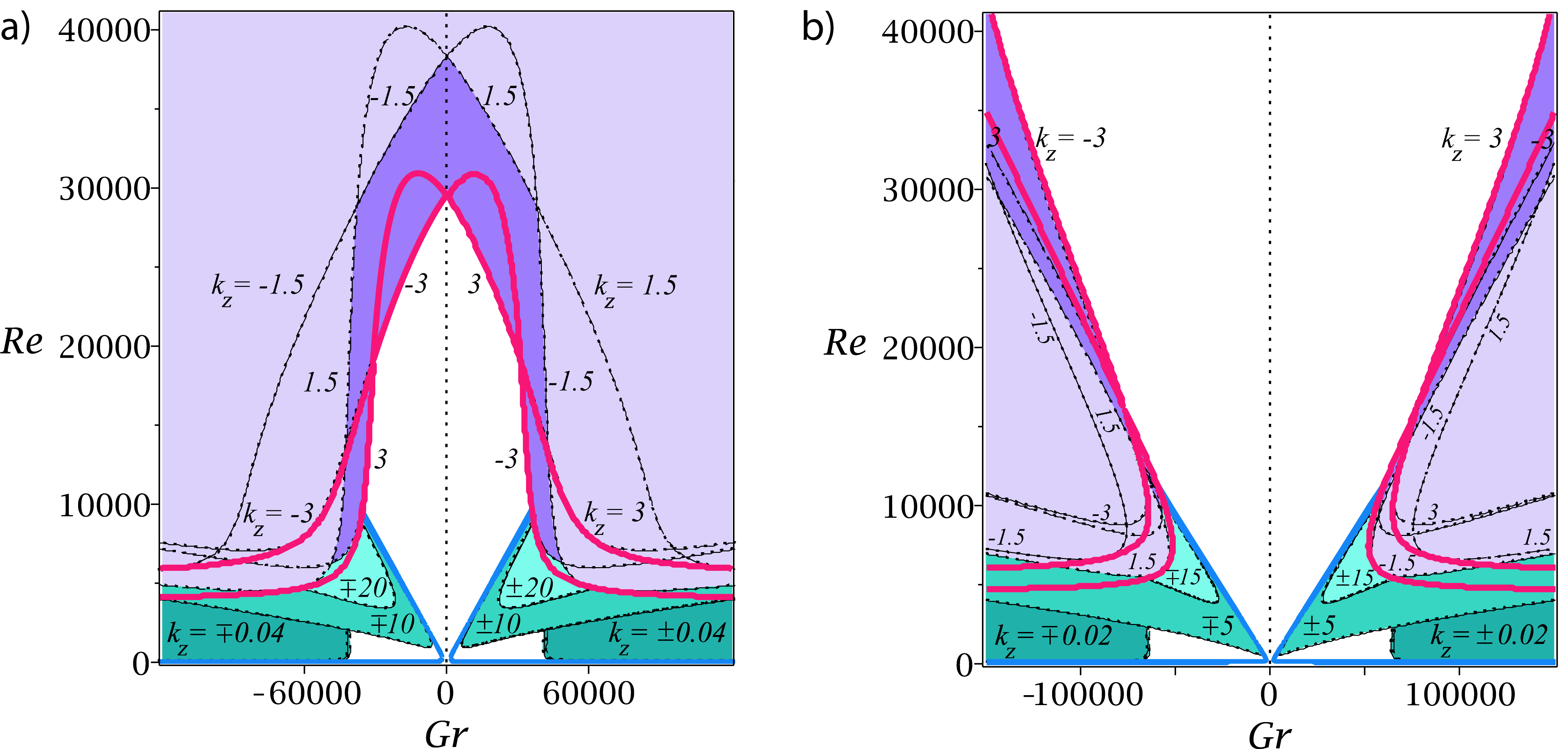}
\caption{Green-shaded regions represent stationary (LELS-GSF) instability domains touching the thick blue envelope, while purple-shaded regions above them correspond to oscillatory (McIntyre) instability domains touching the thick red envelope for the
Rayleigh-stable BCF with $\eta = 0.8$, $k_r = 3\sqrt{2}$, $Pr = 0.03$, inward heating ($\gamma = -0.01$), and (a) $\mu = 0.692$, $k_z = \pm 0.04, \pm 1.5, \pm 3, \pm 10, \pm 20$; and (b) $\mu = 0.8$ and $k_z = \pm 0.02, \pm 1.5, \pm 3, \pm 5, \pm 15$. Dashed black lines indicate neutral stability curves for $k_z > 0$, and solid black lines indicate neutral stability curves for $k_z < 0$.
\label{Figure10}}
\end{figure}

\textcolor{black}{Figure~\ref{Figure9} illustrates the behaviour of the growth rate ${\rm Re}\lambda$ and frequencies ${\rm Im}\lambda$ for the Rayleigh-stable BCF with a quasi-Keplerian azimuthal velocity profile ($\eta = 0.99$, $\mu = \eta^{3/2}$), $k_r = 2\sqrt{2}$, $Pr = 0.5$, inward heating ($\gamma = -0.01$), $Gr = 500$, and $k_z = -1$, corresponding to the stability map shown in figure~\textcolor{red}{\ref{Figure8}d}.}

\textcolor{black}{Figure~\textcolor{red}{\ref{Figure9}a} shows that, as the azimuthal Reynolds number increases, the green curve corresponding to ${\rm Im}\lambda = 0$ rises from negative values, crosses the line ${\rm Re}\lambda = 0$ at some critical $Re$, becomes positive (indicating instability), reaches a maximum, and then decreases again to negative values (restoring stability).}

\textcolor{black}{This interval of instability corresponds to the crossing of the green centrifugal instability domain in figure~\textcolor{red}{\ref{Figure8}d}, which has both lower and upper critical Reynolds numbers. According to \rf{eigp}, ${\rm Re} \lambda = \sigma$, i.e. the growth rate of the instability, and ${\rm Im}\lambda = 0$ corresponds to the frequency of the unstable mode given by $\omega = -m\Omega - \frac{k_zW}{S}$. Thus, the marginal mode is stationary in a reference frame attached to the helical streamlines around the annular axis \citep{KM2024}.}

\textcolor{black}{With a further increase in $Re$, the green curve in figure~\textcolor{red}{\ref{Figure9}a} attains increasingly negative values until it meets another branch, forming a double negative real root (${\rm Re}\lambda < 0$, ${\rm Im}\lambda = 0$). This root then splits into two complex-conjugate $\lambda$-roots with ${\rm Re}\lambda < 0$, which, upon a continued increase in $Re$, eventually intersect the line ${\rm Re}\lambda = 0$, as shown in figure~\textcolor{red}{\ref{Figure9}}.}

\textcolor{black}{Hence, the unstable mode acquires a non-trivial frequency, in contrast to the critical mode of the centrifugal instability, whose frequency is merely the Doppler shift. The growth rate and frequency of the complex-conjugate pair are shown as purple curves in figure~\textcolor{red}{\ref{Figure9}}, and it is evident that the oscillatory instability is weaker than the stationary one, as its growth rate is an order of magnitude smaller, which is typical for dissipation-induced instabilities \citep{M1970,LK2021,K2021}. The growth rate of the oscillatory instability remains positive as long as the system remains within the purple domain of oscillatory instability, see figure~\textcolor{red}{\ref{Figure8}d}.}

\textcolor{black}{As established in the earlier work by \cite{M1970}, the visco-diffusive oscillatory instability is significantly enhanced when the Prandtl number deviates from unity ($Pr \ne 1$). In figure~\ref{Figure10}, a reduction of the Prandtl number to $Pr = 0.03$ reveals a richer and more intricate geometry of the envelope delineating the neutral stability boundaries of oscillatory instability domains, shown in purple. The two thick red branches of this envelope, which define the critical values of the control parameters $Re$ and $Gr$, intersect not only at $Gr = 0$ and $Re = Re'_0$ as in figure~\textcolor{red}{\ref{Figure8}a,c}, but also at two additional nonzero values of $Gr$, symmetrically located with respect to the vertical axis.}

\textcolor{black}{This behavior is illustrated in figure~\textcolor{red}{\ref{Figure10}a}, corresponding to a Rayleigh-stable BCF configuration with $\eta = 0.8$, $k_r = 3\sqrt{2}$, $Pr = 0.03$, inward heating ($\gamma = -0.01$), and $\mu = 0.692$. Moreover, the thick red branches corresponding to oscillatory instability domains intersect the thick blue branches associated with stationary centrifugal instability domains depicted in greenish hues. As a result, the white stability domain enclosed by these envelopes takes on a distinctive diamond-like shape with five codimension-2 points—a phenomenon that, to the best of our knowledge, has not been previously reported in the literature, figure~\textcolor{red}{\ref{Figure10}a}.}

\textcolor{black}{Increasing $\mu$ from 0.692 to 0.8 in figure~\textcolor{red}{\ref{Figure10}b}, while keeping all other parameters identical to those in figure~\textcolor{red}{\ref{Figure10}a}, reveals that the two self-intersection points of the thick red branches—defining the envelope of the oscillatory instability domains—persist even though the central intersection at $Gr = 0$ and $Re = Re'_0$ disappears. In figure~\textcolor{red}{\ref{Figure10}b}, the two thick red branches exhibit distinct asymptotic directions as $Re \rightarrow \infty$. This indicates that one branch governs the stability boundary for lower values of $Re$ before the codimension-2 point, while the other dominates after the intersection, in the limit $Re \rightarrow \infty$. This asymptotic divergence complicates the formulation of an analytical stability criterion analogous to the LELS-GSF criterion for centrifugal instability \rf{elsscf1}, making the development of a corresponding theoretical framework significantly more challenging.}

\textcolor{black}{Indeed, the equation $a_1 a_2 = a_0$, derived from the stability condition (\ref{lsh}\textcolor{red}{c}) and governing the neutral stability curves that bound the domains of oscillatory instability, can be expressed as a polynomial in $k_z$:
\ba{pkz}
c_1k_z^3Re^3 &+& k_z^2\left(c_2k_z^4 + c_3k_z^2 + c_4\right)Re^2 + k_z\left(c_5k_z^6 + c_6k_z^4 + c_7k_z^2 + c_8\right)Re\nn\\
&+& c_9 k_z^{10} + c_{10}k_z^8 + c_{11} k_z^6 + c_{12}k_z^4 + c_{13}k_z^2 + c_{14}=0.
\ea}

\textcolor{black}{By computing the discriminant of the polynomial \rf{pkz}, we find that in the Rayleigh-stable regime, the asymptotic directions of the envelope bounding the domains of oscillatory instability in the $(Gr, Re)$-plane, as $Re \rightarrow \infty$, are determined by the roots of the equation in terms of the scaled parameter $S = \frac{Re}{Gr}$:
\be{eeoid}
6912c_9^3c_1^4c_2^7\left(27c_1^2c_{14}^2 - 18c_1c_4c_8c_{14} + 4c_1c_8^3 + 4c_4^3c_{14} - c_4^2c_8^2\right)=0,
\ee
where
\ba{coeefd}
c_1 &=& 2\overline{DW}\Omega^3Pr^2S\Theta\gamma^2k_r r \left[4\overline{DW}^2\Omega^2\left(1-\gamma\Theta\right) - \left(1+\overline{DW}^2\right)\left(N_{\Omega}^2(1-\gamma\Theta) + N^2 \right)\right],\nn\\
c_2 &=& -\Omega^2 Pr S \gamma r \left(1+\overline{DW}^2\right)^2\nn\\
&\times&\left[\left(1+\overline{DW}^2\right)\left(2 Pr(1-\gamma\Theta)N_{\Omega}^2 + (Pr + 1)N^2\right) - 8\overline{DW}^2\Omega^2 Pr(1-\gamma\Theta)\right],\nn\\
c_4 &=& -\Omega Pr k_r^2 \left\{\Omega S \gamma k_r^2\left(2Pr\left(1+\overline{DW}^2\right)\left(1-\gamma\Theta\right)N_{\Omega}^2\right.\right. \nn\\
&+& \left.\left.\left[4\overline{DW}^2\Omega^2\Theta^2\gamma^2 + (1+\overline{DW}^2)N^2\right]\left(Pr + 1\right) - 8\overline{DW}^2\Omega^2 Pr (1-\gamma\Theta)\right)r + 2N^2\overline{DW}Pr\right\},\nn\\
c_8 &=& -k_r^5\left(Pr + 1\right)\left(2S\Omega^3\Theta\gamma^2\overline{DW}k_r^2\left(3Pr + 1\right)r + N^2Pr\right),\nn\\
c_9 &=& -2S\left(Pr + 1\right)^2\left(1+\overline{DW}^2\right)^5 r\gamma\Omega^2,\nn\\
c_{14} &=& -2S\left(Pr + 1\right)^2 k_r^{10}r\gamma\Omega^2.
\ea}

\textcolor{black}{The prefactors in equation \rf{eeoid} yield the conditions $c_1 = 0$ and $c_2 = 0$, which correspond to possible asymptotic directions. These lead to the following expressions:}
\be{asl1}
\textcolor{black}{
\frac{N^2}{\Omega^2} +(1-\gamma\Theta)\left(\frac{N_{\Omega}^2}{\Omega^2}- \frac{4\overline{DW}^2}{1+\overline{DW}^2}\right) =0
}
\ee
and
\be{asl2}
\textcolor{black}{
\frac{Pr + 1}{2Pr}\frac{N^2}{\Omega^2}+(1-\gamma\Theta)\left(\frac{N_{\Omega}^2}{\Omega^2} - \frac{4\overline{DW}^2}{1+\overline{DW}^2}\right) =0,
}
\ee
\textcolor{black}{respectively. Additional possible asymptotic directions arise from the vanishing of the bracketed term in \rf{eeoid}. Which of these asymptotic directions is realized in a given regime depends on the specific parameter values entering the coefficients defined in \rf{coeefd}.}

\textcolor{black}{This complex structure simplifies considerably in the asymptotic limit $Pr \rightarrow 0$, relevant for astrophysical and geophysical contexts. In particular, such values of $Pr$ are typical in solar and stellar convection zones, where $Pr \sim 10^{-6}$, as well as in certain liquid metals, where $Pr \sim 10^{-3}$ \citep{Kapyla2021}. In this limit, the envelope of the oscillatory instability domains reduces to a single curve in the Rayleigh-unstable case, and to two symmetric, non-intersecting curves in the Rayleigh-stable case. Under these conditions, the bracketed term in \rf{eeoid} determines the unique asymptotic line as $Re \rightarrow \infty$, yielding the following criterion for the onset of the oscillatory McIntyre instability:
\be{pr0bcfoi}
\frac{N_{\Omega}^2}{\Omega^2}-\frac{4\overline{DW}^2}{1+\overline{DW}^2}<\frac{1+\overline{DW}^2}{16\overline{DW}^2(1-\gamma \Theta)\gamma^2\Theta^2}\left[\frac{N^2}{\Omega^2}-\frac{4\overline{DW}^2}{1+\overline{DW}^2}\gamma^2\Theta^2\right]^2.
\ee}

\section{Conclusion}\label{Sec6}

\textcolor{black}{Adapting the geometrical optics method to visco-thermodiffusive flows, we have developed a unified analytical framework for the local linear short-wavelength instabilities of both isothermal and non-isothermal swirling flows.}

\textcolor{black}{Within this framework, we showed that isothermal viscous swirling flows support only stationary centrifugal instabilities. We derived a closed-form expression for the growth rate and an explicit analytical stability criterion that generalizes the classical Ludwieg-Eckhoff-Leibovich-Stewartson  (LELS) criterion by incorporating viscosity. This result recovers known limits, including the Rayleigh and Eckhardt–Yao criteria, and provides a theoretical basis for the observed transition between Rayleigh-stable and Rayleigh-unstable regimes.}

\textcolor{black}{A key innovation of our work lies in identifying and analytically constructing the envelope of the neutral stability curves using the connection between envelopes and polynomial discriminants. This geometric insight enables precise determination of terminal wavenumbers and critical states relevant to both experiments and numerical simulations.}

\textcolor{black}{Extending the analysis to non-isothermal swirling flows, we uncovered the existence of both stationary instability—arising from a combination of the LELS and Goldreich-Schubert-Fricke (GSF) instabilities—and an oscillatory instability. The latter, a visco-thermodiffusive generalization of the McIntyre instability, is governed by a new analytical criterion unifying the effects of radial stratification, rotation, and diffusion. We further developed an algorithm to predict the dominant instability mode as a function of flow parameters, particularly the Prandtl number.}

\textcolor{black}{Our results align well with previous experimental and numerical studies of Spiral Couette, Spiral Poiseuille, and Baroclinic Couette flows, and provide a theoretical tool for guiding future investigations and experimental design.}



\backsection[Funding]{The work of I.M. was supported by the French Space Agency (CNES) and the ANR Programme
d'Investissements d'Avenir LABEX EMC$^3$ through the INFEMA project (Grant No. ANR-10
LABX-0-11) and through the GdR MFA, Grant No. 8103, project INTEHLDI. }

\backsection[Declaration of interests]{The authors report no conflict of interest.}


\backsection[Author ORCIDs]{O.N. Kirillov, https://orcid.org/0000-0002-6150-9308; I. Mutabazi,
https://orcid.org/0000-0001-9863-368X}


\appendix

\section{Isothermal and non-isothermal helical base states}\label{appA}

\begin{figure}
\centering
\includegraphics[width=.31\textwidth]{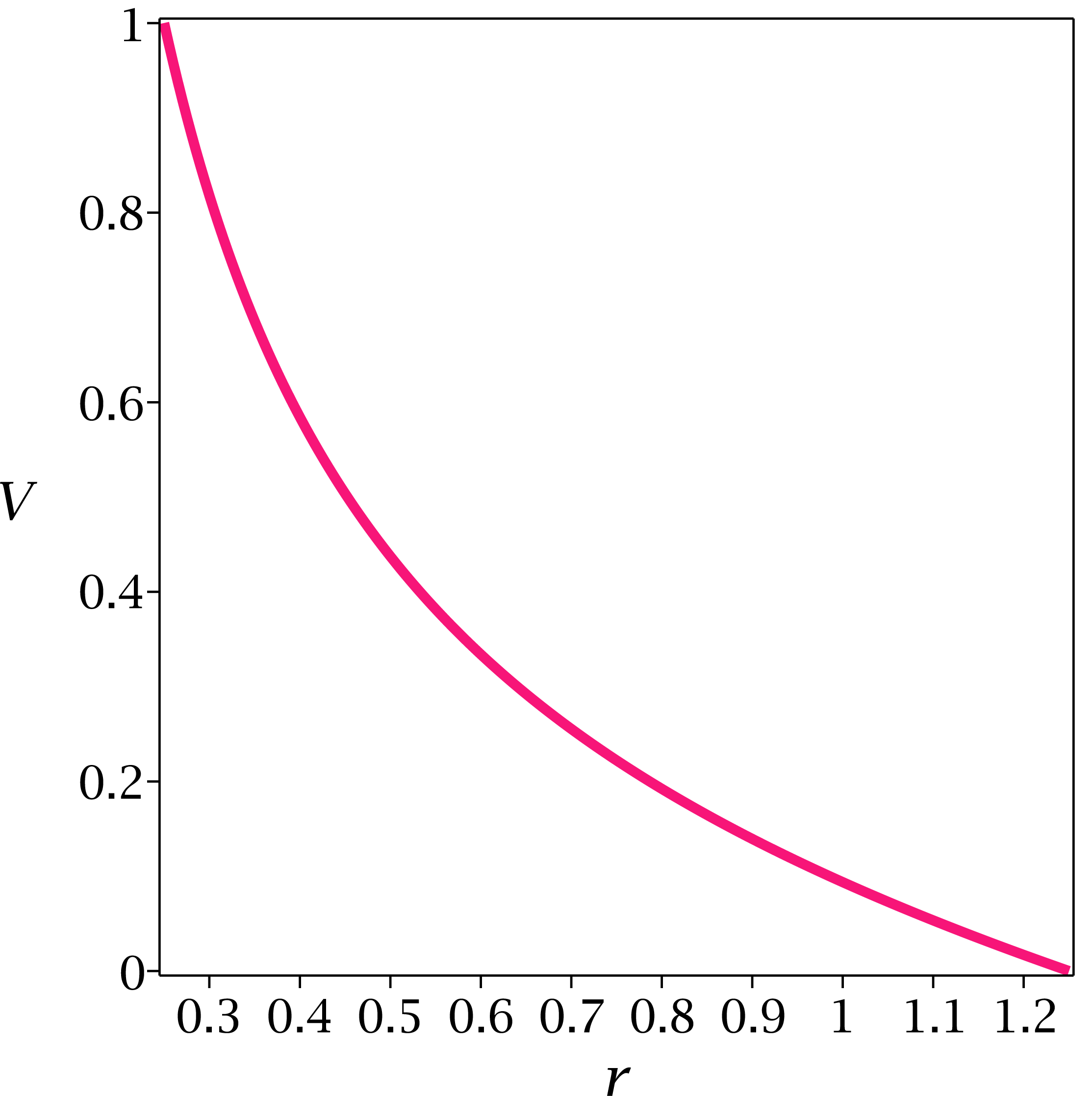}
\includegraphics[width=.31\textwidth]{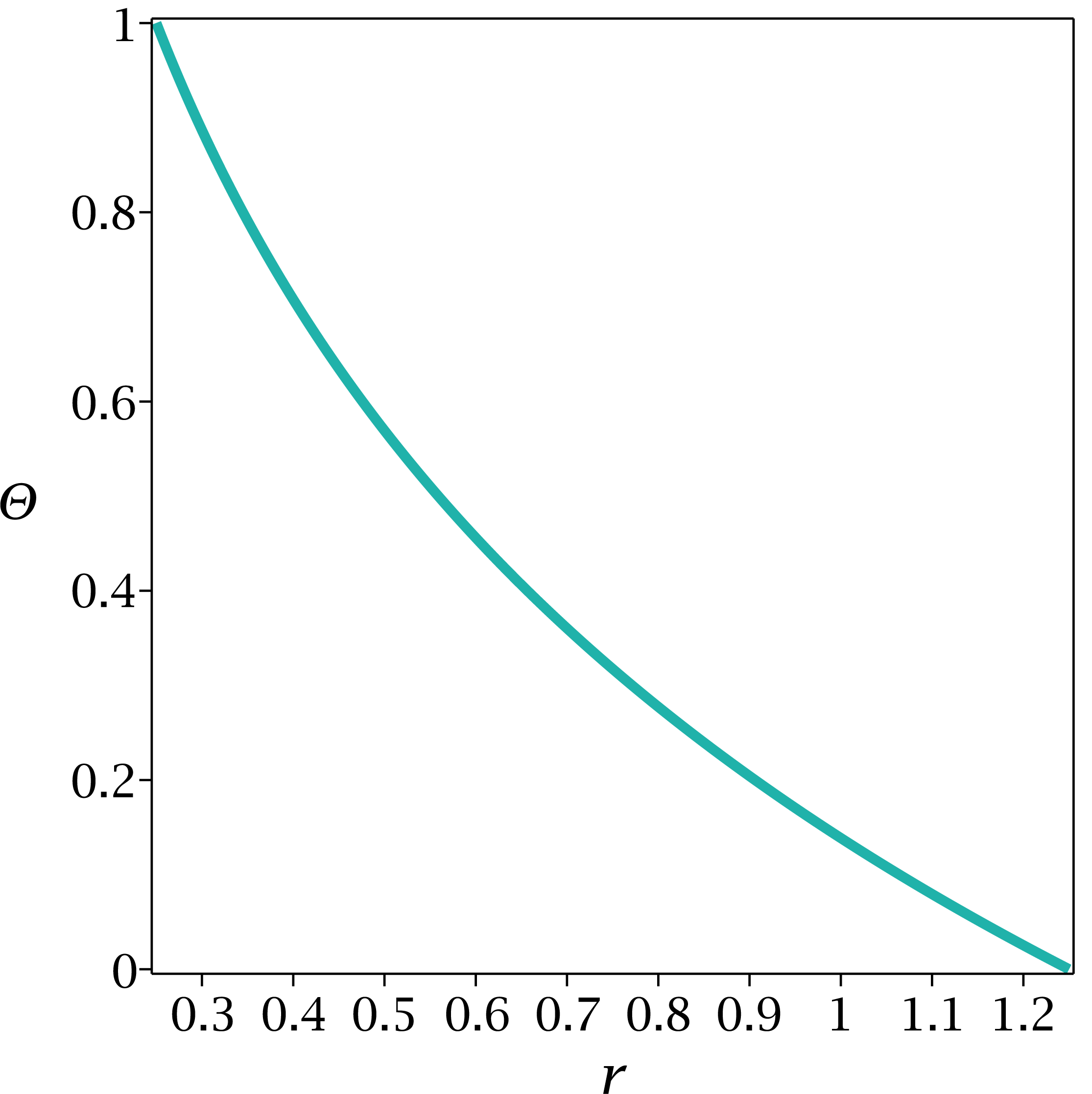}
\includegraphics[width=.305\textwidth]{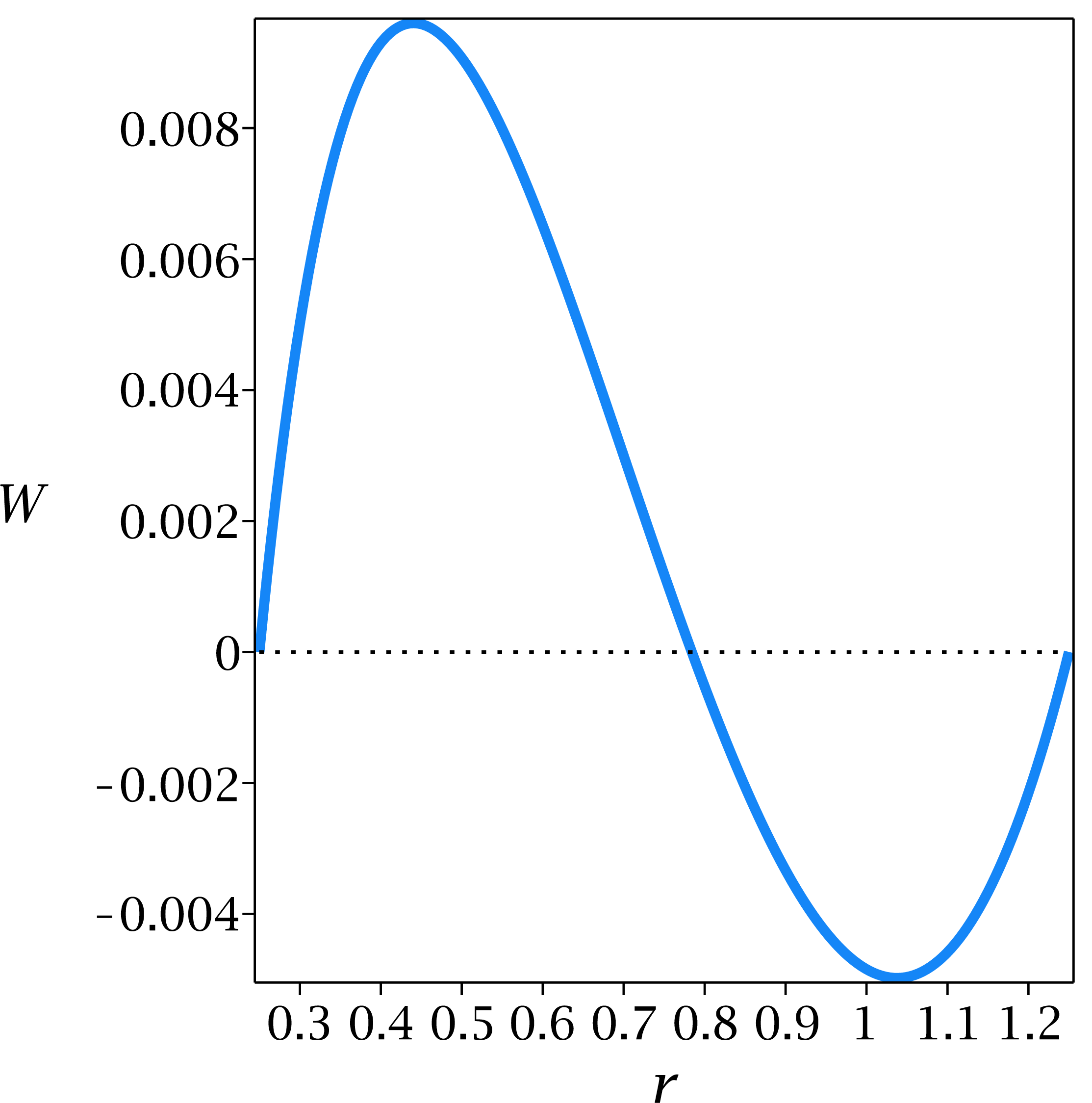}
\caption{Radial profiles of the base baroclinic Couette flow for a rotating inner cylinder ($\mu=0$ and $\eta=0.2$): (a) azimuthal velocity $V(r)$ as described by Eq.~\rf{ssv}, (b) temperature distribution \(\Theta(r)\) from Eq.~\rf{std}, and (c) axial velocity $W(r)$ as given by Eq.~\rf{WBCF}. \label{fig:bcf}}
\end{figure}

We seek the steady-state solution Eq.~\rf{eq3:bse} of the nonlinear governing equations \rf{eq1:nlge}.
From \rf{eq1:nlgeb}, it can be shown that the azimuthal velocity, $V(r)$, must satisfy the following second-order differential equation, subject to standard no-slip boundary conditions:
\be{ssev}
\frac{d^2 V}{d r^2} + \frac{1}{r}\frac{d V}{dr} - \frac{V}{r^2}=0, \quad V(r_1)=1, \quad V(r_2)=\frac{\mu}{\eta}.
\ee
Here, $r_1$ and $r_2$ are determined by the expressions $r_1=\frac{\eta}{1-\eta}$ and $r_2=\frac{1}{1-\eta}$, respectively, where the parameters $\eta$ and $\mu$ are defined according to Eq.~\rf{etamu}. By solving the boundary value problem \rf{ssev}, we obtain the classical Couette-Taylor velocity profile:
\be{ssv}
V(r)=\frac{\eta}{1+\eta}\left( \frac{1-\mu}{(1-\eta)^2}\frac{1}{r}-\frac{\eta^2 - \mu}{\eta^2}r\right),
\ee
see figure~\textcolor{red}{\ref{fig:bcf}a}. As in \citep{KM2017} we provide expressions for the angular velocity $\Omega=\frac{V}{r}$, the Rossby number and their product
\be{omro}
\Omega_g=\frac{1-\eta}{\eta}\frac{\eta+\mu}{1+\eta}, \quad Ro_g= -\frac{\eta}{\eta + \mu}\frac{1 - \mu}{1 - \eta}, \quad
\Omega_gRo_g=\frac{\mu-1}{\eta+1}
\ee
evaluated at the mean geometric radius
\be{mgr}
r_g=\sqrt{r_1r_2}=\frac{\sqrt{\eta}}{1-\eta}.
\ee

Similarly, the steady-state temperature distribution, $\Theta(r)$, is governed by a boundary value problem derived from Eq.~\rf{eq1:nlgec}. The corresponding differential equation, subject to boundary conditions, is given by:
\be{tbvp}
\frac{d^2 \Theta}{d r^2}+\frac{1}{r}\frac{d \Theta}{dr}=0,\quad \Theta(r_1)=1, \quad \Theta(r_2)=0.
\ee
Solving the boundary value problem \rf{tbvp}, we obtain the temperature distribution:
\be{std}
\Theta(r)= \frac{\ln[r(1-\eta)]}{\ln \eta},
\ee
see figure~\textcolor{red}{\ref{fig:bcf}b}.
At the mean geometric radius \rf{mgr} the temperature distribution and its radial derivative are:
\be{tdmgr}
\Theta_{g}=\frac{1}{2}, \quad D\Theta_{g}=\frac{1 - \eta}{\sqrt{\eta}\ln \eta}.
\ee

Finally, the axial velocity distribution, $W(r)$, is governed by the following differential equation, derived from Eq.~\rf{eq1:nlgeb}:
\ba{wbvp1}
\frac{d^2 W}{d r^2} +\frac{1}{r}\frac{d W}{dr}&=&SRe p_2 - \frac{W_T}{W_0} \Theta(r),
\ea
where $\Theta(r)$ is defined by Eq.~\rf{std} for non-isothermal base flows, and $\Theta(r)\equiv 0$ for isothermal flows.
For each specific helical base flow, we must further define the characteristic axial velocity $W_0$, along with the boundary conditions at $r_1$ and $r_2$.
In the subsequent sections, we will derive explicit expressions $W(r)$ for the baroclinic Couette flow (BCF), spiral Couette flow (SCF), and spiral Poiseuille flow (SPF).

\subsection{Baroclinic Couette flow (BCF)}

In this non-isothermal helical base flow, we assume that the characteristic axial velocity, $W_0$, is equal to the thermal velocity, $W_T$, as defined by Eq.~\rf{eq4:wtga}:
\be{w0wt}
W_0=W_T.
\ee
Substituting the temperature distribution $\Theta(r)$, given by Eq.~\rf{std}, into the governing equation \rf{wbvp1}, we can solve for the axial velocity, subject to the boundary conditions:
\be{BCBCF}
W(r_1)=0, \quad W(r_2)=0.
\ee
The resulting axial velocity $W(r)$ depends on an unknown constant pressure gradient $p_2$, which is determined by imposing the zero axial mass flux condition:
\be{fc2}
\int_{r_1}^{r_2} rW(r)dr=0.
\ee
Solving this yields the expression for the axial velocity $W(r)$ as found by \citet{AW1990}:
\ba{WBCF}
W(r)&=& C_1\left[(r_2^2-r_1^2)\frac{\ln(r/r_2)}{\ln \eta} + r^2 - r_2^2\right]-(r^2 - r_1^2)\frac{\ln(r/r_2)}{4\ln\eta},
\ea
where the constant $C_1$ is given by:
\ba{C1}
C_1&=&\frac{ (1-3\eta^2)(1-\eta^2)-4\eta^4\ln \eta }{16( (1-\eta^2)^2+(1-\eta^4)\ln \eta)},
\ea
see figure~\textcolor{red}{\ref{fig:bcf}c}.
Evaluating the axial velocity \rf{WBCF} and its radial derivative at the mean geometric radius \rf{mgr}, yields:
\ba{wgm}
&W_g = \frac{4\eta(\eta^2 + \eta + 1)\ln \eta - (\eta^2 + 4\eta + 1)(\eta^2 - 1)}{32(\eta^2 - 1)(\eta^2\ln\eta - \eta^2 + \ln\eta + 1)},&\nn\\
&DW_g =\frac{ -4\eta(\eta^4 + 1)(\ln \eta)^2 + 2\eta(\eta^2 - 1)(3\eta^2 - 2\eta + 3)\ln \eta + (\eta^2 - 4\eta + 1)(1-\eta^2)^2}{16\sqrt{\eta}(1 - \eta)^2(1 + \eta)((\eta^2 + 1)\ln \eta - \eta^2 + 1)\ln \eta}.&
\ea

\subsection{Spiral Couette flow (SCF)}

\begin{figure}
\centering
\includegraphics[width=.4\textwidth]{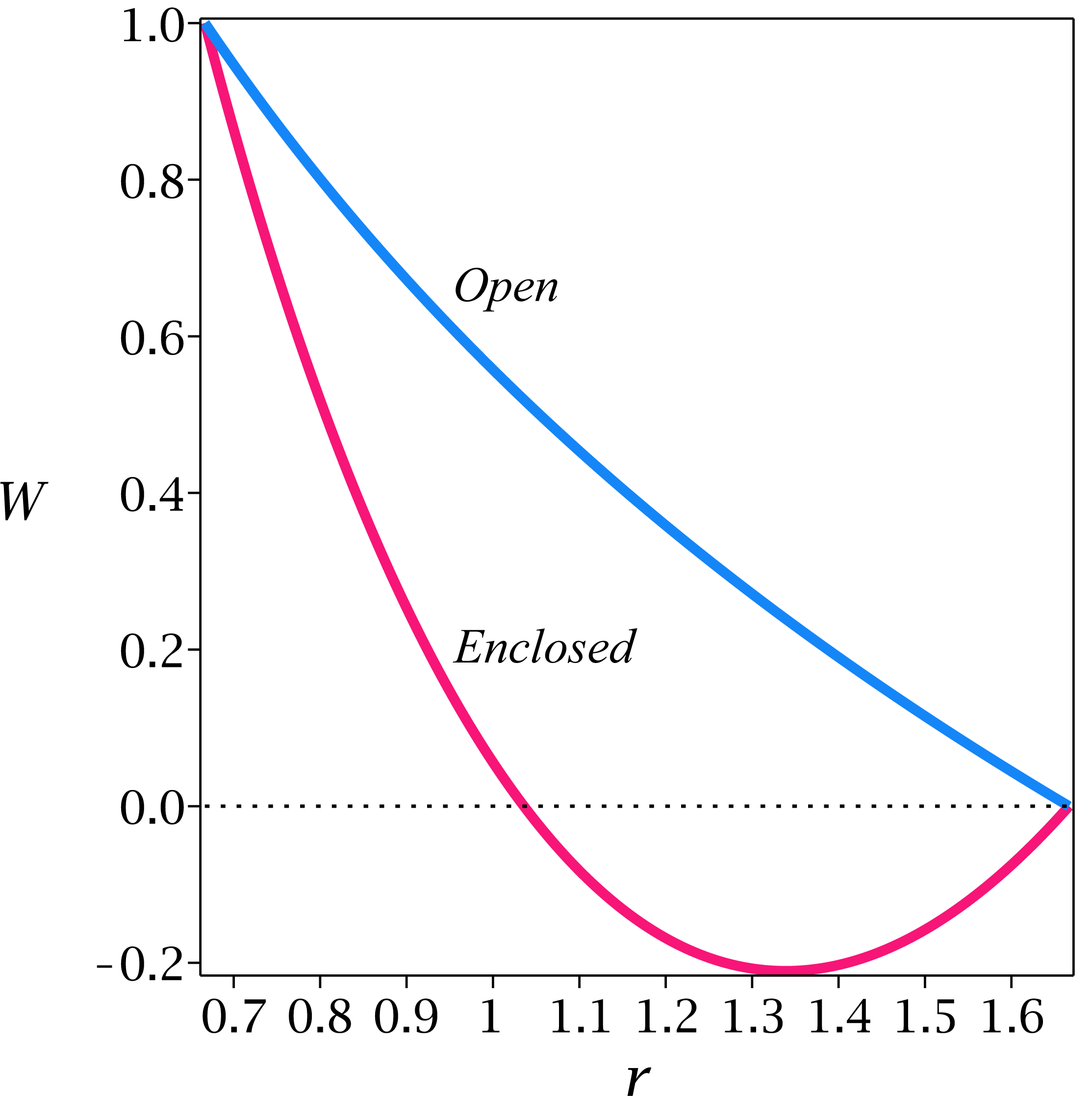}
\includegraphics[width=.4\textwidth]{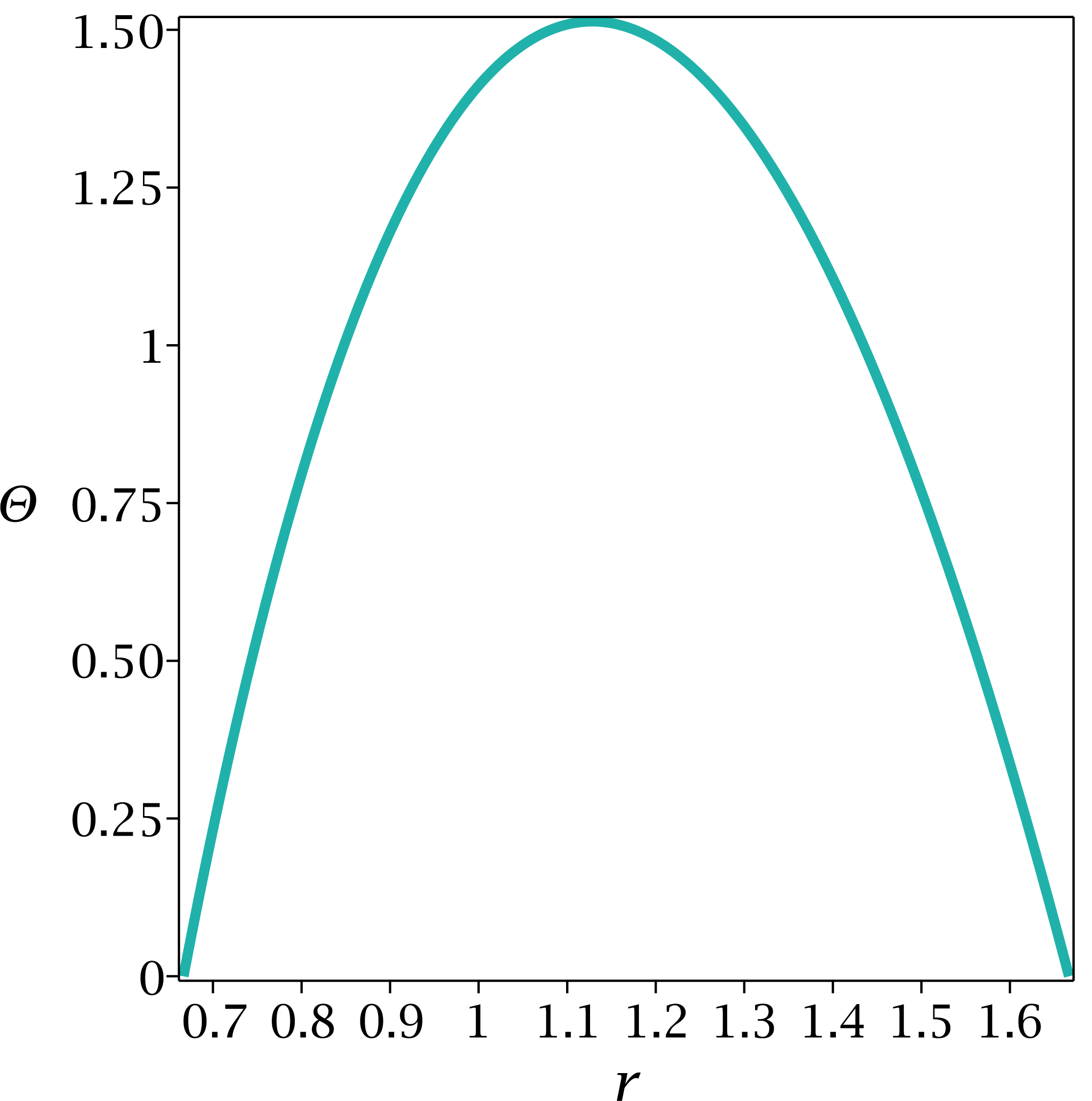}
\caption{Radial profiles of the axial velocity $W(r)$: (a) for open and enclosed spiral Couette flow (SCF) as described by Eq.~\rf{wscf}, and (b) for spiral Poiseuille flow (SPF) as described by Eq.~\rf{spf1}. \label{fig:spcf}}
\end{figure}

In the absence of gravity and a radial temperature gradient, the axial sliding of the inner cylinder generates an isothermal spiral Couette flow. In this scenario, it is reasonable to assume that the characteristic axial velocity $W_0$ equals the velocity of the inner cylinder, $W_1$:
\be{w0w1}
W_0=W_1.
\ee
By substituting the temperature distribution $\Theta(r)\equiv0$ into the governing equation \rf{wbvp1}, we can solve for the axial velocity $W(r)$, subject to the boundary conditions:
\be{BCSCF}
W(r_1)=1, \quad W(r_2)=0.
\ee
The resulting axial velocity $W(r)$ depends on an unknown constant pressure gradient $p_2$, which is determined by enforcing the zero axial mass flux condition \rf{fc2}. This leads to the following expression for the axial velocity:
\be{wscf}
W(r) =   \frac{1+C_2(1+\eta)}{\ln \eta}\ln\left(\frac{r}{r_2}\right)+C_2(1-\eta)(r^2 - r_2^2),
\ee
where the constant $C_2$ is given by:
\be{C2e}
C_2=-\frac{2\eta^2\ln\eta + 1 - \eta^2}{(1+\eta)[(1+\eta^2) \ln \eta  + 1 - \eta^2]}
\ee
for the enclosed SCF flow \citep{AW1993,MM2000}, and
\be{C2o}
C_2=0
\ee
for the open SCF flow \citep{MM2000}, see figure~\textcolor{red}{\ref{fig:spcf}a}. Evaluating the axial velocity \rf{wscf} and
its radial derivative at the mean geometric radius \rf{mgr}, yields:
\be{wgscfe}
W_g=\frac{(1-\eta^3 + 3\eta^2 + \eta)\ln\eta + 2(1-\eta^2)}{2[(1+\eta^2)\ln\eta+1 - \eta^2](1+\eta)}
\ee
and
\be{dwgcfe}
DW_g=\frac{ (\eta^2 - 1)(\eta^2 + 2\eta - 1)-4\eta^3 \ln\eta}{[(1+\eta^2)\ln\eta +1- \eta^2](1+\eta)\sqrt{\eta}}
\ee
for the enclosed SCF. For the open SCF, we obtain:
\be{wgdwgscfo}
W_g=\frac{1}{2}, \quad DW_g=\frac{1 - \eta}{\sqrt{\eta}\ln\eta}.
\ee

\subsection{Spiral Poiseuille flow (SPF)}

In this isothermal helical base flow, the axial velocity is driven by an axial pressure gradient. Substituting the temperature distribution $\Theta(r)\equiv0$ into the governing equation \rf{wbvp1}, we can solve for the axial velocity \(W(r)\), subject to the boundary conditions \rf{BCBCF}:
\be{wspf1}
W(r) = \frac{Re S p_2 r_2}{4}\left[(1-\eta)(r^2 - r_2^2) + \frac{1+\eta}{\ln\eta}\ln\left(\frac{r}{r_2}\right)\right].
\ee
Introducing the mean velocity $W_m$ of the axial flow \citep{MM2002}
\be{mfspf}
\frac{W_m}{W_0}=\frac{2}{r_2^2-r_1^2}\int_{r_1}^{r_2} rW(r)dr=\frac{ReSp_2r_2}{4}\frac{\eta^2 - (\eta^2 + 1)\ln \eta - 1}{2(1 - \eta)\ln \eta}
\ee
we can determine $p_2$ in terms of $W_m$ and write the dimensionless axial velocity as
\be{spf}
W(r)=\frac{W_m}{W_0}\frac{2(1-\eta)^2\ln\eta }{\eta^2-(\eta^2+1)\ln\eta- 1 }\left[r^2 - r_2^2 + \frac{1}{\ln\eta}\frac{1+\eta}{1-\eta}\ln\left(\frac{r}{r_2}\right)\right].
\ee
Choosing the mean velocity $W_m$ as a characteristic axial velocity:
\be{wowm}
W_0=W_m,
\ee
we obtain the final expression for the axial velocity in spiral Poiseuille flow \citep{TJ1981, MM2002, MM2005, CP2004}:
\be{spf1}
W(r)=\frac{2(1-\eta)^2\ln\eta }{\eta^2-(\eta^2+1)\ln\eta- 1 }\left[r^2 - r_2^2 + \frac{1}{\ln\eta}\frac{1+\eta}{1-\eta}\ln\left(\frac{r}{r_2}\right)\right],
\ee
see figure~\textcolor{red}{\ref{fig:spcf}b}. Evaluating the axial velocity \rf{spf1} and its radial derivative at the mean geometric radius \rf{mgr}, yields:
\be{wgspf}
W_g=\frac{(1-\eta)^2\ln\eta}{(1+\eta^2)\ln\eta +1 - \eta^2}
\ee
and
\be{dwgspf}
DW_g=\frac{2(\eta^2-2\eta\ln\eta - 1)(1-\eta)}{[(1+\eta^2)\ln\eta +1- \eta^2 ]\sqrt{\eta}}.
\ee

\subsection{SPF and SCF with a radial temperature gradient (SPFRT and SCFRT)}

By neglecting gravity but retaining the temperature gradient in the governing equations \rf{eq1:nlge}, we can readily identify two additional non-isothermal helical base states. The spiral Poiseuille flow with a radial temperature gradient (SPFRT) has been studied previously, for instance, by \citet{CM2005} and recently revisited by \cite{K2025}. In this flow, the azimuthal velocity is described by \rf{ssv}, the temperature distribution by \rf{std}, the axial velocity by \rf{spf1}, and the characteristic axial velocity by \rf{wowm}. Likewise, the spiral Couette flow with a radial temperature gradient (SCFRT) shares the same azimuthal velocity and temperature distributions. However, it differs in its axial velocity distribution, which is described by \rf{wscf}, and in its characteristic axial velocity, given by \rf{w0w1}.

\section{Connection to \citet{KM2017}}\label{AppB}
In the particular case when $S\rightarrow \infty$, corresponding to a pure azimuthal flow with a radial temperature gradient in the zero-gravity environment, we have $Ri=\frac{W_T}{W_0}\frac{1}{S\widehat{Re}} \rightarrow 0$, $\frac{DW}{SRo} \rightarrow 0$, and $|\boldsymbol{ k}|^2=k_r^2+k_z^2$. With this, \rf{mahc} reduces to
\be{m17}
\mathcal{ H}=\left(
          \begin{array}{ccc}
            -\frac{|\boldsymbol{ k}|^2}{\widehat{Re}} & 2\Omega\frac{k_z^2}{|\boldsymbol{ k}|^2}(1-\gamma \Theta) & -r\Omega^2 \gamma\frac{k_z^2}{|\boldsymbol{ k}|^2}  \\
            -2\Omega\left(1+Ro \right) & -\frac{|\boldsymbol{k}|^2}{\widehat{Re}} & 0 \\
            -D\Theta  & 0 & \frac{|\boldsymbol{ k}|^2}{\widehat{Re}} \frac{Pr-1}{Pr}-\frac{|\boldsymbol{ k}|^2}{\widehat{Re}} \\
          \end{array}
        \right).
\ee

Let us show that \rf{m17} reproduces the result by \citet{KM2017}. First, notice from \rf{eigp} that in the limit $S \rightarrow \infty$ we have $\lambda=s +{\rm i}\Omega k_{\varphi} r$. Then, we write the eigenvalue problem $\mathcal{ H}\boldsymbol{ a}=\lambda \boldsymbol{ a}$ as $\mathcal{ H}_1 \boldsymbol{ a} =s \boldsymbol{ a}$, where
\be{m17a}
\mathcal{ H}_1=\left(
          \begin{array}{ccc}
            -{\rm i}\Omega k_{\varphi} r -\frac{|\boldsymbol{ k}|^2}{\widehat{Re}} & 2\Omega\frac{k_z^2}{|\boldsymbol{ k}|^2}(1-\gamma \Theta) & -r\Omega^2 \gamma\frac{k_z^2}{|\boldsymbol{ k}|^2}  \\
            -2\Omega\left(1+Ro \right) & -{\rm i}\Omega k_{\varphi} r -\frac{|\boldsymbol{ k}|^2}{\widehat{Re}} & 0 \\
            -D\Theta  & 0 & -{\rm i}\Omega k_{\varphi} r -\frac{|\boldsymbol{ k}|^2}{\widehat{Re}Pr} \\
          \end{array}
        \right).
\ee
Taking into account that $D\Theta=\frac{2\Theta Rt}{r}$, $Rt=\frac{rD\Theta}{2\Theta}$, denoting $\beta=k_z/|\boldsymbol{ k}|$, and introducing the Taylor number $Ta=\widehat{Re} \Omega \beta/|\boldsymbol{ k}|^2$, we write \rf{m17a} as
\be{m17b}
\mathcal{ H}_1=\beta \Omega\left(
          \begin{array}{ccc}
            -{\rm i}\frac{k_{\varphi} r}{\beta} -\frac{1}{Ta} & 2\beta(1-\gamma \Theta) & -r\Omega \gamma\beta  \\
            -\frac{2}{\beta}\left(1+Ro \right) & -{\rm i} \frac{k_{\varphi} r}{\beta} -\frac{1}{Ta} & 0 \\
            -2\Theta Rt \frac{1}{r\beta \Omega}  & 0 & -{\rm i} \frac{k_{\varphi} r}{\beta} -\frac{1}{TaPr} \\
          \end{array}
        \right).
\ee
Finally, denoting $n=\frac{k_{\varphi} r}{\beta}$ and introducing the matrix $\mathcal{ R}={\rm diag}(1,1,1/r)$, we get:
\be{m17c}
\mathcal{ R}^{-1}\mathcal{ H}_1\mathcal{ R}=\beta \Omega\left(
          \begin{array}{ccc}
            -{\rm i}n -\frac{1}{Ta} & 2\beta(1-\gamma \Theta) & -\gamma\beta \Omega  \\
            -\frac{2}{\beta}\left(1+Ro \right) & -{\rm i} n -\frac{1}{Ta} & 0 \\
            -2\Theta Rt \frac{1}{\beta \Omega}  & 0 & -{\rm i}n -\frac{1}{TaPr} \\
          \end{array}
        \right),
\ee
which exactly reproduces Eq.~(4.9) in \citep{KM2017}.


\section{Connection to \cite{L1960}, \cite{EC1984}, and \cite{LS1983} criteria (LELS) }\label{AppBa}

\cite{L1960} developed an analytical narrow-gap theory for the stability of inviscid, incompressible SCF, which showed good agreement with his subsequent experiments \citep{L1964}. Within this framework, he derived the following stability criterion (Eq. (7) in \citep{L1960}):
\be{luc}
\frac{d V_{\varphi}}{d r} \frac{r}{V_{\varphi}} - \frac{\left(\frac{d V_{z}}{d r}\right)^{2} \left(\frac{r}{V_{\varphi}}\right)^{2}}{1 - \frac{d V_{\varphi}}{d r} \frac{r}{V_{\varphi}}} > -1,
\ee
where \(V_{\varphi}(r)\) and \(V_z(r)\) are the azimuthal and axial velocity components, respectively, in the \((r,\varphi,z)\) cylindrical frame, treated as arbitrary functions of the radial coordinate \(r\).

Using our notation:
\be{luc2}
\frac{d V_{z}}{d r} = DW, \quad \frac{V_{\varphi}}{r} = \Omega, \quad \frac{d V_{\varphi}}{d r} = \Omega(2 Ro + 1),
\ee
we can rewrite \rf{luc} as:
\be{luc3}
2 Ro + 1 + 2 Ro \frac{DW^2}{4\Omega^2 Ro^2} > -1.
\ee

Introducing the normalized shear parameter:
\be{luc4}
\overline{DW} = \frac{DW}{2\Omega Ro},
\ee
we further transform \rf{luc3} into:
\be{luc5}
1 + Ro(1 + \overline{DW}^2) > 0.
\ee

Finally, expressing the Rossby number \(Ro\) in terms of the squared epicyclic frequency \(N_{\Omega}^2\) as defined in \rf{rd}, we reduce the stability condition \rf{luc5} to:
\be{luc6}
\frac{N_{\Omega}^2}{\Omega^2} - \frac{4\overline{DW}^2}{1 + \overline{DW}^2} > 0,
\ee
which is simply the reversed inequality \rf{lels}, corresponding to the inviscid and incompressible LELS instability criterion.

\cite{ES1978} extended the instability criterion \rf{lels} to inviscid and compressible flows using geometric optics stability analysis. Later, \cite{EC1984} derived an elegant formulation of this criterion in the incompressible limit:
\be{echoff}
\frac{V}{r} \left( DV - \frac{V}{r} \right) \left( DV^2 + DW^2 - \frac{V^2}{r^2} \right) < 0,
\ee
where \(DV\) and \(DW\) are the radial derivatives of the azimuthal \((V(r))\) and axial \((W(r))\) velocity components, respectively.

Using \rf{luc2}, where we set \(V_{\varphi} = V\), the inequality \rf{echoff} simplifies to:
\be{echoff2}
2\Omega^2 Ro \left( 4\Omega^2 Ro (Ro + 1) + DW^2 \right) < 0,
\ee
which, when expressed in terms of \(\overline{DW}\) \rf{luc4} and \(N_{\Omega}^2\) \rf{rd}, precisely reproduces \rf{lels} and, with the reversed inequality, \rf{luc6}.

\cite{LS1983} derived a criterion for the instability of columnar vortices in an inviscid, incompressible fluid:
\be{lstew}
2V D\Omega \left[ D(rV) D\Omega + DW^2 \right] < 0,
\ee
where \( V = \Omega r \), \( D(rV) = \frac{d(rV)}{dr} \), and \( V(r) \) and \( W(r) \) are the azimuthal and axial velocity components, respectively, treated as arbitrary functions of the radial coordinate.

By computing \( D(rV) \) and \( D\Omega \) and expressing these quantities in terms of \( V \) in \rf{lstew}, we reduce this criterion to \rf{echoff}, and consequently, to \rf{lels}.

\textcolor{black}{\section{Connection to \cite{DPA2010}}\label{AppCC}}

\textcolor{black}{\cite{DPA2010} derived the following criterion for centrifugal instability in an inviscid swirling flow with radial density stratification but without mass diffusivity:
\be{dpa1}
G^{2}\left(W'^{2}+r^{2} \Omega'^{2}\right) -2 r \Omega \Omega^{\prime}\left(r \Omega^{\prime} \Theta+W'^{2}\right)>0,
\ee
where $r$ denotes the radial coordinate, $' = \frac{d}{dr}$, $W(r)$ is the axial velocity profile of the base flow, $V(r)$ its azimuthal velocity, $\Omega = \frac{V}{r}$ the angular velocity, $\Theta = V^{\prime} + \Omega$, and
\be{dpa2}
G^{2} = -\frac{V^{2}}{r} \frac{\rho_{b}^{\prime}}{\rho_{b}}
\ee
is the squared Brunt–V\"ais\"al\"a frequency, with $\rho_b(r)$ denoting the radial density profile of the base flow.}

\textcolor{black}{Using \rf{ro} and denoting $DW = \frac{d W}{d r}$, we transform \rf{dpa1} into
\be{dpa3}
G^{2}\left(D W^{2}+4 \Omega^{2} Ro^{2}\right) -4 \Omega^{2} Ro \left(4 \Omega^{2} Ro(Ro+1)+DW^2\right)>0.
\ee
Subsequently, by applying \rf{rd} and \rf{luc4}, we obtain from \rf{dpa3}:
\be{dpa4}
G^{2}\left(1+\overline{DW}^{2}\right) - \left(N_{\Omega}^{2} + 4 \Omega^{2} Ro \overline{DW}^{2}\right) > 0,
\ee
and further reduce it to
\be{dpa5}
G^{2}\left(1+\overline{DW}^{2}\right) - \left(N_{\Omega}^{2} + \left(N_{\Omega}^{2} - 4 \Omega^{2}\right)\overline{DW}^{2}\right) > 0.
\ee
Rearranging terms in \rf{dpa5}, we obtain:
\be{dpa6}
\left(G^{2} - N_{\Omega}^{2}\right)\left(1 + \overline{DW}^{2}\right) + 4 \Omega^{2} \overline{DW}^{2} > 0,
\ee
which finally leads to:
\be{dpa7}
\frac{N_{\Omega}^{2}}{\Omega^{2}} - \frac{G^{2}}{\Omega^{2}} - \frac{4 \overline{DW}^{2}}{1 + \overline{DW}^{2}} < 0.
\ee}

\textcolor{black}{In our study, we define the Brunt–V\"ais\"al\"a frequency as in \cite{KM2017}:
\be{dpa8}
N^{2} = \frac{d \rho}{d r} \frac{V^{2}}{r},
\ee
which, for $\rho(r) = 1 - \gamma \Theta(r)$, yields \rf{bv}. Thus, $G^{2}$ has the opposite sign to $N^{2}$, and in terms of $N$, equation \rf{dpa6} becomes:
\be{dpa9}
\frac{N_{\Omega}^{2}}{\Omega^{2}} + \frac{N^{2}}{\Omega^{2}} - \frac{4 \overline{DW}^{2}}{1 + \overline{DW}^{2}} < 0.
\ee}

\textcolor{black}{Comparing \rf{dpa9} with our LELS-GSF instability criterion \rf{elsscf1}, we find that both share the same structure, differing only by terms specific to thermal density stratification, viscosity, and diffusivity.}


\bibliographystyle{jfm}


\end{document}